\documentclass[aps]{revtex4}
\usepackage{graphicx}
\DeclareGraphicsExtensions{.jpg,.jpeg,.pdf,.png,.mps,.eps,.ps,.gif}
\usepackage{amsthm, amssymb,amsmath}
\usepackage{subfigure}
\usepackage{float}
\usepackage{amsfonts}
\usepackage{enumerate}

\newcommand{\bminil}[1]{\begin{minipage}[l]{#1 \textwidth}}
\newcommand{\bminir}[1]{\begin{minipage}[r]{#1 \textwidth}}
\newcommand{\bminic}[1]{\begin{minipage}[c]{#1 \textwidth}}
\newcommand{\emini}{\end{minipage}}
\newcommand{\EQL}{\begin{equation}\label}
\newcommand{\EQ}{\begin{equation}}
\newcommand{\EN}{\end{equation}}
\newcommand{\BFG}{\begin{figure}}
\newcommand{\EFG}{\end{figure}}
\newcommand{\ITM}{\begin{itemize}}
\newcommand{\ITN}{\end{itemize}}
\newcommand{\EEM}{\begin{enumerate}}
\newcommand{\EEN}{\end{enumerate}}
\newcommand{\BEA}{\[\begin{array}}
\newcommand{\EEA}{\end{array}\]}
\newcommand{\EQAL}{\begin{eqnarray}\label}
\newcommand{\EQA}{\begin{eqnarray}}
\newcommand{\ENA}{\end{eqnarray}}

\newcommand{\btriangle}{\mbox{\boldmath$\triangle$}}

\newcommand{\bS}{\mbox{\boldmath$S$}}

\newcommand{\br}{\mbox{\boldmath$r$}}

\newcommand{\bu}{\mbox{\boldmath$u$}}

\newcommand{\bx}{\mbox{\boldmath$x$}}

\newcommand{\bomega}{\mbox{\boldmath$\omega$}}

\newcommand{\bxi}{\mbox{\boldmath$\xi$}}

\newcommand{\ppto}[1]{\frac{\partial #1}{\partial t}}

\newcommand{\half}{\mbox{$\frac{1}{2}$}}

\newcommand{\quart}{\mbox{$\frac{1}{4}$}}

\usepackage{amsthm, amssymb,amsmath}
\newcommand{\oldon}[1]{}%(OFF)  % create something to save memory
\newcommand{\bpurp}[1]{}
\newcommand{\bcyan}[1]{}
\newcommand{\etall}{{\it et al.}}
\newcommand{\biband}{~and~}
\newcommand{\authone}[2]{#1 #2,}
\newcommand{\authtwo}[4]{#1 #2 \biband~#3 #4,}
 \newcommand{\auththr}[6]{#1 #2, #3 #4 \biband~#5 #6}
\newcommand{\authmanytwo}[4]{#1 #2, #3 #4,}
\newcommand{\authmanythr}[6]{#1 #2, #3 #4, #5 #6,}
\newcommand{\private}[2]{ (private communication).}
\newcommand{\yjour}[6]{ #6 {#2} {\bf #3}, #4 (#1).}

\newcommand{\yproc}[7]{#1~ #4. In {\em #5} (ed. #6), pp. #2-#3. #7.}

\newcommand{\ybook}[3]{{\em #2}. #3 (#1).}

\begin{document}
\title{\vspace{-6mm}Dependence of trefoil vortex knots upon the initial vorticity profile.}
%Evolution of helicity density and enstrophy in trefoil vortex knots before reconnection}
\author{\vspace{-3mm}Robert M. Kerr}
%\thanks{Email address for correspondence: Robert.Kerr@warwick.ac.uk}
\affiliation{Department of Mathematics, University of Warwick, Coventry CV4 7AL, United Kingdom.email: Robert.Kerr@warwick.ac.uk}
%\date{}
\begin{abstract} 
Six sets of Navier-Stokes trefoil vortex knots in $(2\pi)^3$ domains show how the shape of the initialprofile influences the evolution of the enstrophy 
$Z$, helicity ${\cal H}$ and dissipation-scale. Significant differences 
develop even when all have the same three-fold symmetric trajectory, the same 
initial circulation and the same range of the viscosities $\nu$. Maps of the 
helicity density $h=u\cdot\omega$ onto vorticity isosurfaces patches show 
where $h\lesssim0$ sheets form during reconnection.
For the Gaussian/Lamb-Oseen profile helicity ${\cal H}$ grows significantly, 
with only a brief spurt of enstrophy growth as thin braids form then decay 
during reconnection.  The remaining profiles are algebraic. For the 
untruncated algebraic cases, $h<0$ vortex sheets form in tandem with
$\nu$-independent convergence of $\sqrt{\nu}Z(t)$ at a common $t_x$.
For those with the broadest wings, enstrophy growth accelerates
during reconnection, leading to approximately $\nu$-independent convergent
finite-time dissipation rates $\epsilon=\nu Z$.  By mapping terms from the 
budget equations onto  centerlines, the origins of the divergent behavior are 
illustrated.  Lamb-Oseen has six locations of centerline
convergence form with local negative helicity dissipation,
$\epsilon_h<0$, and small, but positive $h$.  Later, the sum of these
localized patches of  $\epsilon_h<0$ leads to a positive
increase in the global ${\cal H}$ and suppression of enstrophy production.
For the algebraic profiles: There are only three locations of centerline
convergence, each with spans of less localized $\epsilon_h<0$ and some $h<0$.
Spans that could be the seeds for the $h<0$ vortex sheets that form in the
lower half of the trefoil as the $\sqrt{\nu}Z(t)$ phase begins and can 
explain accelerated growth of the enstrophy and evidence for finite-time 
energy dissipation $\Delta E_\epsilon$.  Despite the initial symmetries.
\end{abstract} 
\maketitle
\vspace{-12mm}
\section{Background \label{sec:back}}
\vspace{-2mm}

For the incompressible, three-dimensions Navier-Stokes equation the three 
significant quadratic integrated diagostics of the velocity $u$ and vorticity 
$\omega$ are: the kinetic energy with $E\sim 0.5 u^2$; the enstrophy with 
$Z\sim\omega^2$; and the helicity ${\cal H}$. ${\cal H}$ is the global integral of the 
helicity density $h=\bu\cdot\bomega$ and can take either sign. Equations representing 
their budgets are defined in section \ref{sec:governing}.

The robust relationship between the energy $E$ and enstrophy $Z$ is 
well-known.  Given a viscosity $\nu$, the energy dissipation rate 
$dE/dt=\epsilon$ with $\epsilon=\nu Z$.  The importance of $\epsilon$ for 
turbulent flows is that irregularity of the vorticity can lead to very large 
enstrophy and a energy dissipation rate $\epsilon$ that is large enough to 
support a finite, Reynolds number-independent energy dissipation. This is known 
as a {\it dissipation anomaly}, defined as the finite integral
\EQL{eq:dissanom}\Delta E_\epsilon=\int_0^{T_\epsilon} \epsilon\,dt>0\quad
\mbox{in a finite-time}~T_\epsilon \,.\EN
This is observed in many laboratory and environmental turbulent flows. 
This relation between irregular vorticity and turbulent decay is robust, 
but has this caveat: Can a smooth initial state far from boundaries 
numerically generate $\nu\to0$ finite $\Delta E_\epsilon$ without either 
forcing or a parameterized dissipation $\epsilon$? 

Could a better understanding of the helicity density $h$ help?  What is known is that without 
viscosity, that is for the inviscid $\nu=0$ Euler equations, the global helicity ${\cal H}$
is preserved, in addition to the energy $E$.  And on that basis 
it has been proposed that ${\cal H}$ can constrain nonlinear Euler
growth of the enstrophy $Z$.  However, could the formation 
of local $h<0$ along a vortex lead to a alternative scenario?

Trefoil vortex knots are an initial state that is inherently helical, 
self-reconnecting, and mathematically compact, meaning that they can be 
isolated far from boundaries.  The goal of this paper is to revisit recent 
trefoil knots simulations 
\cite{KerrFDR2018,KerrJFMR2018,YaoYangHussain2021,ZhaoScalo2021}
to ascertain why different initial vorticity profiles generate 
starkly contrasting answers to those questions. 

Before the results in papers \cite{KerrFDR2018,KerrJFMR2018,YaoYangHussain2021,ZhaoScalo2021},
the most that numerics has been able to  tell us about the role of
helicity is that for single-signed helical Fourier modes, energy dissipation can be 
suppressed for a short time \cite{AlexakisBiferale2018}. These flows then evolve
into traditional decaying numerical turbulence: without any
further insight into whether $h$ has a role in either achieving, or suppressing,
finite energy dissipation as the viscosity decreases.

Could trefoil vortex knots robustly overcome those limitations? Robustly meaning,
are the numerics adequate to reach consistent conclusions? One conclusion coming
from comparing the recent trefoil papers is that the results are not robust.
With different initial states or numerics, different trends 
are observed for the evolution of the enstrophy $Z(t)$ and helicity ${\cal H}(t)$, 
particularly as reconnection begins and immediately afterward. 

To illustrate the differences, figures \ref{fig:Gd05dm1ZHnus} and \ref{fig:r1ZHnu} 
compare $Z(t)$ and ${\cal H}(t)$ for two sets of calculations with the same circulation
$\Gamma=1$ \eqref{eq:Gamma} and same three-fold symmetric trajectories, but
representing different initial core profiles. % from section \ref{sec:config}. 
Respectively, evolution using a Gaussian/Lamb-Oseen \eqref{eq:Gauss}  core profile, as 
recently reported \cite{YaoYangHussain2021}, and from a $p_r=1$ algebraic core 
\eqref{eq:Rosenh} that has already provided 
evidence for a dissipation anomaly, finite $\Delta E_\epsilon$ 
\eqref{eq:dissanom} \cite{KerrJFMR2018}. Another difference in their initialization is
the vortex core width. 

How do $Z(t)$ and ${\cal H}(t)$ evolve for these cases?  At very early times and for all the 
profiles, $Z(t)$ decreases, meaning more enstrophy dissipation than production. This 
similarity between the two continues only until $t=0.4$.  
After which $Z(t)$ and ${\cal H}(t)$ diverge slowly 
until the innermost (centerline) vorticity isosurfaces begin to reconnect at a common time of 
$t_r\approx4$. Then as $t\to t_r$, the differences become dramatic. For Lamb-Oseen, 
after some enstrophy growth at $t\sim t_r$, its enstrophy
$Z(t)$ decreases again while the helicity ${\cal H}$ grows. With thin vortex bridges and 
braids forming, as previously 
observed \cite{YaoYangHussain2021} and discussed in section \ref{sec:Greconnect}.
%Although major differences between 
%$Z(t)$, ${\cal H}(t)$ and the innermost (centerline) vorticity isosurfaces do not appear until 
%$t\geq t_r\approx4$, defined as when reconnection begins.
% and the last time that a single centerline vortex can be identifed. 

In contrast, for the three-fold symmetric trefoils with a $p_r=1$  algebraic profile, 
while reconnection begins at the same $t_r$, it is not completed until a somewhat later
time of $t_x$. Figure \ref{fig:r1sqnuZ}a defines $t_x$ as when there is 
$\nu$-independent convergence of $\sqrt{\nu}Z(t)$, a `reconnection-enstrophy'.  Convergence 
that has previously been associated with the formation of vortex sheets 
\cite{KerrJFMR2018}. Figure \ref{fig:3DT3p6} in section \ref{sec:mid}
goes further: showing that the vortex sheets have $h<0$. 

However convergence of $\sqrt{\nu}Z(t)$ is not convergence of the dissipation rates 
$\epsilon(t)=\nu Z(t)$. What has been found for algebraic trefoils with perturbations, in
far larger domains, is that convergence of $\epsilon(t)=\nu Z(t)$ in a finite time 
is possible \cite{KerrJFMR2018}. Can the algebraic calculations reported here
develop finite-time convergence of $\epsilon(t)=\nu Z(t)$: despite the 
three-fold symmetry and a tighter domain?

They do, with figure \ref{fig:r1sqnuZ}b providing evidence for weak convergence of the 
dissipation rates $\epsilon(t)=\nu Z(t)$ at $t_\epsilon\approx 2 t_x$. In figure \ref{fig:r1ZHnu}
this is accompanied by a modest increase in ${\cal H}(t)$ at the 
higher Reynolds numbers before ${\cal H}$ decays. This is discussed in section 
\ref{sec:reconnect}. %III D

%Note that convergent $\epsilon$ despite the extra symmetries and with a tighter domain 
%than was used by the earlier calculations with convergent $\epsilon$ 
%\cite{KerrJFMR2018}.  A wider ($r_o>0.015$) version of the $p_r=1$ profile is used by 
%all my post-2010 calculations \cite{Kerr2013a,Kerr2013b,Kerr2013c,KerrFDR2018,KerrJFMR2018,KerrJFM2018c}.

To complete the discussion of profiles, a set calculations using the
$p_r=2$ Rosenhead regularized profile \eqref{eq:Rosenh} 
of a point vorticity \cite{Rosenhead1931} is discussed in section \ref{sec:KSRdiss}. 
The mathematics community calls this the Kaufman-Sculley profile and it will be 
designated as the K-S-R profile here.  The shape of the central core
is intermediate between the two others, but its overall behavior is closer to that of
the $p_r=1$ algebraic profile.

Given these differences in the $Z(t)$ and ${\cal H}(t)$ evolution, these
questions can be asked (tentative answers in parentheses). 
\vspace{-1mm}
\ITM\item Can the $t\sim0$ origins of the divergent behavior be identified? 
(The Rayleigh inflection-point instability discussed in 
section \ref{sec:Rayleigh}.)
\vspace{-1mm}
\item What are the differences in the post-reconnection $t>t_x$ dissipative 
structures? (Sheets lead to a {\it dissipation anomaly}, braids and bridges
do not.)
\vspace{-1mm}
\item Are there diagnostics for identifying the intervening, divergent $0<t<t_r$ 
dynamics?  (Mapping terms in the enstrophy and helicity budgets onto
vortices' centerlines.)
\ITN
\vspace{-1mm}

To reduce the number of possible sources for those differences, all of the new 
calculations are three-fold symmetric and run in $(2\pi)^3$ periodic domains. 
This ensures that the only differences between each set of trefoils are the choices of 
their initial vorticity profiles and their widths.  

\begin{figure}[H]
\bminic{0.5}\subfigure{
\includegraphics[scale=.20,clip=true,trim=0 0 0 105]{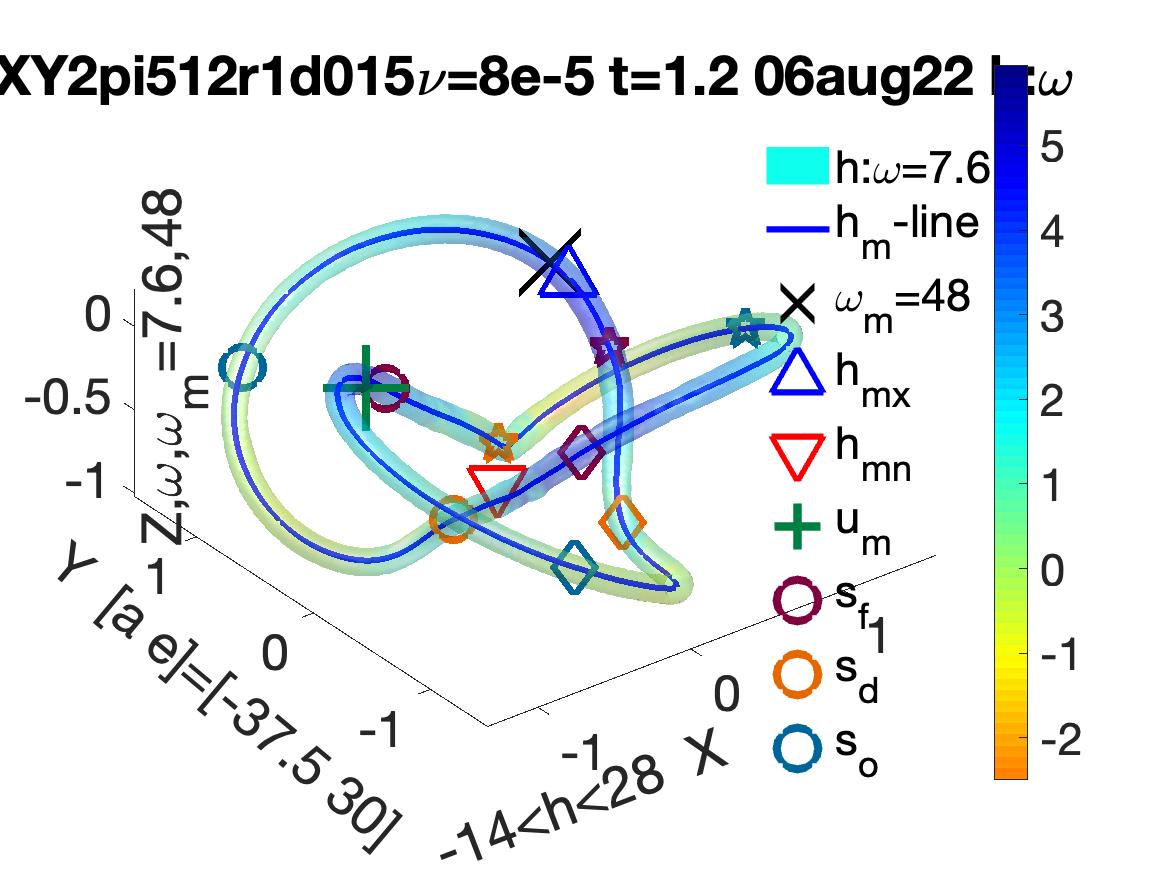} 
\begin{picture}(0,0)\large\put(-200,46){(a)}\put(-200,149){\bf r1d015 $\nu$=8.4e-5  t=1.2}
\end{picture} }
\emini\bminic{0.5}\subfigure{
\includegraphics[scale=.20,clip=true,trim=0 0 0 100]{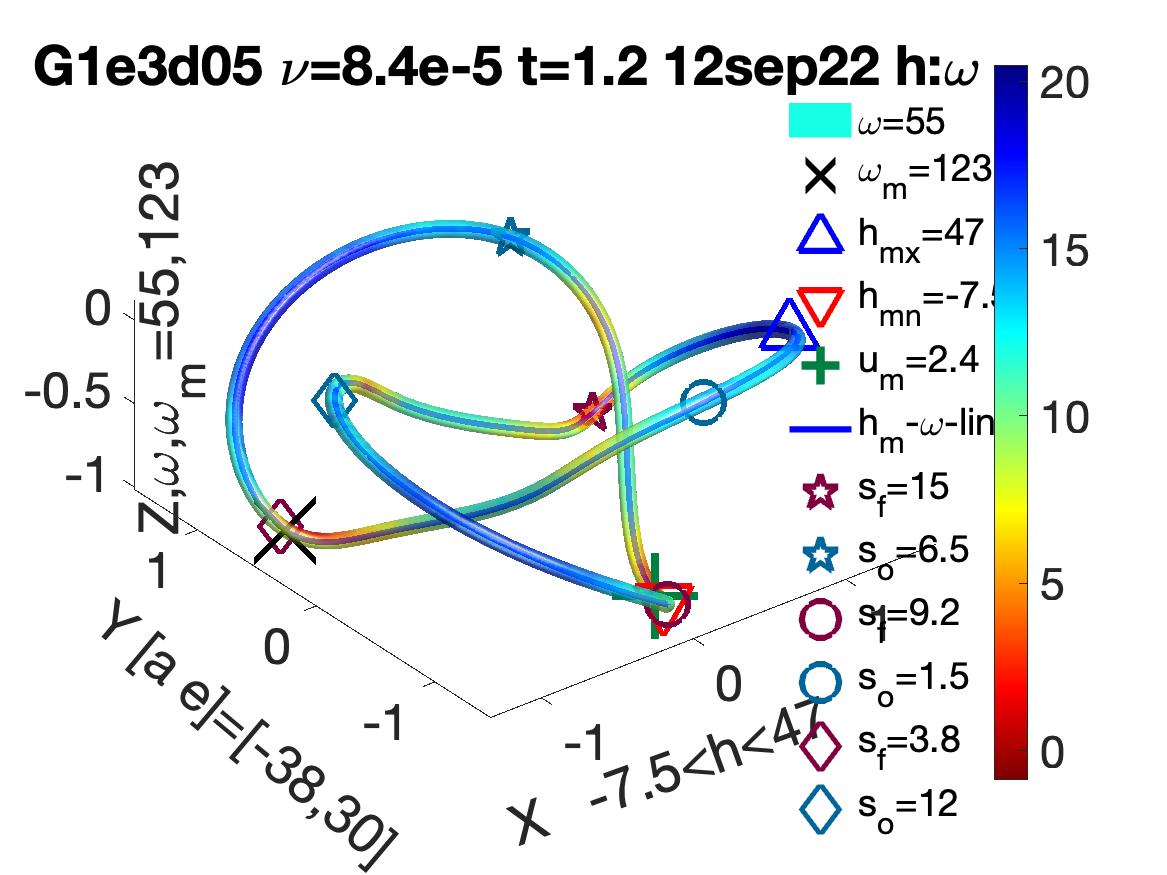} 
\begin{picture}(0,0)\large\put(-200,46){(b)}\put(-210,146){\bf Gd05 $\nu$=8.4e-5  t=1.2}
\end{picture} }
\emini 
\caption{\label{fig:3Dr11p2G2pi} 
Three-dimensional vorticity isosurfaces with mapped helicity at $t=1.2$ for
two of the  three-fold symmetric trefoils. (a) From the
$p_r=1$, $r_o=0.015$ algebraic \eqref{eq:Rosenh} calculation (r1d015). 
(b) Lamb-Oseen profile \eqref{eq:Gauss} (Gd05). 
The primary extrema of interest: Maximum vorticity, minima and maxima of the 
helicity, and the maximum velocity are indicated in both frames, with symbols in 
the legends. In addition, each frame indicates the three-dimensional positions of 
the $s_f$, local $\min(h_f)$, and their opposing $s_o$ points, 
closest points in 3D on their opposite loops. For the algebraic, the $s_d$, local
$\min(\epsilon_h)$.  These are also marked on the $t=1.2$ centerline budget profiles in figures
\ref{fig:T1p2uuoo} and  \ref{fig:GuuooT1p2} and will be used for reference at later times.
%Note the different positions of $\omega_m$ on the frames. 
%In (a), case r1d015, $\omega_m$ is at the blue triangle, $h_{mx}$ helicity maxima. 
%In (b), case Gd05, $\omega_m$ is at the maroon diamond, a local $\min(h_f)$, 
%a minima of the helicity flux.
}
\end{figure}
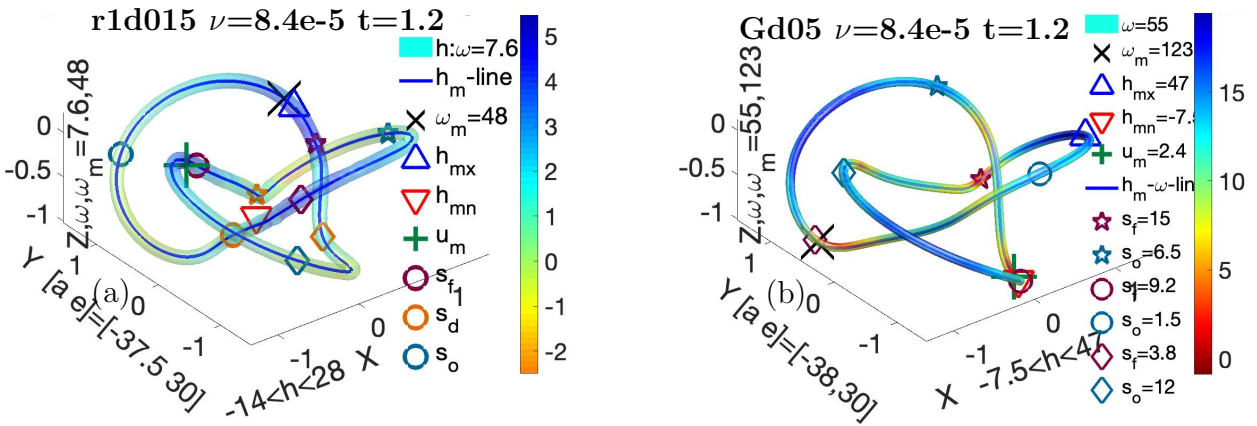

Figure \ref{fig:3Dr11p2G2pi} provides an early time, three-dimensional perspective on the 
vorticity isosurfaces at $t=1.2$ for algebraic case r1d015 and Lamb-Oseen Gd05. In terms of 
the overall structure they are almost identical. Perhaps the only identifiable difference 
is the different positions of the maximum of vorticity $\omega_m=\|\omega\|_\infty$, 
indicated by {\bf X}. For the algebraic case on the left, $\omega_m$ is co-located with the 
blue triangle, maximum of helicity $h_{mx}$. For Lamb-Oseen on the right, $\omega_m$ is at 
the maroon diamond, a local minima of the helicity flux \eqref{eq:helicity}, $\min(h_f)$. 
However, on the centerlines their respective enstrophy and helicity density budgets 
are quite different.

The paper is organized as follows. After the introduction of the profile-dependent 
evolution of the primary global diagnostics, and their early vorticity isosurfaces,
the governing and budget equations are given. Next are the steps required to 
initialize the vortices, including how the raw, unbalanced mapped vorticity fields 
are made incompressible. Once the initial profiles are defined, recent mathematics 
for determining their stability is referenced and a new set of diagnostics are 
defined that map the terms from the enstrophy and helicity budget equations 
(\ref{eq:enstrophy},\ref{eq:helicity}) onto the evolving centerline trajectories. 
Up to $t=3.6$, both helicity-mapped vorticity isosurfaces and mapped centerline budgets 
are used in the comparisons between the evolution of cases Gd05 (Gaussian/Lamb-Oseen) and 
r1d015 ($p_r=1$, $r_o=0.015$ algebraic). The $t<t_r=4$ differences in the budget terms 
lead to profound differences in the $t\gtrsim 4$ dissipative structures and dissipation
rates $\epsilon(t)$. For Lamb-Oseen at and after reconnection: thin bridges, then braids and
decaying dissipation rates. While for all of the algebraic calculations: vortex sheets start
to form with $\sqrt{\nu}Z(t)$ convergence for $t_x\lesssim 1.5t_r$; and for the widest initial 
algebraic profiles, $\nu$-independent dissipation rates $\epsilon$ 
that approximately converge at $t_\epsilon\approx 2.5 t_r$.  \\

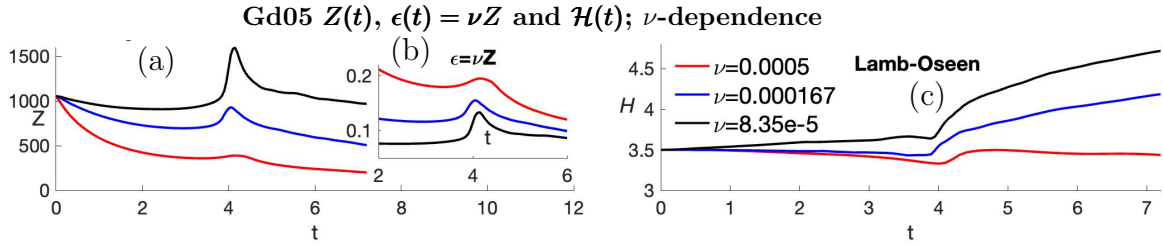
\begin{figure}[H] 
\includegraphics[scale=0.21,clip=true,trim=0 0 0 76]{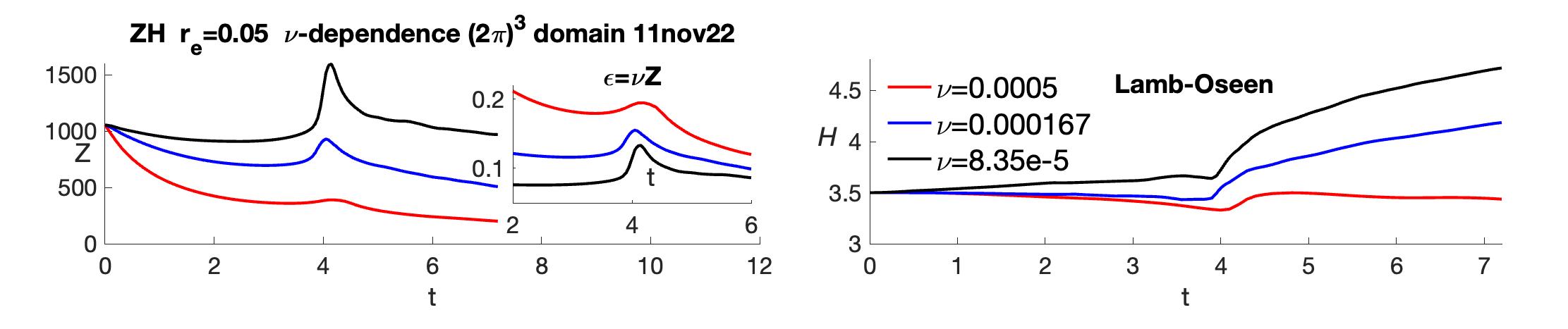}
\begin{picture}(0,0)\put(-410,71){\large(a)}\put(-314,75){\large(b)} \put(-118,58){\large(c)}
\put(-370,92){~}
\put(-370,88){\normalsize\bf Gd05 $\pmb{Z(t)}$, $\pmb{\epsilon(t)=\nu Z}$ and 
$\pmb{{\cal H}(t)}$; $\nu$-dependence}
\end{picture} 
\caption{\label{fig:Gd05dm1ZHnus}
Time dependence of (a) the enstrophy $Z(t)$, dissipation rate $\epsilon(t)=\nu Z$ (inset) 
and (c) global helicity ${\cal H(t)}$ for case Gd05, a three-fold 
symmetric trefoil with a Gaussian/Lamb-Oseen profile \eqref{eq:Gauss}. Three
viscosities (in legend) are given, whose Reynolds numbers are [2000 6000 12000]. 
Similar to figure 3 of \cite{YaoYangHussain2021}.
All calculations are in $(2\pi)^3$ periodic boxes.} 
\end{figure}
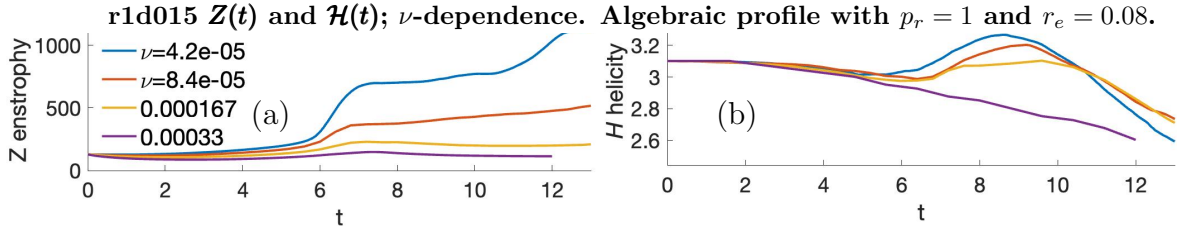
\begin{figure}[H]
%%\graphics[scale=0.19]{3fold/figjpg3fold/XYylnu16rZH30oct21.jpg}
\includegraphics[scale=0.21,clip=true,trim=0 0 0 80]{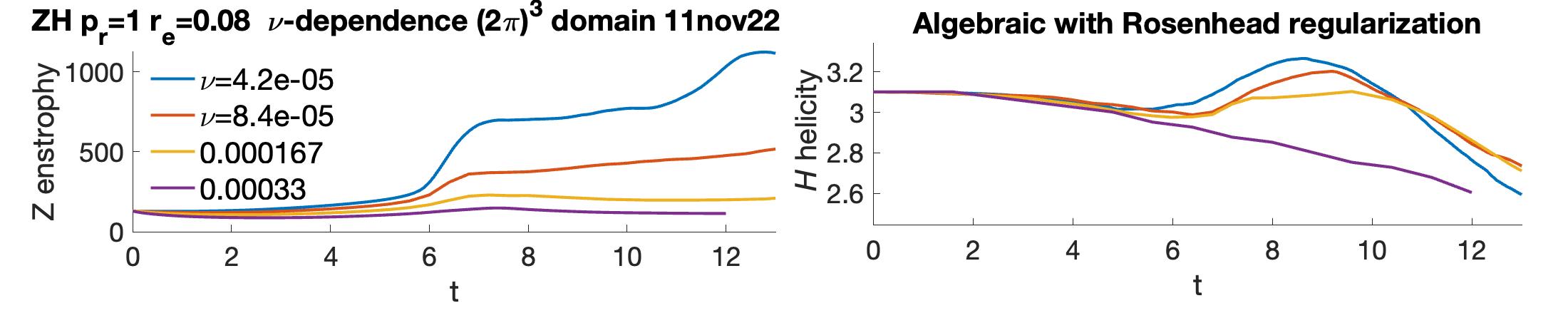}
\begin{picture}(0,0)\put(-366,46){\large(a)} \put(-190,46){\large(b)}
\put(-420,84){\normalsize\bf r1d015 $\pmb{Z(t)}$ and $\pmb{{\cal H}(t)}$; $\nu$-dependence.
Algebraic profile with $p_r=1$ and $r_e=0.08$.}\end{picture} 
\caption{\label{fig:r1ZHnu}
Time dependence of (a) the enstrophy $Z(t)$ and 
(b) the global helicity ${\cal H}(t)$ for algebraic \eqref{eq:Rosenh} case r1d015, with 
$p_r=1$, $r_o=0.015$ and $r_e=0.08$, at several viscosities (in legend) with 
Reynolds numbers [24000 12000 6000 3000].  } \end{figure}

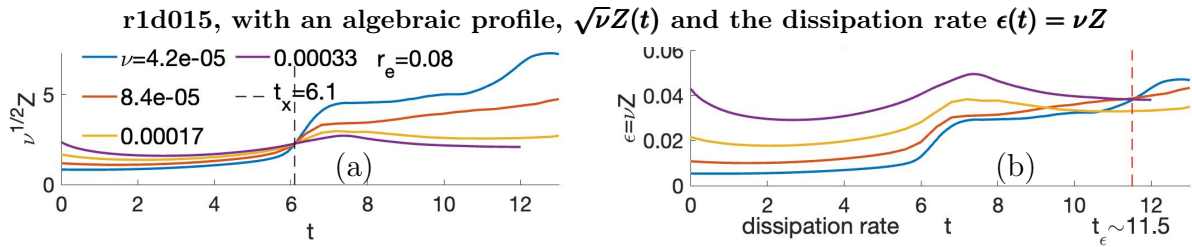
\begin{figure}[H]
\includegraphics[scale=0.21,clip=true,trim=0 0 0 51]{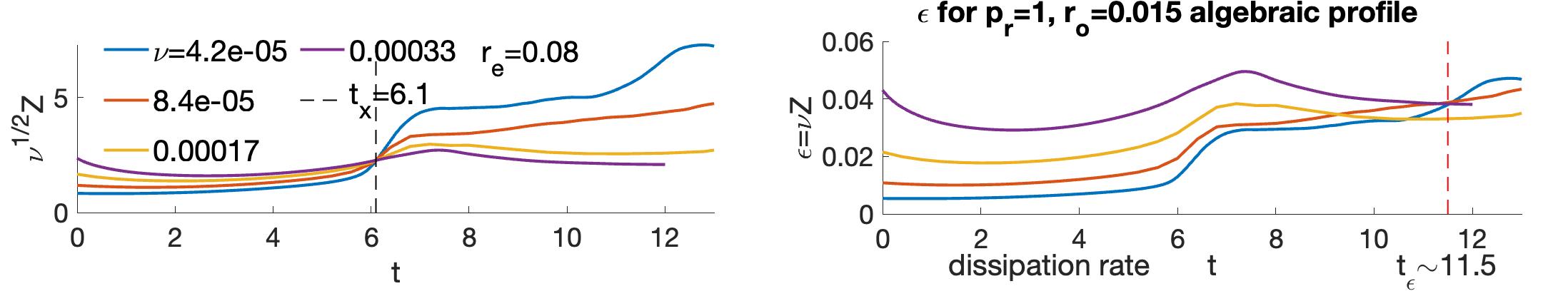}
\begin{picture}(0,0)\put(-340,28){\large(a)} \put(-90,28){\large(b)}
\put(-420,84){\normalsize\bf r1d015, with an algebraic profile, 
$\pmb{\sqrt{\nu}Z(t)}$ and the dissipation rate $\pmb{\epsilon(t)=\nu Z}$}
\end{picture}
% \bminic{0.5}\graphics[scale=0.19,clip=true,trim=0 450 0 0]{3fold/
% figjpg3fold/XYyld015nusqnuZ30apr22.jpg}
%\emini \bminic{0.5}
%\graphics[scale=0.19,clip=true,trim=0 0 0 400]{3fold/figjpg3fold/XYyld015nusqnuZ30apr22.jpg} \emini 
\caption{\label{fig:r1sqnuZ}
For the case and viscosities in figure \ref{fig:r1ZHnu}: (a) time dependence of the 
reconnection-enstrophy $\sqrt{\nu}Z(t)$, with convergence at $t_x=6$ that is used to define the 
end of the first reconnection; (b) the dissipation rate $\epsilon(t)=\nu Z$, whose
convergence at $t\approx10$ is used to define the dissipation anomaly $\Delta E_\epsilon$
\eqref{eq:dissanom}.}
\end{figure}

\section{Equations, numerics, initial conditions, centerline maps,
stability. \label{sec:governing}}

The governing equations are the incompressible Navier-Stokes equations: for the velocity 
\EQL{eq:NS} \hspace{-8mm} \ppto{\bu} + ({\bu}\cdot\nabla){\bu} = -\nabla p+ 
\underbrace{\nu\triangle{\bu}}_{\rm viscous drag}, \qquad 
\nabla\cdot{\bu}=0\,; \EN
and the vorticity $\bomega=\nabla\times\bu$
\EQL{eq:omega} \hspace{-8mm} \ppto{\bomega} + ({\bu}\cdot\nabla){\bomega} = 
({\bomega}\cdot\nabla){\bu} + \nu\triangle{\bomega},\qquad \nabla\cdot{\bomega}=0\,.\EN
%The numerical method will be a 2/3rds-dealiased pseudo-spectral code with a very high 
%wavenumber cut-off filter \citeppp{Kerr2013a}.
\noindent{\bf Numerics.} All of the calculations are done in $(2\pi)^3$ periodic boxes with 
a 2/3rds-dealiased pseudo-spectral code and a high-wavenumber cutoff filter 
\cite{BustaKerr2008,Kerr2013b}. These features remove aliasing errors and absorb 
high-wavenumber fluctuations that would otherwise be  reflected (in Fourier space) from the 
abrupt high-wavenumber cut-off. 
Extensive tests showed that with these features the calculations do at least as well as a 
calculation on a mesh that is 1.5 times greater. Some tests, such as doubling the mesh and 
comparing the maximum vorticities, have been  repeated here.

Based on this past experience, the evolution of the global helicity and enstrophy shown for 
all cases can be trusted.  For the more detailed analysis on vortex lines and 
three-dimensional graphics, the algebraic r1d015 $\nu=1.6$e-4 statistics are reliable 
for all times, but those with $\nu=8.4$e-5 are given only to $t=3.6$. 
The detailed results for case G1e3d05 $\nu=8.4$e-4 can 
be trusted up to $t=4.4$, but not for $t\geq4.8$.

Five initial profiles are discussed, each run for at least three viscosities. 
A larger number of profiles were done before choosing these five, so in the interest of 
economy and ease of use, the vorticity graphics for cases other than Gd015 and r1d015 use
$512^3$ meshes.  Several of the smallest viscosity calculations, 
and all of the Lamb-Oseen calculations, are from $1024^3$ mesh calculations.

The continuum equations for the densities of the energy, enstrophy and helicity,
$e=\half|\bu|^2$, $\zeta=|\bomega|^2$ and $h=\bu\cdot\bomega$, with their production,
flux and dissipation rates are:
\EQL{eq:energy} \hspace{-15mm} \ppto{e}+ ({\bu}\cdot\nabla)e = -\nabla\cdot(\bu p) 
+\nu\triangle e
-\underbrace{\nu(\nabla\bu)^2}_{\epsilon={\rm dissipation}=\nu Z},\qquad 
E=\half\int\bu^2dV\,;\EN
\EQL{eq:enstrophy} \hspace{-15mm} \ppto{\zeta}+ ({\bu}\cdot\nabla)|\bomega|^2 = 
\underbrace{2\bomega\bS\bomega}_{\zeta_p={\rm production}}
+\nu\triangle|\bomega|^2
-  \underbrace{2\nu(\nabla\bomega)^2}_{\epsilon_\omega=Z-{\rm dissipation}},\qquad 
Z=\int\bomega^2dV\,;\EN
\EQL{eq:helicity} \hspace{-15mm} \ppto{h}+ ({\bu}\cdot\nabla)h = 
\underbrace{-\bomega\cdot\nabla\Pi}_{h_f=\omega-{\rm transport}}
+\underbrace{\nu\btriangle h}_{\nu-{\rm transport}} -\underbrace{
2\nu{\rm tr}(\nabla\bomega\cdot\nabla\bu^T)}_{\epsilon_h={\cal H}-{\rm dissipation}}
\qquad
{\cal H}=\int\bu\cdot\bomega dV\,.\EN
$\Pi=p-\half\bu^2\neq p_h$ is not the pressure head $p_h=p+\half\bu^2$. \medskip

While the global energy $E$ and helicity ${\cal H}$ are inviscid invariants 
\cite{Moffatt2014}, their inviscid Lagrangian local densities $e$ and $h$ can change 
due to the pressure gradient $-\nabla p$ and the $\omega$-transport $h_f$ respectively. 
Under $\nu\neq0$ Navier-Stokes, both the helicity flux $h_f$ and dissipation $\epsilon_h$
can generate local negative helicity $h\!<\!0$. 
Note that $h$ is not locally Galilean invariant due to $h_f$. \medskip

{\bf Role for $\pmb{h<0}$?} Can local $h\!<\!0$ break helicity's constraint upon the 
nonlinear growth of the enstrophy $Z$?
Section \ref{sec:mapping} shows how this question can be addressed by mapping the 
budget terms onto the vorticity centerlines.

For short times another set of inviscid short-time conservation laws are the 
circulations $\Gamma_i$ for closed loops ${\cal C}_i$ about those trajectories:
\EQL{eq:Gamma} \Gamma_i=\oint_{{\cal C}_i} \bu_i\cdot \br_i 
\quad{\rm where}\quad \br_i~~
\mbox{is a closed loop about}~~{\cal C}_i\,. \EN
With the appropriate choice of the closed loop, %path ${\cal C}_i$, 
$\Gamma_i$ can be preserved during Navier-Stokes reconnection for very short times.  
Could this constraint that have additional consequences?
\bpurp{And during this short-lived
period ($t\sim4-6$) of circulation conservation, figure \ref{fig:r1ZHnu} also shows
minimal change in the global helcity as $\nu\to0$. Similar to what has been previously 
suggested for anti-parallel reconnection \cite{Laingetal2015}.  Could this short-lived
Navier-Stokes constraint be used to relate the formation of vortex sheets with the 
$\sqrt{\nu}Z$ scaling regime that all of algebraic profiles are generating?}

\subsection{Initial conditions \label{sec:config}}

Four elements are used to define an incompressible vortex knot. 
\ITM\item[1)] The $\bx(\phi)$ trajectory of the centerline of the vortex knot
\eqref{eq:trefoil}.
\bpurp{with a characteristic size $r_f$.}
\item[2)] The vorticity profile $|\omega(\rho)|$, with the distance 
$\rho$ defined as the distance between a given mesh point $\bx$ and the
nearest point on the trajectory $\bx(\phi)$: $\rho=|\bx-\bx(\phi)|$. 
\ITM\item[a)] The profiles are either algebraic \eqref{eq:Rosenh}, with a chosen 
power-law $p_r$, or Gaussian/Lamb-Oseen \eqref{eq:Gauss}.
\item[b)] Each $|\omega(\rho)|$ has two parameters: A radius $r_o$ 
and the centerline vorticity $\omega_o$.
\item[$\bullet$] The final $\omega_o$ are chosen so that the circulation $\Gamma\equiv1$ \eqref{eq:Gamma} after step 4.
\item[$\bullet$] In this paper $\Gamma=1$ and $r_f=1$ are fixed so the 
nonlinear timescale for all the calculations is $t_{NL}=1$ \eqref{eq:trefoil}.
\ITN
\item[3)] 
The chosen profile is mapped onto a Cartesian mesh using previous algorithms \cite{KerrFDR2018}, 
with the direction of vorticity given by 
the centerline direction: $\hat{\omega}(\rho)=\hat{\omega}(\bx(\phi))$.
\item[4)] Finally, we need to remove the non-solenoidal components of the 
raw vorticity field by projection. This also makes the velocity field 
incompressible. Except for the Lamb-Oseen profile, this operation
invariably leads to reductions in the values of the maximum vorticity $\omega_m$ 
and the enstrophy $Z$. %Further details are in appendix \ref{sec:Istep4Rosenh}.
\ITN

The initial trajectory $\bxi_0(\phi)=[x(\phi),y(\phi),z(\phi)$ 
of all the trefoils in this paper is defined over $\phi=1:4\pi$ by this 
closed double loop, with $r_f=1$ and $r_1=0$:
\EQL{eq:trefoil}\begin{array}{rrl} & x(\phi)= & r(\phi)\cos(\alpha) \\
& y(\phi)= & r(\phi)\sin(\alpha) \qquad z(\phi)= a\cos(\alpha) \\
{\rm where} & r(\phi) =& r_f+r_1a\cos(\phi) +a\sin(w\phi+\phi_0)\\
{\rm and} &  \alpha=& \phi+a\cos(w\phi+\phi_0)/(wr_f) \\
{\rm with} & t_{NL}  =&r_f^2/\Gamma \mbox{ the nonlinear time-scale,} \\
{\rm and} & r_e=& (\Gamma/(\pi\omega_m))^{1/2} \mbox{ the effective radius.}
\end{array} \EN

The four algebraic Rosenhead regularized profiles $\omega_{\mbox{raw}}(\rho)$ 
are parameterized by a radius $r_o$, maximum/centerline vorticity $\omega_o$ and
a power law $p_r$.
\EQL{eq:Rosenh} \omega_{\mbox{raw}}(\rho)=
\omega_o\frac{(r_o^2)^{p_r}}{(\rho^2+r_o^2)^{p_r}}\,. \EN
For a columnar vortex, \eqref{eq:Raystable} suggests that
the $p_r=2$ K-S-R profile is stable unless there are perturbations with 
high azimuthal wavenumber $m$ \eqref{eq:mkmode}. % Section \ref{sec:Rayleigh} 
\bcyan{and appendix \ref{sec:instability}.}
The `broader' $p_r=1$ algebraic profile 
has been used as the second initialzation step of several earlier papers
\cite{KerrFDR2018,KerrJFMR2018,KerrJFM2018c}.

\begin{figure}[H]
\includegraphics[scale=.40,clip=true,trim=40 0 0 0]{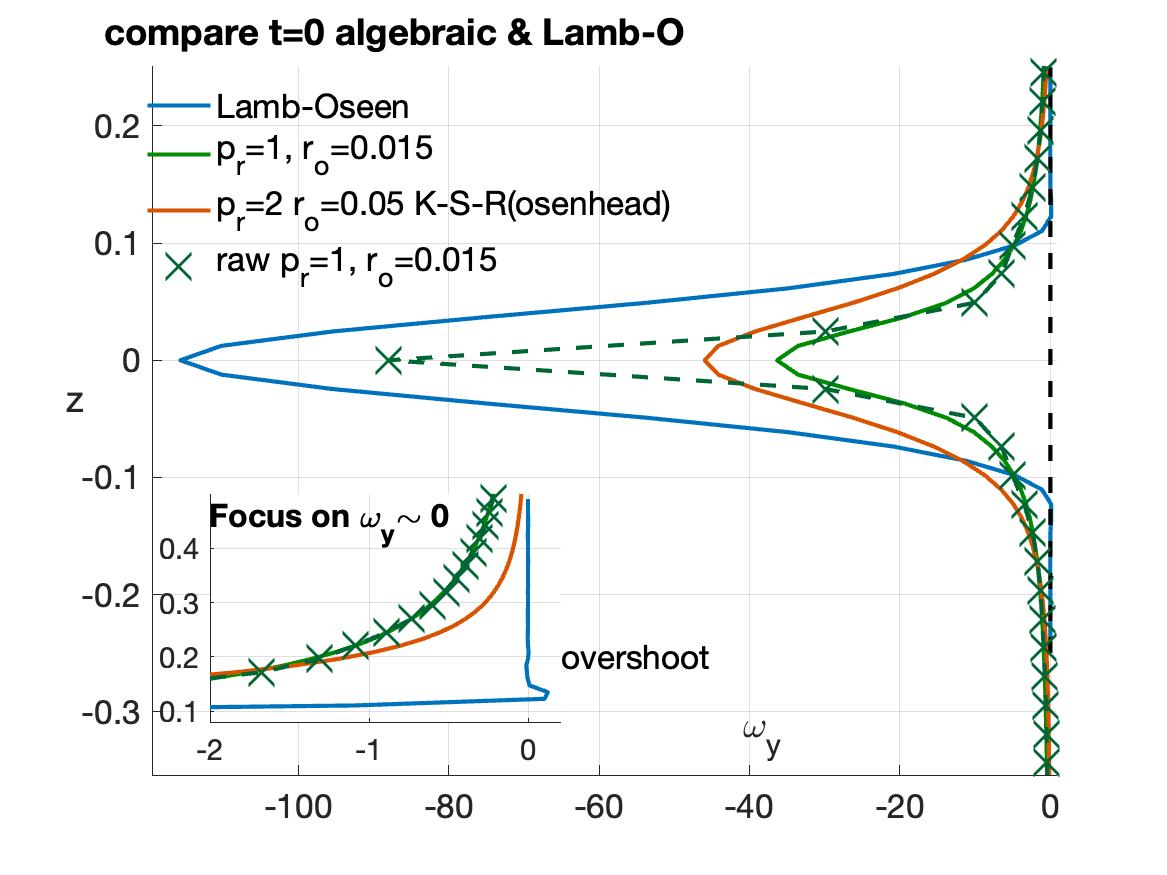}
\begin{picture}(0,0)\large\put(-240,256){\large(a)} \put(-380,116){\large(b)}
\end{picture} 
%\graphics[scale=.30]{3fold/G2pi/Gt0Rht0-omegay-z-08feb22.jpg}
\caption{\label{fig:Gt0Rht0omyz} $t$=0, $\omega_y(z)$ profiles 
through the $\min(\omega_y)$ of the $y=0$ $x\!-\!z$ plane for three of the cases from
table \ref{tab:radiiZ}.
All except one curve are taken after the non-solenoidal Fourier components have been 
removed. The profiles are for the $r_o=0.05$ Lamb-Oseen case \eqref{eq:Gauss} (Gd05)
and two of the algebraic profiles that use the Rosenhead regularization 
\eqref{eq:Rosenh}. r2d05: $p_r=2$, $r_o=0.05$, referred to K-S-R, and 
r1d015: $p_r=1$, $r_o=0.015$. 
The other curve is the `raw' $p_r=1$, $d=0.015$ curve, taken
through its pre-Fourier-projected $\omega_y$ field. 
%The total circulation $\Gamma=1$ and except for $p_r=1$, $r_o=0.015$ with 
%$r_e\approx0.07$, the rms radii of the vortices are $r_e\approx0.05$. 
(a) The primary figure shows the full profiles in $z$. (b) The lower-left inset
focuses upon $z>0.1$ wings with small $\omega_y$. 
Note the slight $\omega_y>0$ overshoot at the boundaries of the Lamb-Oseen 
profile.  This is the likely seed for the oscillations about $\omega_y=0$
in figure \ref{fig:Go2xzT1p2}.}
% The upper-left inset shows the contributions to the vortices' circulation % $\delta\Gamma(z)=2\pi\omega_y(z)$, showing that for the Rosenhead profile % the circulation primarily comes from $z>r_o=0.015$ for this case. 
%Based upon $(\Gamma/(\pi\omega_m))^{1/2}$ $r=0.071$ for d006.
%For Gt0 based on r=1./sqrt(pi*0.5.*(206+172)) $r=0.041$.\\
\end{figure}
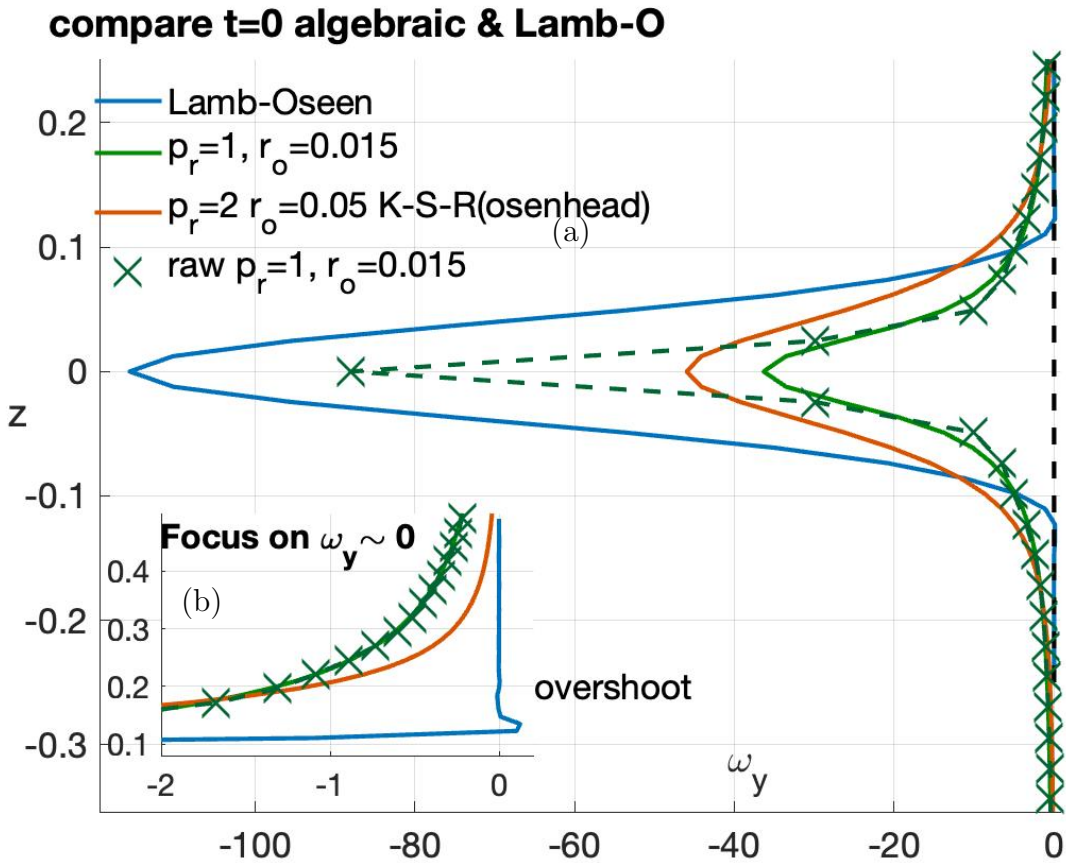

The Gaussian/Lamb-Oseen profile is
\EQL{eq:Gauss} \omega_{\mbox{raw}}(\rho)=\omega_o\exp(-(\rho/r_o)^2)\quad\text{for}\quad
\rho<\rho_+\,.\EN
This definition of the Lamb-Oseen profile has these advantages: 
$\omega_m=\omega_o$ and the effective 
radius $r_e=r_o$, without the factor of 2 required by the Lamb-Oseen profile in
current use \cite{YaoYangHussain2021}.  The only difference between that profile and 
\eqref{eq:Gauss} is that the core in figure \ref{fig:Gt0Rht0omyz} is
$\sqrt{2}$ wider. This, along with a different definition of the enstrophy $Z$ 
\eqref{eq:enstrophy} (a factor of 2), yields enstrophy and helicity evolution 
that are (in appearance) nearly identical to theirs \cite{YaoYangHussain2021}.

\begin{table}
  \begin{center}
\begin{tabular}{cllllllllll}
Cases & $p_r$ & $n^3$& $r_o$ & $\omega_o$ & $Z_o$ & $r_e$ & $\omega_m$ & $Z(0)$ & $\nu$'s & t-3D-$\omega$ \\
Gd05 & $-$ & $1024^3$ & 0.05 & 130 & 1057 & 0.05 & 130 & 1055 & 5e-4 1.7e-4  8.4e-5 & $t\leq4.4$\\
% d06 & $256^3$ & 0.06 & 13 & 0.117 & 0.18 & 9.7 & 0.13 \\
% VSTRGTH(1)=10.0,DIAM0(1)=0.06,rpow=1, XY2pi256d06Mar22 sqrt(oom)=9.9971
r2d05 & 2 & $512^3$ & 0.05 & 64.3 & 326 & 0.07 & 62 & 306 & 3.3e-4 1.7e-4  8.4e-5 & $t\leq5.2$ \\
%r2d07 & $p_r=2$ & & $512^3$ 0.0707 & 64.3 & 180 & 0.099 & xx & xxx \\
r2d1 & 2 & $512^3$ & 0.1 & 17.85 & 97.1 & 0.14 & 17.3 & 96.5 & 3.3e-4 1.7e-4  8.4e-5 & all times \\
% XY2pi256r2d05may2922 Z=2.438558E+00 XYyl2pi256r2d05sm2e-12
% XY2pi256r2d05raw3D29may22.jpg
%XY2pi256r2d05may2922.inVSTRGTH(1)=71.4,DIAM0(1)=0.05,ICURVTYPE(1)=3,RPOW(1)=2
r1d006 & 1 & $1024^3$ & 0.006 & 554 & 333 & 0.053 & 138 & 229 & 1.7e-4  8.4e-5 & only $Z$,${\cal H}$  \\
r1d015 & 1 & $1024^3$ & 0.015 & 100 & 138 & 0.078 & 56 & 124 & 1.7e-4 & $t\leq6$\\
r1d015 & 1 & $512^3$ & 0.015 & 100 & 138 & 0.078 & 56 & 124 & 3.3e-4 4.2e-5 & only $Z$,${\cal H}$  \\
r1d015 & 1 & $1024^3$ & 0.015 & 100 & 138 & 0.078 & 56 & 124 & 8.4e-5 & $t\leq 3.6$ \\
r1d015dm025 & 1 & $512^3$ & 0.015 & 182 & 362 & 0.056 & 102 & 325 & 8.4e-5 & $t\leq10$
% KE=1.039e-2 Z=1.3; Z(2pi)^3=325; omegam=102; H=0.0127*(2pi)^3=3.15
\end{tabular}
\caption{Raw core radius $r_o$ and vorticity $\omega_o$ parameters, 
resulting enstrophy $Z_o$, then effective radii $r_e$\eqref{eq:trefoil},  
maximum vorticity $\omega_m$ and enstrophy $Z$ after fields are made divergent-free. 
The t-3D-$\omega$ column is the last time for which detailed three-dimensional graphics
were made for those cases. The global enstrophy $Z(t)$ and helicity ${\cal H}(t)$ are
reliable for all cases listed.
The only Lamb-Oseen case is labeled Gd05 and the algebraic cases are labeled by 
the power-law $p_r$ as in r1d015: (r1$\equiv p_r$=1) and raw core radii (d015=$r_o=0.015$).
Last is r1d015dm025: (r1$\equiv p_r$=1) with radii (d015=$r_o=0.015$) and 
a $\rho_+=0.025$ cut-off. }
  \label{tab:radiiZ}
  \end{center}
\end{table}

Table \ref{tab:radiiZ} gives the details of the 5 initial profiles: 
The parameters, $r_o$ and $\omega_o$ for the profile formulae 
(\ref{eq:Rosenh},\ref{eq:Gauss}), the generated raw enstrophies $Z_o$.
Then the divergence-free $t=0$ values: the effective radii $r_e$ \eqref{eq:trefoil}, 
vorticity maxima $\omega_m$ and enstrophies $Z(0)$. 
The viscosities are given in the figure legends.  

An additional, inherent parameter is the maximum radius $\rho_+$ used to map 
$\omega_{\rm raw}(\rho)$ onto the Cartesian mesh in step 3. Empirically, the 
trefoils’ evolution is independent of $\rho_+$ so long as the circulation 
$\Gamma=1$ and $\rho_+\sim0.5-1$ (trefoil radius is $r_f=1$), with 
$\rho_+\geq0.75$ for all cases here except one in the appendix. Case 
r1d015dm025 with $\rho_+=0.025$ and evolution that is similar to Lamb-Oseen.

{\bf Initial profiles.} The specific profiles listed are: Lamb-Oseen (case Gd05), 
two broad algebraic $p_r=1$ cases (r1d015, r1d006) and 
two K-S-R $p_r=2$ cases (r2d05, r2d1). 
With most of the analysis figures are taken from the highest Reynolds number calculations 
of the Lamb-Oseen (Gd05) and the $p_r=1$, $r_o=0.015$ ‘broad’ algebraic profile (r1d015).
Figure \ref{fig:3Dr11p2G2pi} compares their 
slightly evolved $t=1.2$ three-dimensional helicity-mapped vorticity isosurfaces. 

Figure \ref{fig:Gt0Rht0omyz} compares the $t=0$ profiles of $\omega_y(z)$ for three of 
the profiles in table \ref{tab:radiiZ}. Each taken through the $\min(\omega_y)$ positions in 
their $y=0$, $x-z$ planes, as in figure \ref{fig:r1d015T2p4}. 
Both the main figure$^\dagger$ %\footnote
and the `wings' inset show that all of the $t=0$ algebraic profiles 
have smooth extended wings that never overshoot the $\omega_y=0$ axis. 
In contrast, on the outer edge of the Lamb-Oseen profile there is some overshoot. 
Consistent with what has been seen before when Gaussian-like 
profiles are used for anti-parallel reconnection 
\cite{Kerr2013a,BustaKerr2008}.

The source of the Lamb-Oseen overshoot arises from the combined effects of the steepness of the 
outer edge of the L-O profile and a limitation of the algorithm 
(here and \cite{YaoYangHussain2021}) that is used map the $\omega_{\rm raw}(\rho)$ field onto the 
Cartesian mesh in step 3). The mapping problem arises when the directions $\hat{\omega}$  of 
neighboring mesh points come from different positions on the centerline. 
Common when the distance $\rho$ from the centerline is large. The steepness problem arises when 
finite $|\omega|$ points are next to points with $|\omega|\approx0$. The mapped field sees these 
as finite jumps. Combined, in step 4) the projection of the mapped field can generate overshoots 
to negative values on the profile’s edge.  Overshoots whose magnitude is a function of the 
curvature of the centerline and the outer, $\rho\sim\rho^+$+, steepness of $|\omega|(\rho)$.

It has been claimed that a curved coordinate system that accommdates internal twist 
\cite{XiongYang2019} can yield divergence-free fields. That is the trajectory \eqref{eq:trefoil} 
used here, with zero internal twist and because the vortices are thinner than in my earlier
papers, trajectory source points $\bx(\phi)$ are adjacent for neighbouring $\bx$, so that the raw 
vorticity fields are divergence-free. However, these are not the $t=0$ initial fields of the 
simuations.  This is because the profiles have sharp cut-offs at $\rho=\rho_+$ and when imported 
into a Fourier code, those interfaces generate Gibbs fluctuations. Leaving the investigator two
choices. Either remove those fluctuations with a Fourier filter. Or continue with that
background noise. Figure \ref{fig:Gd05dm1ZHnus} quantifies that noise.

To demonstrate the importance of excessive steepness, one can decrease the maximum radius $\rho^+$ 
on an otherwise smooth profile. In section \ref{sec:r1d015dm} a $\rho_+=0.025$ variant of the 
$p_r=1$, $r_o=0.015$ case is given whose $Z(t)$ and ${\cal H}(t)$ evolution has similarities with
that of Lamb-Oseen in figure \ref{fig:Gd05dm1ZHnus}.  
Further implications of this could be the topic of another paper.

\begin{figure}[H]
\includegraphics[scale=.40,clip=true,trim=0 420 0 0]{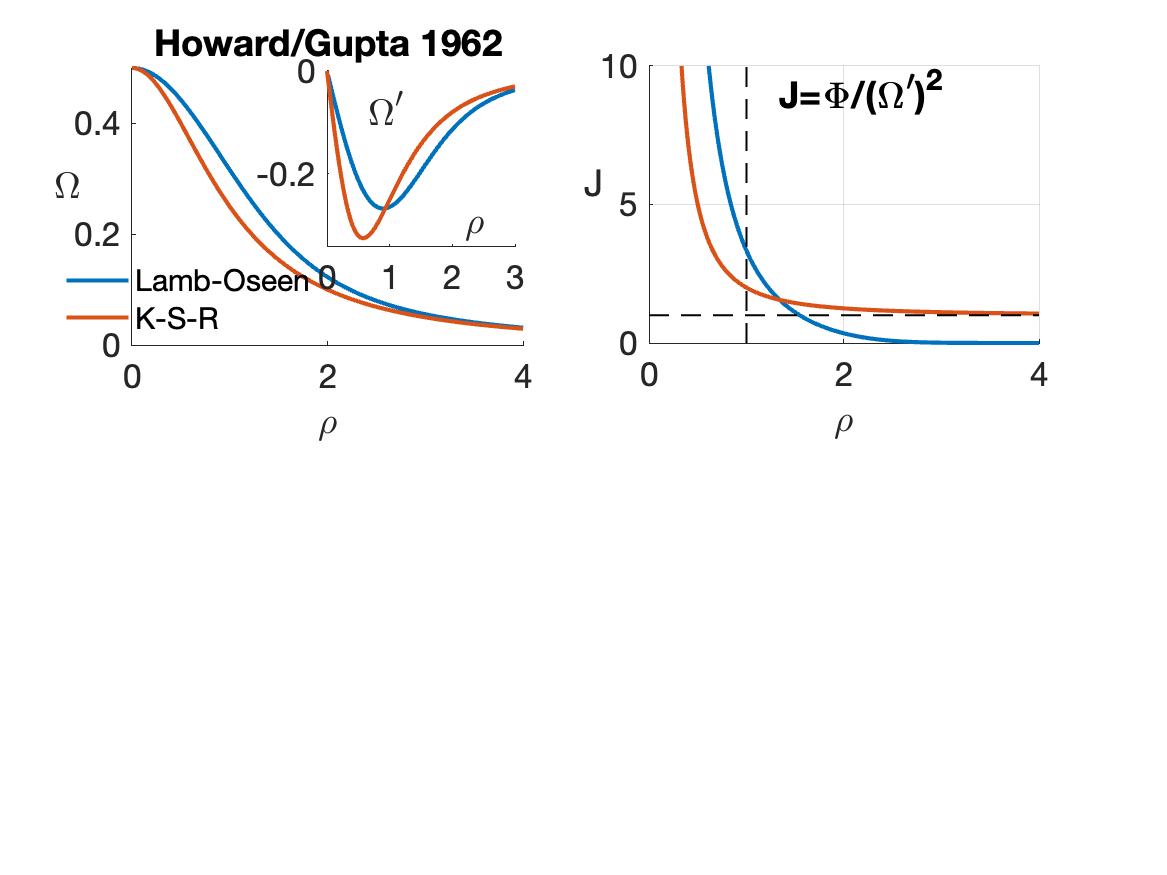}
\begin{picture}(0,0)\put(-380,136){\large(a)}
(0,0)\put(-150,86){\large(b)}\end{picture}
\caption{\label{fig:HGJ} To show how stability is determined using the 
$t$=0 Richardson functions $J(\rho)$ \eqref{eq:Jrr} for the Lamb-Oseen 
\eqref{eq:Gauss} and K-S-R \eqref{eq:Rosenh} profiles with $r_o=1$. 
(a) First, their $\Omega(\rho)$ and $\Omega'(\rho)$ profiles are similar.
(b) What is important is how $J(\rho)$ asymptotes as $\rho\to\infty$. For Lamb-Oseen
its $J(\rho)\to0$ from \eqref{eq:JG}, suggesting instability. While K-S-R, it is almost
always stable by \eqref{eq:Raystable} as $J(\rho)\to r_o^2$, finite.} 
\end{figure}
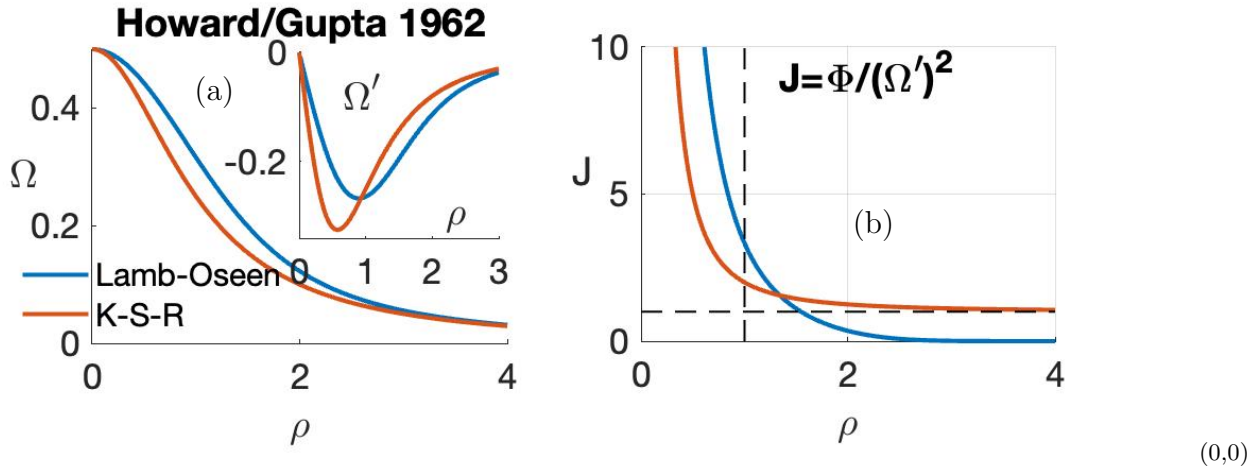
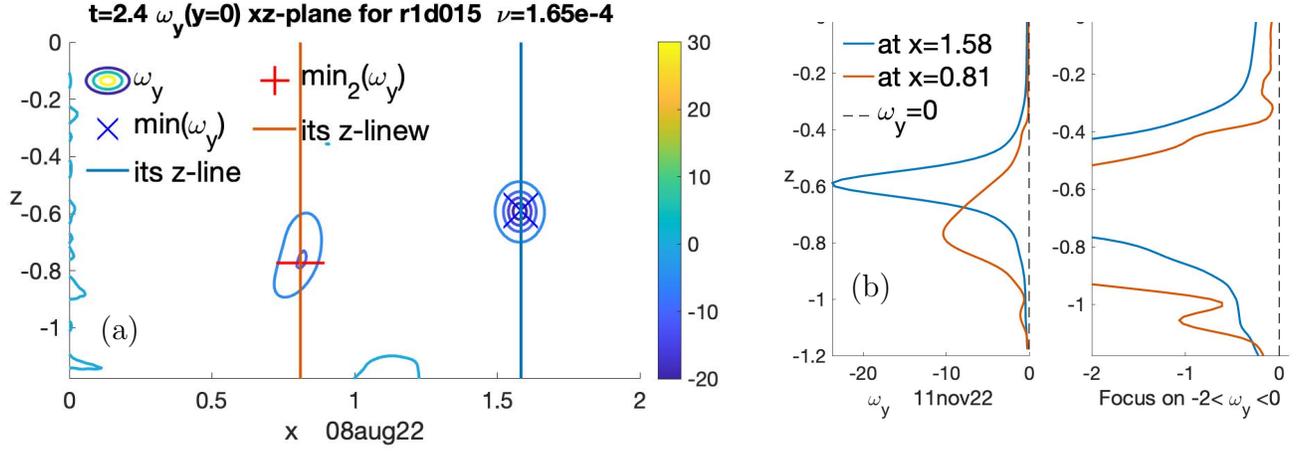
\begin{figure}[H]
\hspace{-3mm}\bminic{0.58} \subfigure{
\includegraphics[scale=.26,clip=true,trim=0 90 0 90]{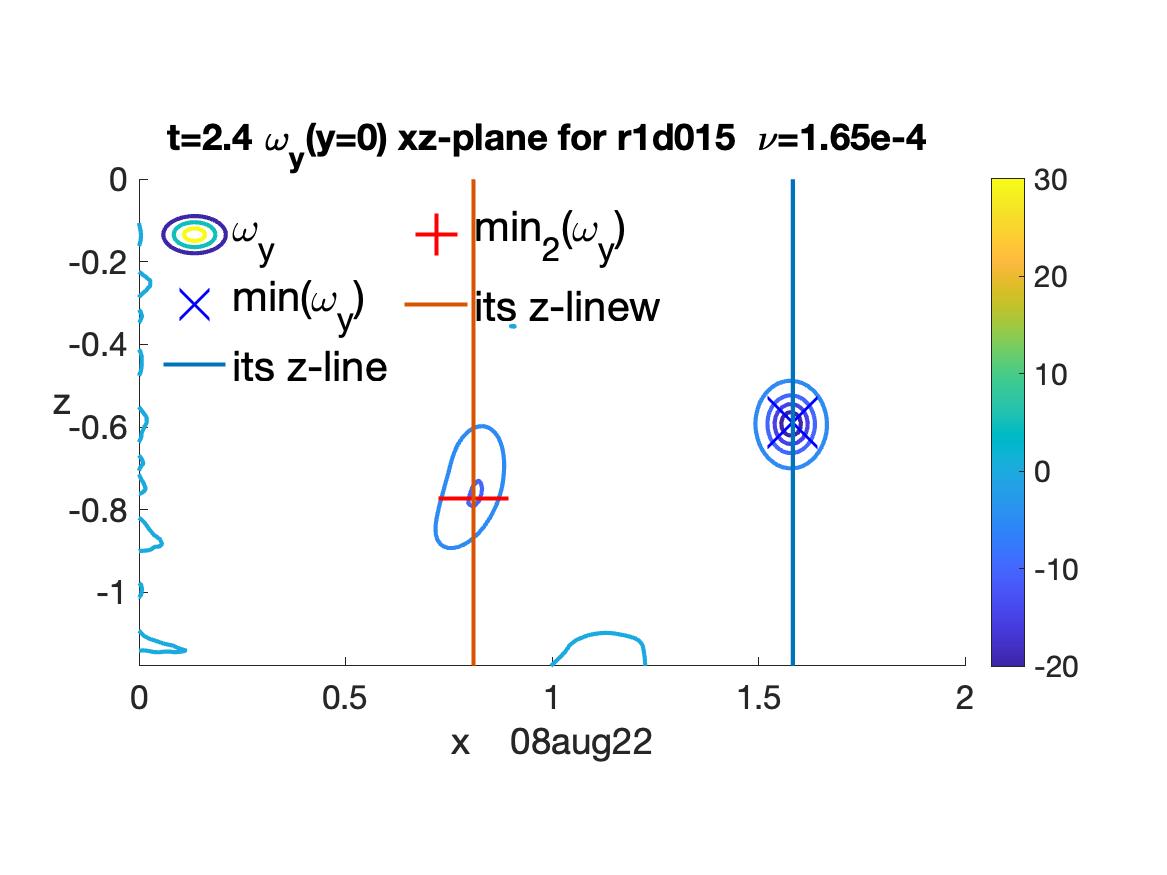}
\begin{picture}(0,0)\put(-260,46){\large(a)}\end{picture} }
\emini \bminic{0.30} \subfigure{
\includegraphics[scale=.19,clip=true,trim=0 0 0 90]{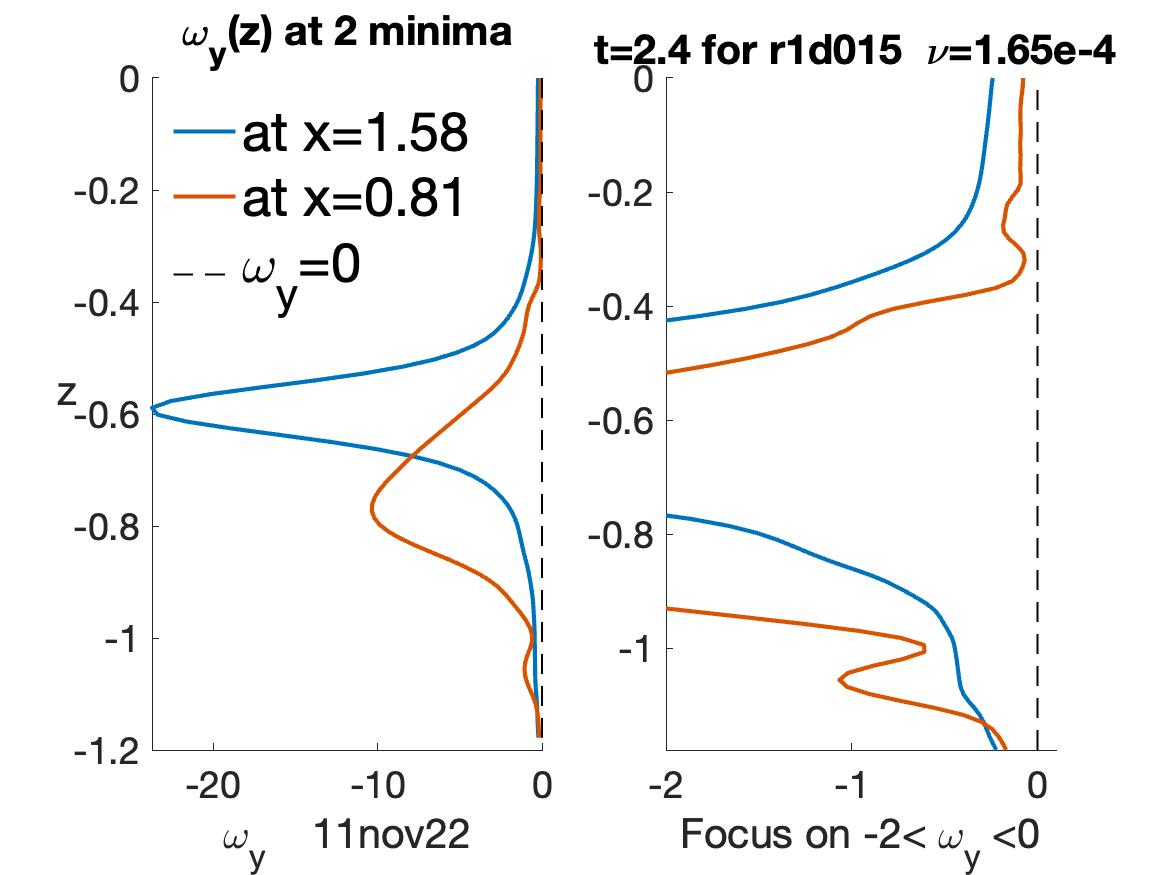}
\begin{picture}(0,0)\put(-190,46){\large(b)}\end{picture} }
\emini
\caption{\label{fig:r1d015T2p4} $\omega_y$ at $t=2.4$ on the $y=0$, $x\!-\!z$ plane 
from algebraic case r1d015 with $p_r=1$ and $r_o=0.015$ \eqref{eq:Rosenh}.
(a) Contour plot with local $\min(\omega_y)$ indicated. $|\omega_y|\sim0$ 
contours do not appear.
(b) $\omega_y(z)$ profiles through those minima at $x=1.58$ and $x=0.81$. 
First full $\omega_y(z)$, then focus on small $\omega_y$. Contours and 
profiles at $t=1.2$ are similar.} 
\end{figure}
\begin{figure}[H]
\bminic{0.61} \subfigure{
%\graphics[scale=0.26,clip=true,trim=0 90 10 100]{Gd05dm1nu16m1e3o2xzT1p2-15jul22.jpg}
\includegraphics[scale=0.26,clip=true,trim=0 90 10 100]{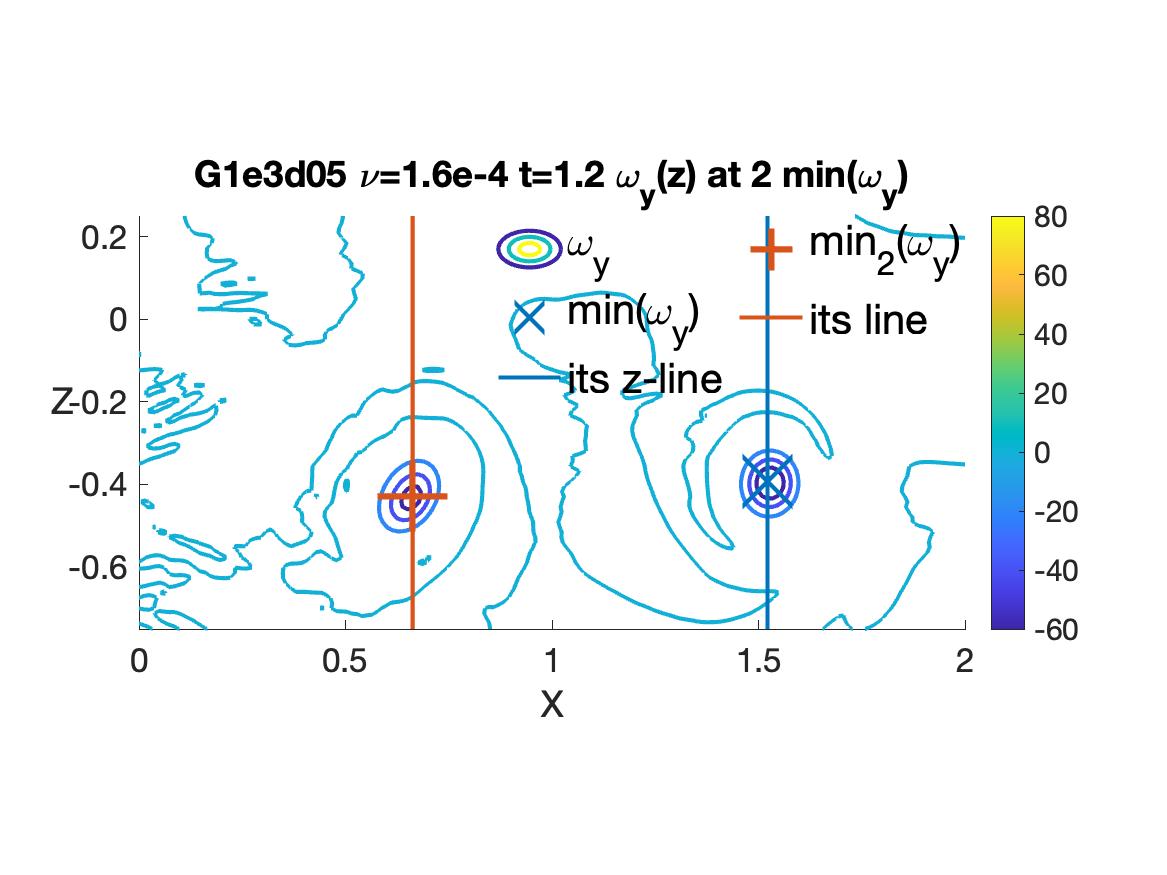}
\begin{picture}(0,0)\put(-246,86){\large(a)}\end{picture} }
\emini \bminic{0.32} \subfigure{\vspace{-3mm}
\includegraphics[scale=0.18,clip=true,trim=10 0 0 0]{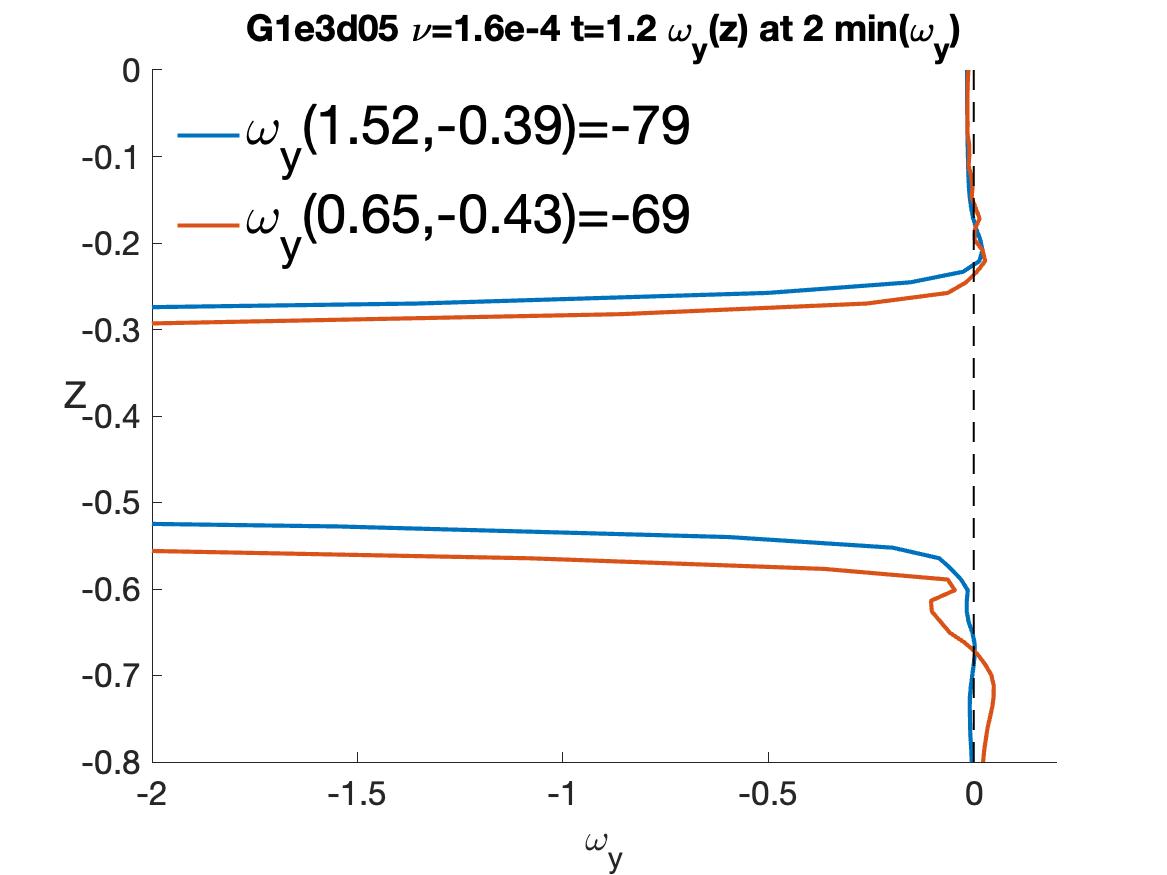}
\begin{picture}(0,0)\put(-180,76){\large(b)}\end{picture} }
\emini
\caption{\label{fig:Go2xzT1p2} $\omega_y$ at $t=1.2$ on the $y=0$, $x\!-\!z$ plane 
from the Lamb-Oseen profile \eqref{eq:Gauss} Gd05 calculation. 
(a) Contour plot with local $\min(\omega_y)$ indicated. 
A few $|\omega_y|\sim0.001$ contours are included.
(b) $\omega_y(z)$ profiles through those minima at $x=1.52$ and $x=0.65$. 
The positive overshoots of $\omega_y(z)$ 
show the magnitude of the $|\omega_y|\sim0$ contours on the left.\\
$\dagger$ Note that the $y=0$, $x-z$ plane 
negative $\omega_y$ extrema are not at the positions of 
the global $\max(|\omega|)$ for these fields.}
\end{figure}
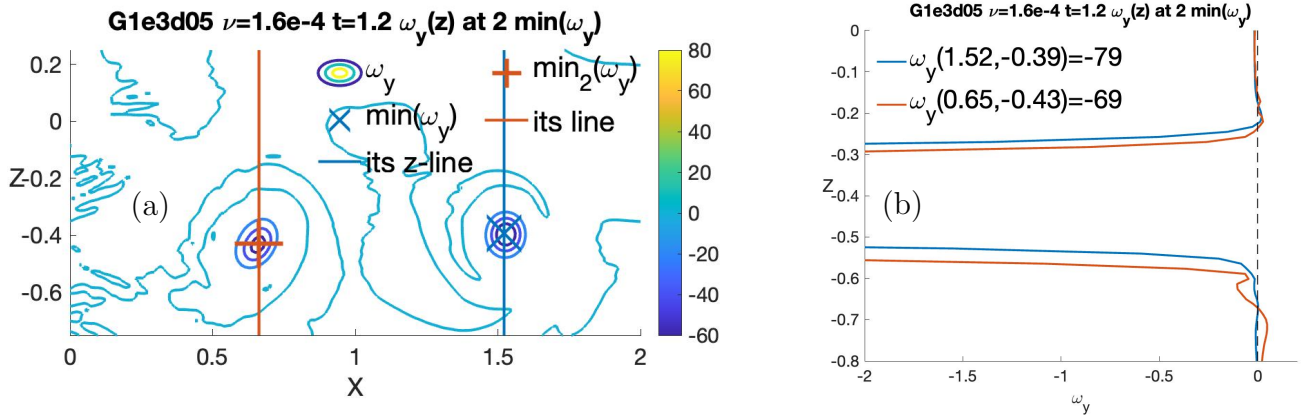
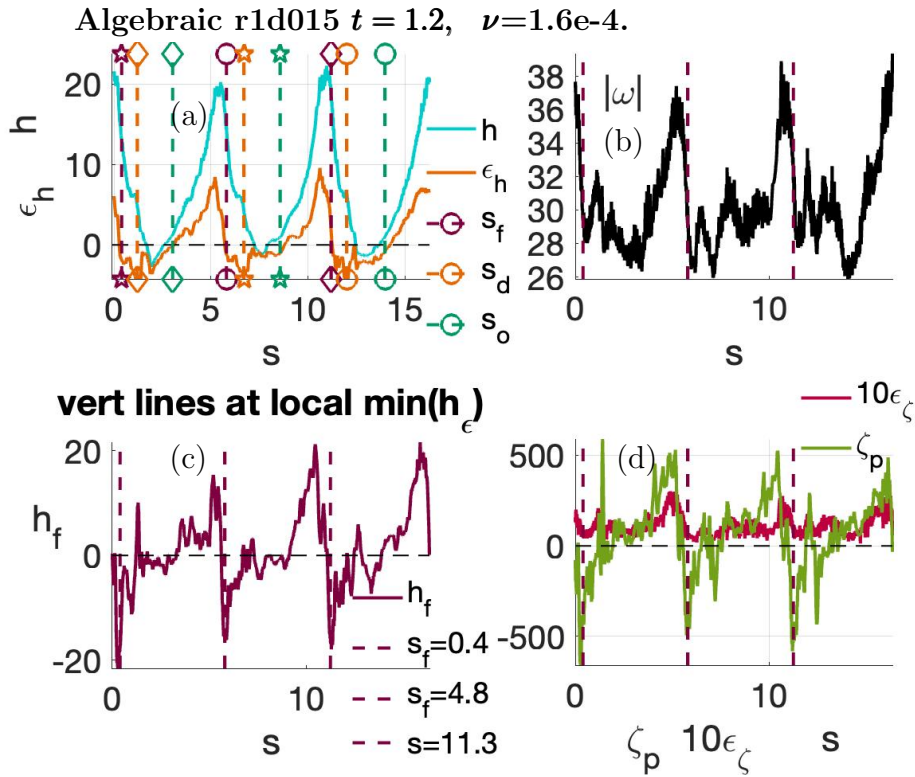
\begin{figure}[H]
\bminic{0.76}\hspace{-8mm}
\includegraphics[scale=.30,clip=true,trim=10 10 0 100]{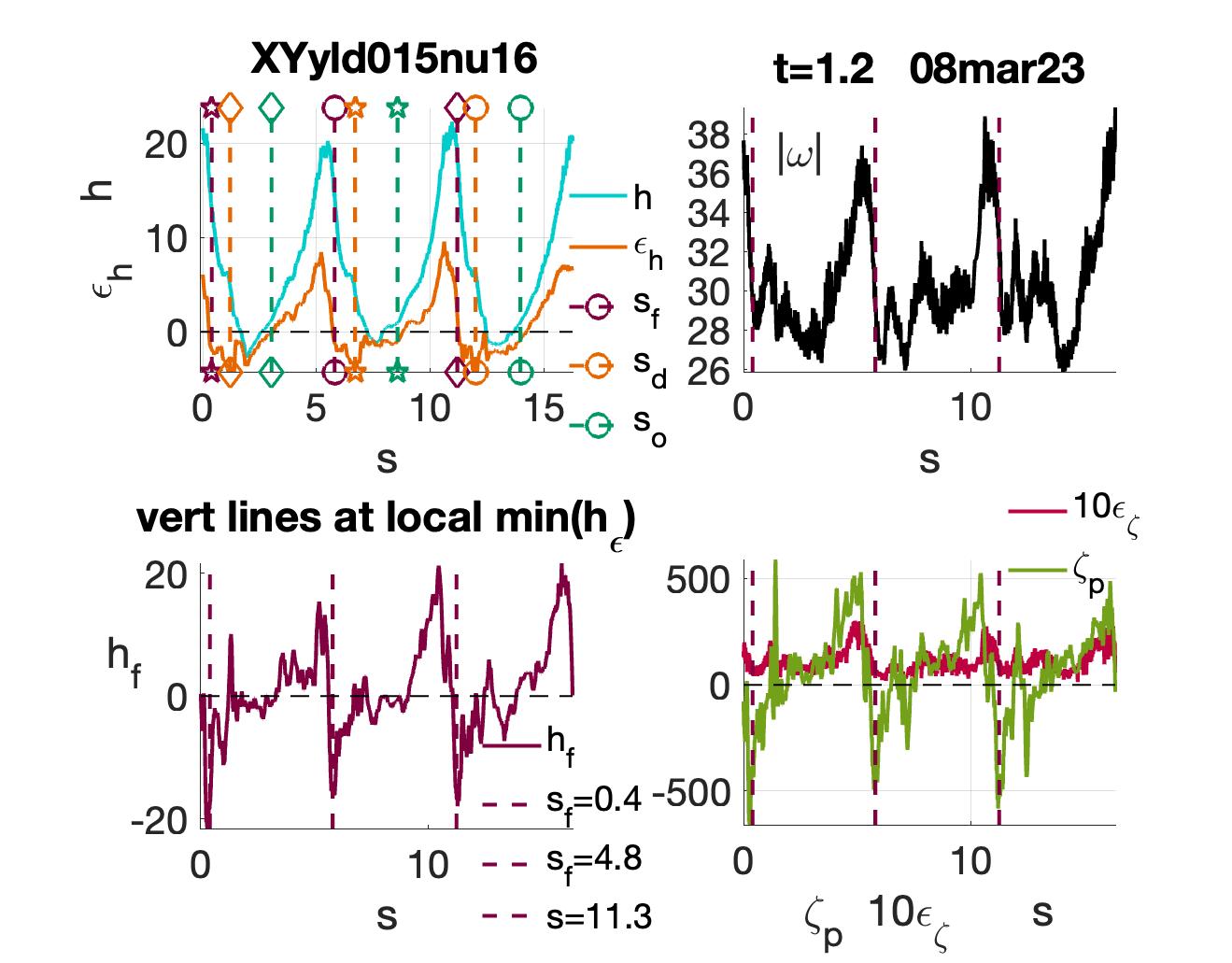} 
\begin{picture}(0,0)\put(-350,282){\large\bf Algebraic r1d015 $\pmb{t=1.2,~~\nu}$=1.6e-4.}
\put(-314,246){\large(a)}\put(-150,236){\large(b)}
\put(-314,116){\large(c)}\put(-145,116){\large(d)}
\end{picture} 
\emini\bminic{0.24}
\caption{\label{fig:T1p2uuoo} $t=1.2$ centerline budget profiles for algebraic case 
r1d015, $p_r=1$ with $r_o=0.015$, $\nu=1.6$e-4 of $h$, $\epsilon_h$, $|\omega|$,
$h_f$, $\epsilon_\zeta$ and $\zeta_p$.
(a) $h$ and $\epsilon_h$ \eqref{eq:helicity}.
(b)  $|\omega|$.  (c) $h_f$. 
(d) Production $\zeta_p$ and dissipation of 
$\epsilon_\zeta$ of the enstrophy \eqref{eq:enstrophy}.
Each frame has three vertical maroon lines at the $s_f$ positions of the local $\min(h_f)$. 
Frame (a) has two additional sets: $s_d$ positions of the local $\min(\epsilon_h)$;
$s_o$ positions that oppose the $s_f$.
All of the algebraic $0.4<t\lesssim 2.4$ budget profiles are similar to these.} 
\emini\end{figure}

\begin{figure}[H]
\bminic{0.74}\includegraphics[scale=.30,clip=true,trim=0 10 0 56]{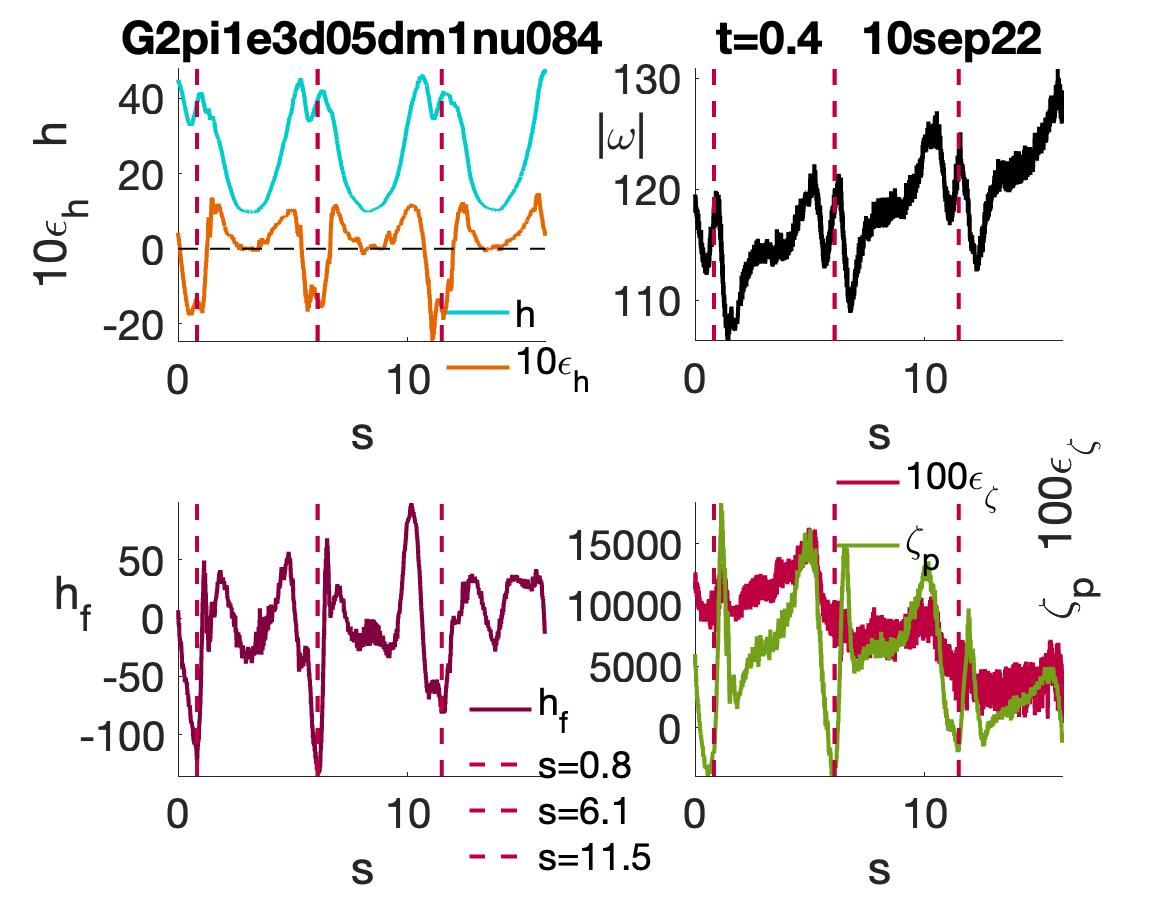} 
\begin{picture}(0,0)\put(-340,262){\large\bf Lamb-Oseen Gd05 $\pmb{t=0.4,~~\nu}$=8.35e-5.}
\put(-290,246){\large(a)}\put(-138,236){\large(b)}
\put(-290,116){\large(c)}\put(-138,116){\large(d)}
%\put(-300,186){(a)}\put(-150,186){(b)}
%\put(-300,46){(c)}\put(-150,46){(d)}
\end{picture} 
\emini\bminic{0.26}
\caption{\label{fig:GuuooT0p4} Early $t\!=\!0.4$ \\Lamb-Oseen $r_o=0.05$ centerline 
budget profiles of $h$, $\epsilon_h$, $|\omega|$, $h_f$, $\epsilon_\zeta$ and $\zeta_p$.
These are similar, but not identical, to the algebraic profiles at $t$=1.2 in 
figure \ref{fig:T1p2uuoo}.  Significant positions: (c) $s_f$ positions of local $\min(h_f)$.
These are co-located with: (a) Local $\max(h)$ and $\min(\epsilon_h)$. (b) Secondary local
$\max|\omega|$.  (d) Local $\min(\zeta_p)$, meaning at points of maximum centerline compression.}
\emini 
\end{figure}
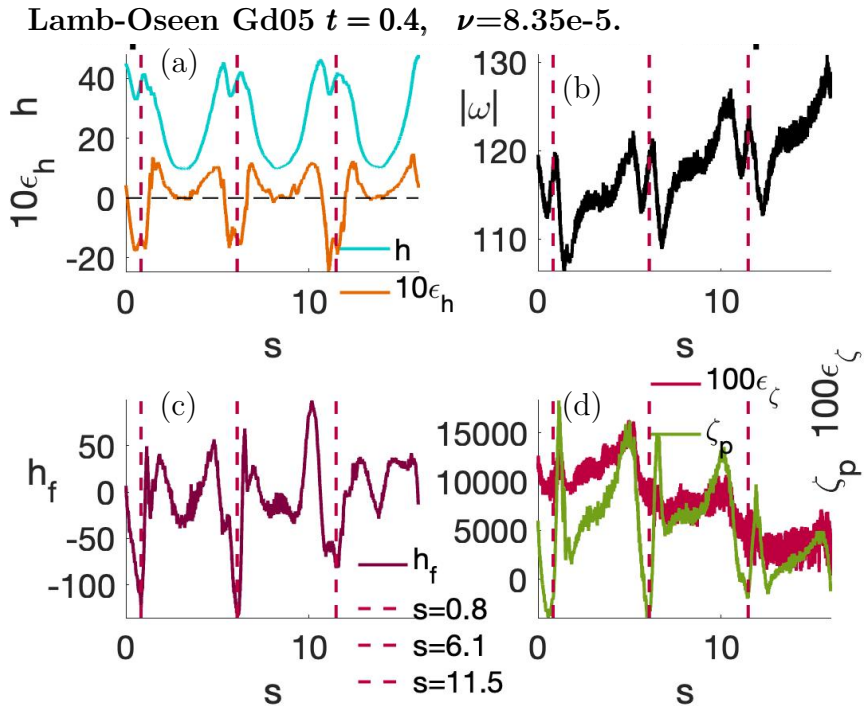
\begin{figure}[H]
%Gm1e3d5dm1nu0835T0p4uuoosuo11sep22.jpg}
%\graphics[scale=.30,clip=true,trim=0 40 0 0]{Gm1e3d5dm1nu0835T1p2uuoosuo11sep22.jpg}
\bminic{0.5}\subfigure{
\includegraphics[scale=.32,clip=true,trim=0 380 600 68]{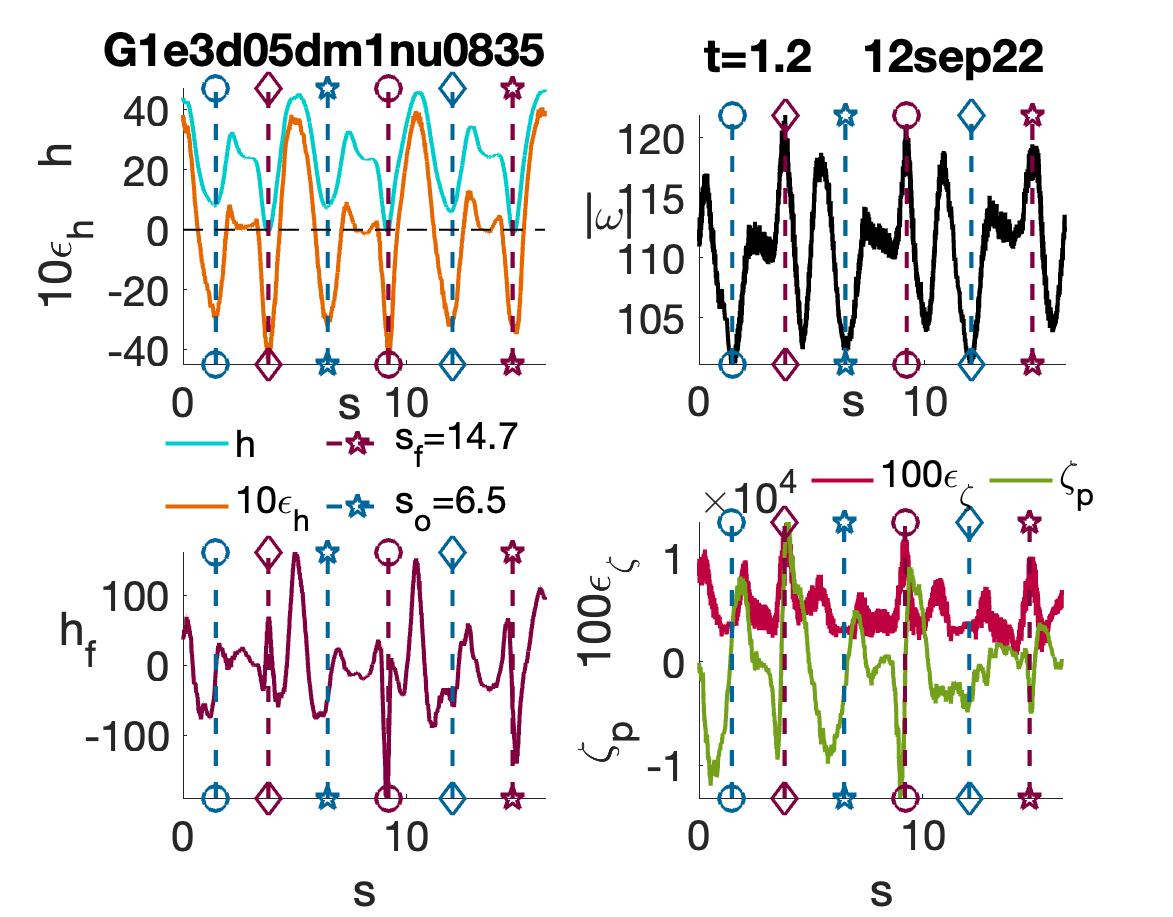}
\begin{picture}(0,0)\put(-170,36){\large(a)}
\put(-140,152){\large\bf Lamb-Oseen Gd05 $\pmb{t}$=1.2.}
\end{picture} }
\emini \bminic{0.4} \subfigure{
\includegraphics[scale=.32,clip=true,trim=567 10 0 460]{Gm1e3d5dm1nu0835T1p2uuoosuo-18feb23.jpg}
\begin{picture}(0,0)\put(-180,36){\large(b)}\end{picture} 
}
\emini
\caption{\label{fig:GuuooT1p2} $t=1.2$ $r_o=0.05$ Lamb-Oseen budget profiles. These 
are very different than the $t=1.2$ algebraic budget profiles in figure 
\ref{fig:T1p2uuoo}. In (a) there are six positions with strong negative helicity 
dissipation, local $\min(\epsilon_h)$ and local $\min(h)$.  
The positions are separated into two sets of three.  
The $s_f$ in maroon are at the strongest $\min(\epsilon_h)$, adjacent to the local $\min(h_f)$ 
($h_f$ panel is not shown).  The $s_o$ in turquoise are the points that oppose the $s_f$ in
3D figure \ref{fig:3Dr11p2G2pi}.
In (b), all six positions are at very large positive gradients of $\zeta_p$ between 
local $\min(\zeta_p)$ and $\max(\zeta_p)$. Strong local $\min(\zeta_p)$ means strong local
centerline compression. The $s_f$ are also at $\max(\epsilon_\zeta)$ positions, maxima of the
enstrophy dissipation.
} \end{figure}

%Gm1e3d5dm1nu0835T2p4uuoosuo12aug22.jpg}
\begin{figure}[H]
%\graphics[scale=.30,clip=true,trim=0 40 0 0]{Gm1e3d5dm1nu0835T1p2uuoosuo11sep22.jpg}
\bminic{0.5}\subfigure{
\includegraphics[scale=.32,clip=true,trim=42 428 565 49]{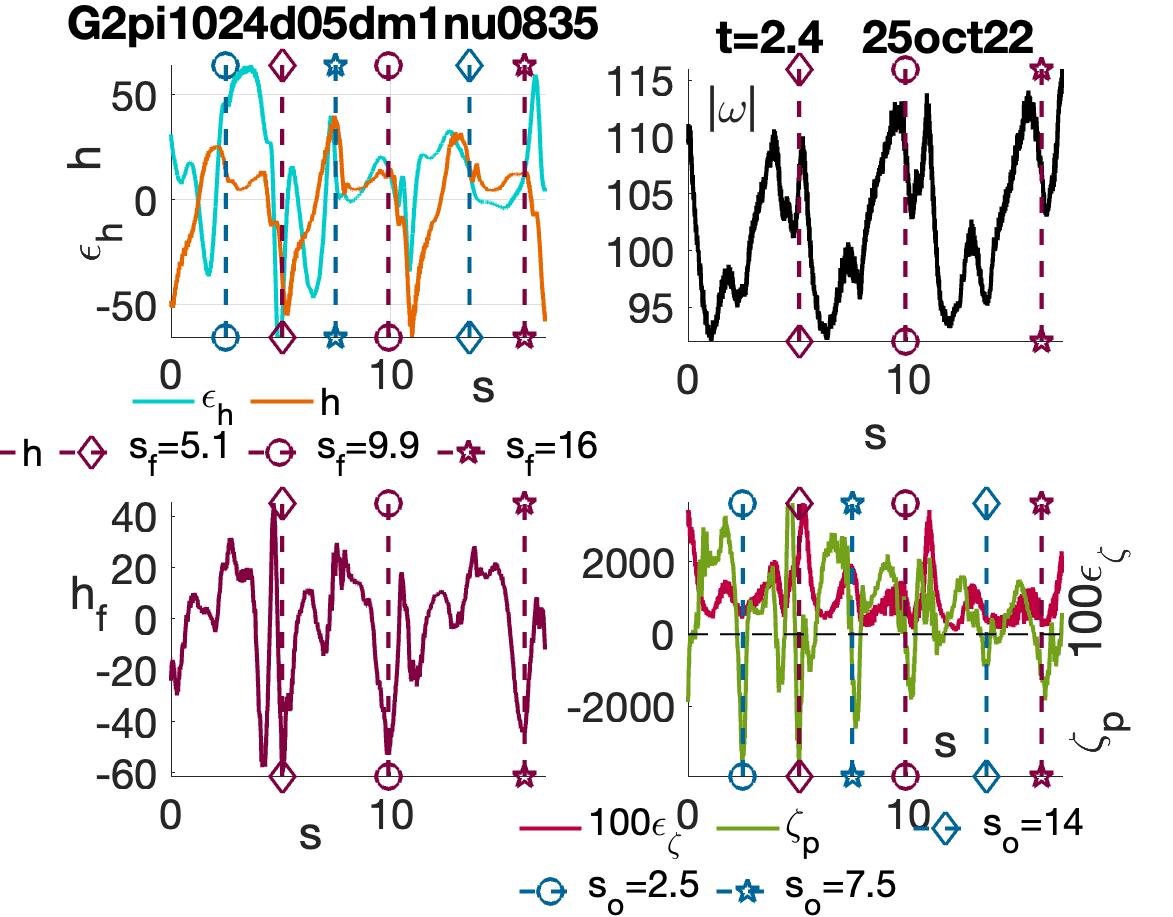}
%\graphics[scale=.32,clip=true,trim=0 400 600 30]{Gm1e3d5dm1nu0835T1p2uuoosuo12sep22.jpg}
\begin{picture}(0,0)\put(-170,36){\large(a)}
%\put(-400,210){\large\bf Lamb-Oseen Gd05 $\pmb{t}$=1.2.}
\put(-140,146){\large\bf Lamb-Oseen Gd05 $\pmb{t}$=2.4.}
\end{picture} }
\emini \bminic{0.4} \subfigure{
\includegraphics[scale=.32,clip=true,trim=535 0 0 470]{Gm1e3d5dm1nu0835T2p4uuoosuo20feb23.jpg}
%\graphics[scale=.32,clip=true,trim=567 10 0 460]{Gm1e3d5dm1nu0835T1p2uuoosuo12sep22.jpg}
\begin{picture}(0,0)\put(-180,39){\large(b)}\end{picture} 
}
\emini
\caption{\label{fig:GuuooT2p4} $t=2.4$ $r_o=0.05$ Lamb-Oseen centerline budget 
profiles. (a) $h(s)$, $\epsilon_h(s)$, $s_f$ (maroon) for local $\min(h_f)$  and the $s_f$'s 
opposing $s_o$ (turquoise) are marked. The
$\epsilon_h(s)$ profiles are three-fold symmetric again and more like the 
algebraic profiles at $t=1.2$ and $t=2.4$ and Lamb-Oseen at $t=0.4$. (b) However, there
are still six positions of local $\min(\zeta_p)<0$ compression: The three $s_f$ and three $s_o$.
Having this many local compression locations is why 
the post-reconnection Lamb-Oseen vortex structures in section \ref{sec:Greconnect} are
braids, not the sheets generated by the algebraic profiles.}
\end{figure}

\begin{figure}[H]
%\bminic{0.74}
\includegraphics[scale=.37,clip=true,trim=0 10 0 110]{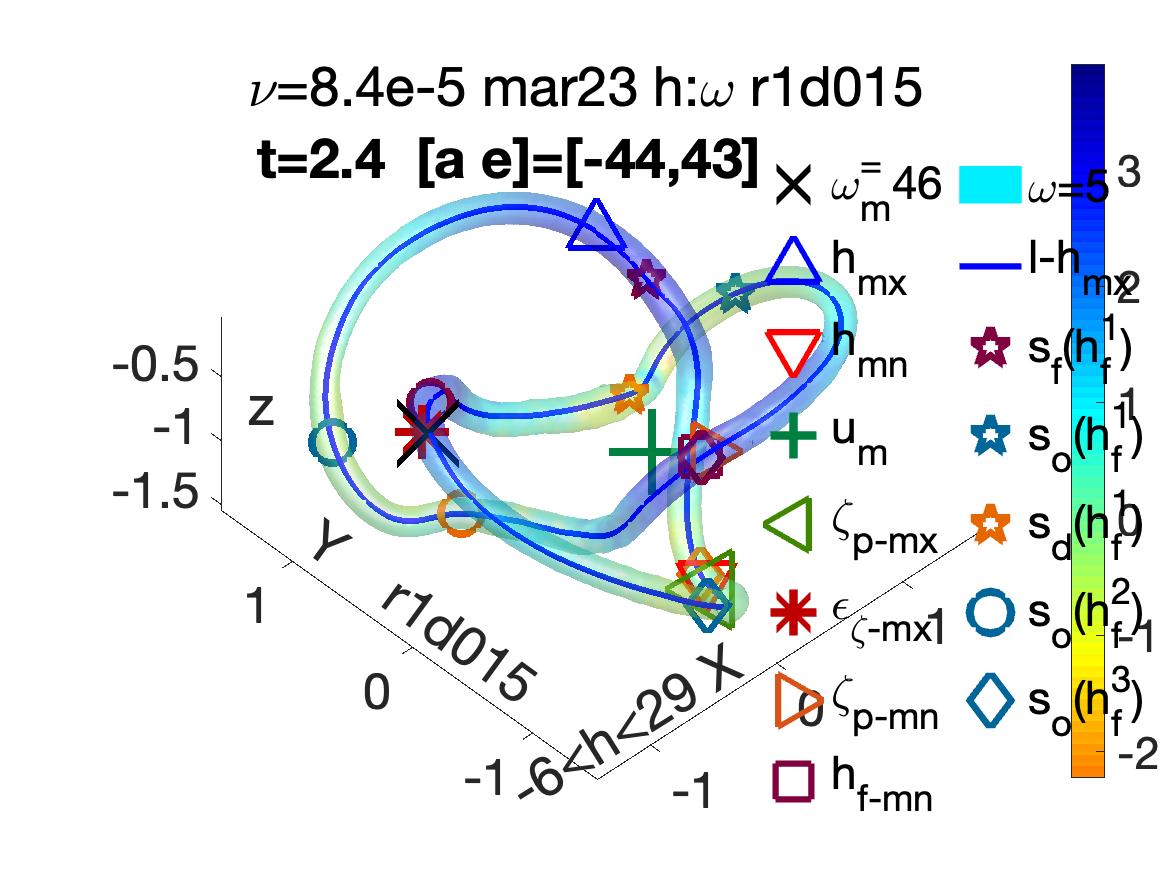} %\emini\bminic{0.26}
\caption{\label{fig:3DT2p4r1d015} 
A $t=2.4$ mapped-helicity $\omega$-isosurface from the r1d015 $p_r=1$, $r_o=0.015$ 
algebraic \eqref{eq:Rosenh} calculation at the beginning of the initial phase of reconnection. 
Symbols (from legend) show the three-dimensional positions of the 
basic $u$, $\omega$ and $h$ extrema as well as extrema from the enstrophy and helicity budget 
equations (\ref{eq:enstrophy},\ref{eq:helicity}). 
Plus, from their centerline positions in figure \ref{fig:T2p4uuoo}, 
the $s_f$ (maroon) positions of local $\min(h_f)$, the $s_o$ (turquoise) positions that
oppose the $s_f$ and the $s_d$ (yellow) positions of the local $\min(\epsilon_h)$. Each
in sets of three associated with the 1st, 2nd and 3rd local centerline $\min(h_f)$ positions.
There is a cluster of $\omega_m$ ({\bf X}), $\max(\epsilon_\zeta)$ and 
$s_f(h_f^2)=5.9$ on the left.  Another cluster is next to 
$u_m$ with $\min(h_f)$, $\min(\zeta_p)$ and $s_f(h_f^3)=11.7$. 
And one at the bottom with $\min(h)$ and $\max(\zeta_p)$ with $s_d(h_f^3)=2.3$ and $s_o=3.2$, 
both {\large$\diamond$}'s. The $s_d$ and $s_o$ with the same symbols are approaching one
another on the same centerline spans of the trefoil. 
The best diagnostic for the Biot-Savart evolution of the vortex centerlines over this period is 
the separation of the three color-coded {\large$\circ$}'s on the left from $t=1.2$, to 2.4 
then 3.6.
%in 
}%\emini
\end{figure}

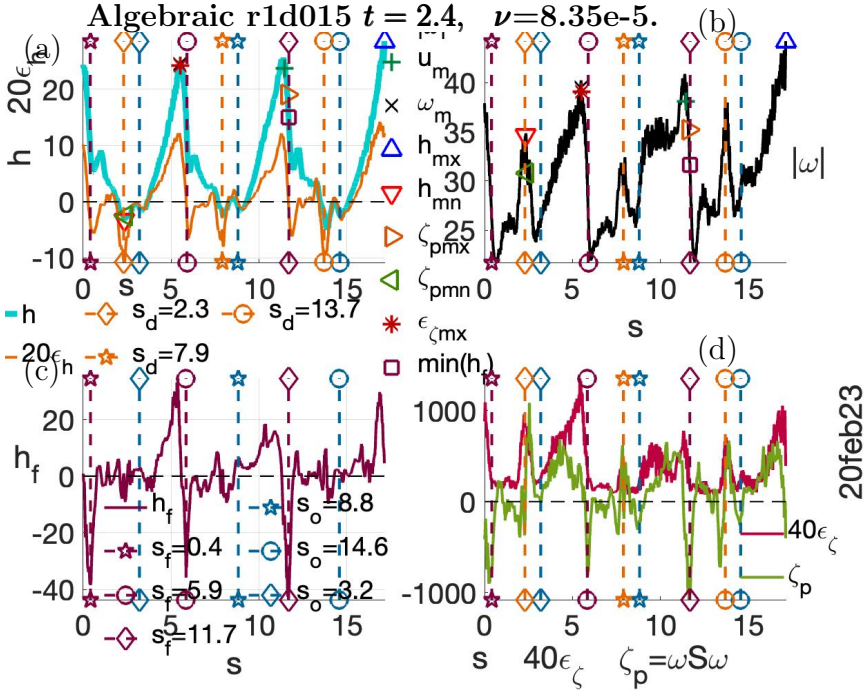
\begin{figure}[H]
\bminic{0.70}\includegraphics[scale=.26,clip=true,trim=64 10 0 65]{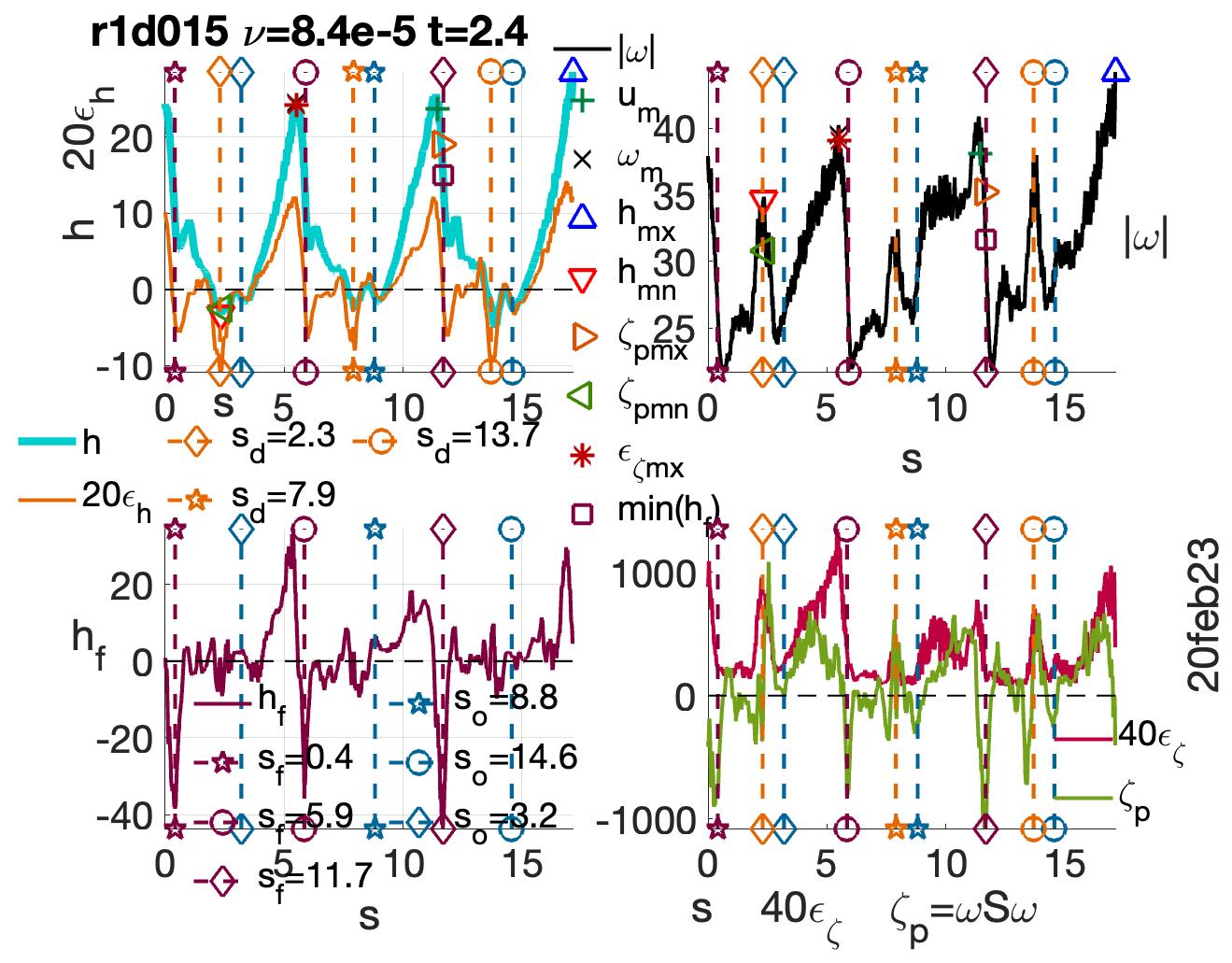}
%\begin{picture}(0,0)\put(0,50)
\begin{picture}(0,0)\put(-300,252)
{\large\bf Algebraic r1d015 $\pmb{t=2.4,~~\nu}$=8.35e-5.}
\put(-324,240){\large(a)}\put(-69,250){\large(b)}
\put(-324,118){\large(c)}\put(-69,126){\large(d)}\end{picture}
\emini\bminic{0.30}
\caption{\label{fig:T2p4uuoo} 
Vorticity centerline budget profiles at $t=2.4$ of $h$, $\epsilon_h$, $|\omega|$,
$h_f$, $\epsilon_\zeta$ and $\zeta_p$, case r1d015. %broad algebraic profile.
%Upper left: $h$ and $\epsilon_h$ 
%\eqref{eq:helicity}. Upper right: $|\omega|$.  Lower left: $h_f$. 
%Lower right: Production $\zeta_p$ and dissipation of 
%$\epsilon_\zeta$ of the enstrophy \eqref{eq:enstrophy}. 
Added to each panel are three sets of three vertical lines.
Maroon lines at the local $\min(h_f)$. Yellow for local $\min(\epsilon_h)$ and
turquoise for the $s_o$, the points opposing the $s_f$.
The $s_f$ points are on one side of each reconnection, with the 
$s_d-s_o$ zones representing the other side of those reconnections.} \emini
\end{figure}

\subsection{Rayleigh stability criterion \label{sec:Rayleigh}}
The stability of different core profiles $\omega(\rho)$ can be determined using the
$J(\rho)$ \eqref{eq:Jrr} stability functions.  The $J(\rho)$ are a type of Richardson number and
derived for columnar vortices \cite{HowardGupta1962} by extending an earlier result for shears 
on boundary layers. 

Recent analysis \cite{GallaySmets2020} that determines and uses the $J(\rho)$ begins with the
azimuthal profiles of the velocity $u(\rho)$, vorticity $\omega(\rho)$ and the pressure $p$:
\EQL{eq:VWP2} u=V(\rho)e_\theta,\quad\omega=W(\rho)e_z,\quad p=P(\rho) \,. \EN
$P$ is determined up to an additive constant by centrifugal balance 
$rP'(\rho)=V^2(\rho)$. Then by introducing, the angular velocity
$\Omega(\rho) = V(\rho)/\rho$ and $\Phi(\rho)=2\Omega(\rho)W(\rho)=-P$,
one can define these ${\cal C}^\infty$ and ${\cal C}^1$ functions:
\EQL{eq:Jrr} \Phi(\rho)=2\Omega(\rho)\omega(\rho)\quad\text{and}\quad
J(\rho)=\frac{\Phi(\rho)}{\Omega'(\rho)^2},\quad \rho>0\,. \EN
Next, consider a small, but not tiny, perturbation of one Fourier mode:
\EQL{eq:mkmode} \bu(\rho,\theta,z,t)=u_{m,k}(\rho,t)e^{im\theta}e^{ikz},\quad 
\bomega(\rho,\theta,z,t)=\omega_{m,k}(\rho,t)e^{im\theta}e^{ikz}\,, \EN
stability is determined by 
\EQL{eq:Raystable} \frac{k^2}{m^2}J(\rho)\geq\quart\quad\mbox{for all}~\rho>0 \EN
Figure \ref{fig:HGJ} shows $J(\rho)$, and how it is determined, for the Lamb-Oseen 
\eqref{eq:Gauss} and $p_r=2$ algebraic \eqref{eq:Rosenh} profiles for the 
same $\omega_o=1$ and $r_o=1$. What is important are their different $\rho\to\infty$ 
behavior. For the Lamb-Oseen profile 
\EQL{eq:JG} J_G(\rho)\to \frac{\rho^4}{r_o^2}e^{-(\rho/r_o)^2}\to 0\,, \EN
implying that the inequality \eqref{eq:Raystable} is always violated as $r\to\infty$. 

Whereas for the K-S-R $p_r=2$ algebraic profile, 
\EQL{eq:JR} \frac{k^2}{m^2}J(\rho)\to \frac{(k^2 r_o^2)}{m^2}\quad\mbox{as}\quad \rho\to\infty. \EN
This says that unless $m$ is large for $kr_o\sim1$, that is its azimuthal wavelength is 
small, then for all $\rho$, $(k^2/m^2)J(\rho)\geq\quart$ can be satisfied. 
With an example of small being the Lamb-Oseen perturbation in the inset of 
figure \ref{fig:Gt0Rht0omyz}, %5, 
probably generated by the solenoidal projection in 
initialization step 4 in section \ref{sec:config}.

Can the respective algebraic and Lamb-Oseen $J(\rho)$ stability curves in figure 
\ref{fig:HGJ} %6 
foretell whether their evolution diverges at early times? The first test 
in figures \ref{fig:r1d015T2p4} (r1d015, $t=2.4$) and \ref{fig:Go2xzT1p2} (Gd05 $t=1.2$). 
considers  vertical profiles of $\omega_y$ taken though $y=0$, $x-z$ slices.

For K-S-R, $J(\rho)\to r_o^2>0$, 
so stability is expected if $m$ is large. And demonstrated by the $\omega_y$ contours in 
figure \ref{fig:r1d015T2p4}. %7 
And for Lamb-Oseen $J(\rho)\to0$ ($<\frac{1}{4}$) and because there is a small perturbation, 
instability is possible. And demonstrated by the irregular $\omega_y\sim0$ contours in 
figure \ref{fig:Go2xzT1p2}. %8. 
What is less clear for Lamb-Oseen is how tiny the perturbations must be to create 
instability \cite{GallaySmets2020}. As discussed in section \ref{sec:discuss}.

%In contrast, for these and other %configurations \cite{Kerr2013b,KerrJFM2018c} 
%using vortices with the $p_r=1$ algebraic 
%Rosenhead regularized profile \eqref{eq:Rosenh}, the generation of $t\gtrsim0$ 
%unphysical instabilities has not been observed. 
%\newpage
\subsection{Effect of being stable or unstable \label{sec:effect}}

Do the stability differences indicated by figures \ref{fig:HGJ}, \ref{fig:r1d015T2p4} and
\ref{fig:Go2xzT1p2} yield differences in 
the subsequent evolution of the Lamb-Oseen and algebraic cases? 

One difference between the respective $x-z$ slices (figures \ref{fig:r1d015T2p4} and 
\ref{fig:Go2xzT1p2}) is that the algebraic contours in figure \ref{fig:r1d015T2p4} do not 
generate oppositely-signed contours. In contrast, Lamb-Oseen in figure \ref{fig:Go2xzT1p2} does:
as shown by the $|\omega_y|\sim0$ contours and the $\omega_y(z)$ slice on the right.
These fluctuations of oppositely-signed $\omega_y$ are a source of local interactions.
Local interactions that could be the source for the $t=1.2$ differences between the algebraic and 
Lamb-Oseen centerline budget profiles in figures \ref{fig:T1p2uuoo},%9
and \ref{fig:GuuooT1p2} respectively. %11, 
This is discussed further in section \ref{sec:early}. %III-A

\subsection{Mapping budgets terms onto centerline vortices \label{sec:mapping}}

While single-color helicity isosurfaces \cite{KerrFDR2018} suggested that helicity
has a role in reconnection, the mapped $h$-vorticity isosurfaces used 
by two 2021 trefoil papers \cite{YaoYangHussain2021,ZhaoScalo2021} are a better 
tool.  In particular, small values of localized oppositely-signed helicity $h<0$ 
indicated where reconnection was forming. 

There are similar yellow to red $h<0$ patches at $t=1.2$ in figure \ref{fig:3Dr11p2G2pi}.
For both algebraic and Lamb-Oseen.
%With yellow indicating weak $h<0$ and stronger $h<<0$ reddish patches are on the larger
%$|\omega|$ isosurfaces inside these, as for $t=3.6$ in figure \ref{fig:T3p6uuoo}.
And for all cases, up to $t=3.6$ there are similar $h<0$ patches on their inner, higher $\omega$ 
isosurfaces. However, are the observed $t\leq3.6$ differences sufficient for identifying the 
origins of the post-reconnection differences in the evolution of the algebraic and Lamb-Oseen
calculations? Given how small those $t\leq3.6$ inner isosurface differences are, they are not. 

Why are the surface helicities of the different cases qualitatively similar?
Likely because before reconnection begins, similar long-range Biot-Savart terms dominate the 
surface helicity dynamics for all cases .
Therefore, what is needed are new diagnostics related to what is within the isosurfaces to 
explain the major differences in the $T>3.6$ enstrophy and helicity evolution in 
figures \ref{fig:Gd05dm1ZHnus} and \ref{fig:r1ZHnu}.  % figures 1 and 2.  
Meaning another set of pre-reconnection diagnostics is required.

Because these are questions about the evolution of local helicity $h(\bx,t)$, 
which is controlled by its budget equation \eqref{eq:helicity}, one alternative
set of diagnostics is to instead map the primary terms 
from the enstrophy and helicity density budget equations 
(\ref{eq:enstrophy},\ref{eq:helicity}) onto the isosurfaces.  
The variations of these terms upon the isosurfaces are very small, so
are not useful for analysing the dynamics by themselves. However,
this exercise indicated that the local variations are strongest near the
centerlines. 

Suggesting that a better way to visualize the budget terms would be to map them onto 
the vorticity centerlines directly, if the centerlines can be identified. 
If successful, this 
would provides us with an analysis tool that is both local (at a point) and global 
(between distant points on the centerline).

To identify centerlines one must first choose appropriate seed points $\bx_\omega(0)$
within a vorticity isosurface, then trace the vortex lines emanating from those
points using a streamline function, giving
trajectories $\bx_\omega \in{\cal C}$ obeying
\EQL{eq:vortexlines} \bxi_\omega(s)=\frac{d\bx_\omega(s)}{ds}=\bomega(\bx_\omega(s))\,,
~~\mbox{whose lengths are}~~L_\omega=\oint\left|\bxi_\omega(s)\right|ds\,. \EN
In \cite{KerrJFMR2018,KerrJFM2018c} the position of the maximum vorticity was 
used as the seed. With more experience, it has been found that seeding at 
either maximum or minimum of helicity, then using $-\bomega(\bx)$ direction in
\eqref{eq:vortexlines}, yields trajectories that stay within the observed 
isosurfaces. This is the practice in this paper.

In all cases, the trajectories do not close upon themselves perfectly, which is
only relevant for determining the topological numbers, 
twist, helicity and self-linking as in earlier work \cite{KerrJFMR2018,KerrJFM2018c}. 
That is not an objective of this paper.

Once the trajectories have been defined, the profiles of important dynamical
terms are mapped onto those curves to determine how those properties are
related to one another.  

Note that because these vortex lines are almost closed upon themselves, 
initially the integral of the stretching $u_{s,s}=d\bu/ds\cdot\hat{\omega}$ 
on the $\omega$-line is identically zero: 
\EQL{eq:linestretch} \oint_0^{L_\omega} u_{s,s}ds=u(L_\omega)-u(0)\equiv0\,.\EN 
Due to this, any stretching along this line at $t=0$ is balanced by equal compression 
somewhere else.  And for these vortices, 
that compression also immediately yields an increase in the local enstrophy dissipation and
negative helicity dissipation rates, $\epsilon_\zeta$ and $-\epsilon_h$. 
As well as a very early decrease in the enstrophy and increase in the helicity: 
$dZ/dt|_{t=0}<0$ and $d{\cal H}/dt|_{t=0}>0$ as seen in figures 
\ref{fig:Gd05dm1ZHnus} (Lamb-Oseen) and \ref{fig:r1ZHnu} (algebraic). %1 and 2 
More for the larger $\nu$ Lamb-Oseen calculations than the others.

\subsection{Using these tools as time progresses. \label{sec:use}}

\begin{figure}[H]
\bminic{0.55} 
\includegraphics[scale=.36,clip=true,trim=0 482 360 0]{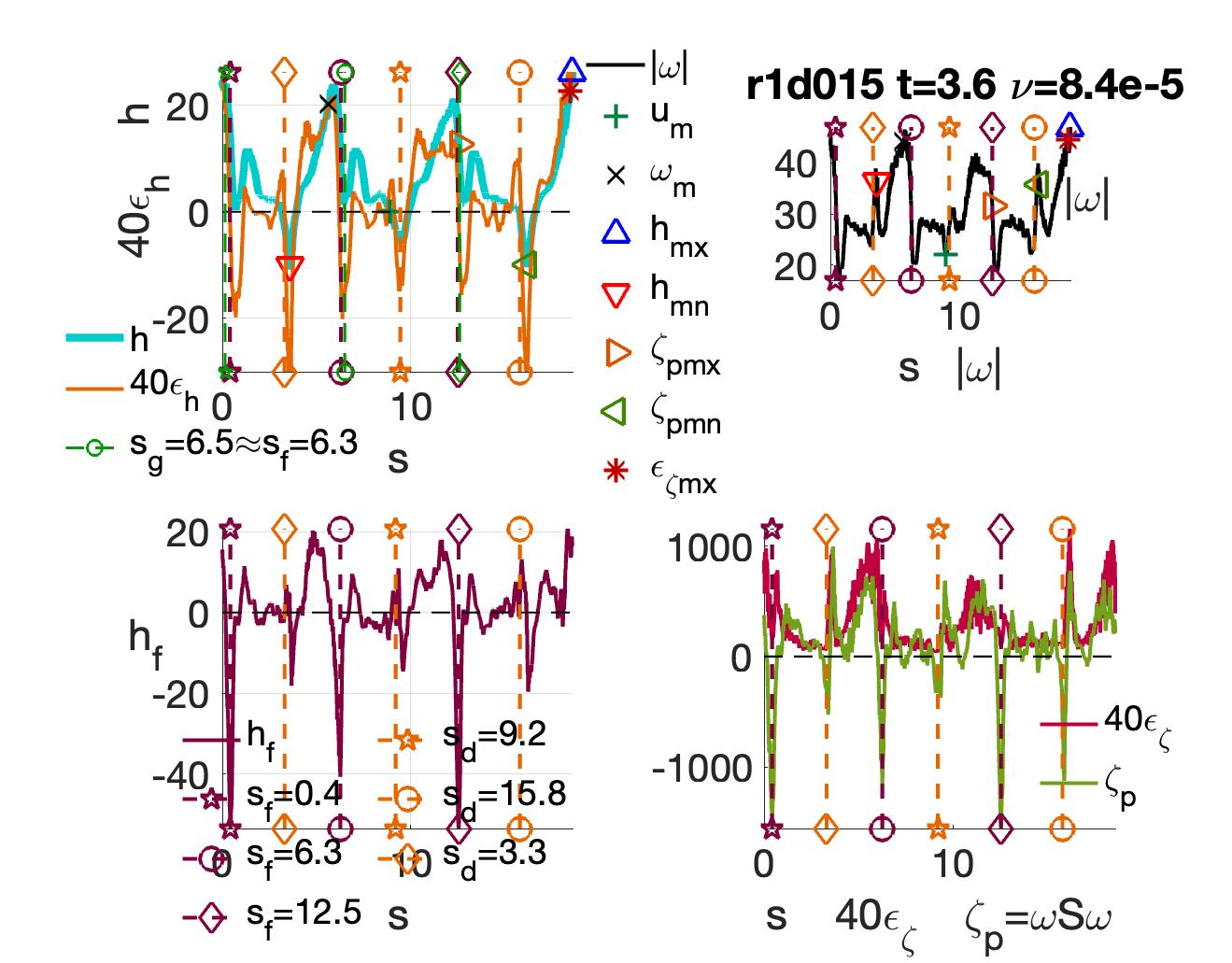} 
\emini\bminic{0.5} 
\includegraphics[scale=.20,clip=true,trim=60 40 30 90]{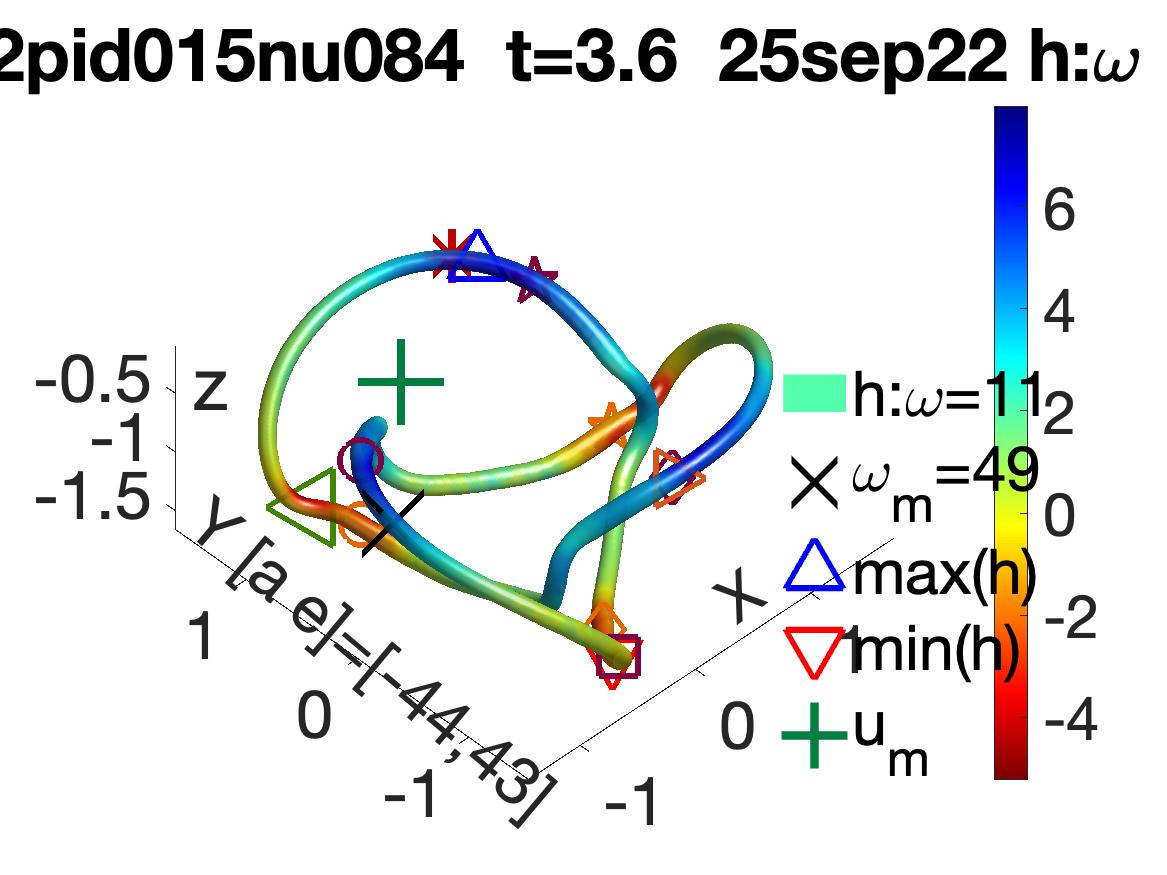}  
\begin{picture}(0,0)
\put(-440,170){\large(a)}
\put(-192,120){\large(b)}
\put(-410,170){\bf\Large{r1d015~~$\nu$=8.4e-5}}
\put(-150,162){\bf\Large t=3.6}
\end{picture} 
\emini \\ \vspace{4mm}
\includegraphics[scale=.36,clip=true,trim=0 0 0 550]{XYyl512d015nu084T3p6uuoosuo03mar23cut.jpg}
\begin{picture}(0,0) \put(-384,174){\large(c)}\put(-177,174){\large(d)} \end{picture} \\
\vspace{-12mm}
\caption{\label{fig:T3p6uuoo} %15
Vorticity centerline profiles and an isosurface plot at $t=3.6$ for case r1d015. 
Budget profiles: $h$, $\epsilon_h$, $h_f$, $\epsilon_\zeta$ and $\zeta_p$, with %broadalgebraic
added vertical dashed lines in each panel for these local positions: $s_f$ (maroon, $\min(h_f)$); 
$s_d$ (yellow, $\min(\epsilon_h))$; with in the upper-left panel $s_g$ (green) for the $s_d$ 
opposing points.
The $s_f$ are also at $\min(\zeta_p)$ and are at two of the $\max(\epsilon_\zeta)$ positions,
local enstrophy dissipation peaks.
The $s_d$ are also at the local minima of 
the helicity $\min(h)<0$, at cross-overs between secondary local $\min(\zeta_p)$ to 
$\max(\zeta_p)$ and at two of the local $\max(\epsilon_\zeta)$ positions.  And are co-located 
with the opposing  positions to the $s_f$. The $s_g$ oppose the $s_d$ and nearly coincide with
the $s_f$. Where might reconnection form? The positioning of the $s_f$ and $s_d$, plus
their opposing points, suggests that reconnection would form between the $s_f$ and $s_d$.
Consequences: Local $\zeta_p<0$ means that $du_s/ds<0$ and due to incompressibility this implies 
the existence of stretching perpendicular to the vorticity at these points. The stretching 
needed to needed to create the $h<0$ vortex sheets. The upper-right panel uses 
a larger vorticity ($\omega=0.2\omega_m$) isosurface than
in figure \ref{fig:3DT3p6} to show continuity with the earlier inner isosurface 
evolution. The labels for the  auxiliary symbols are in figure \ref{fig:3DT3p6}.
}
\end{figure}

The six terms from enstrophy and helicity budget terms that are mapped onto the 
centerlines are arranged into four panels:
\ITM\item[a)] The helicity density $h$ (cyan) 
and its dissipation rate $\epsilon_h$ (yellow). 
\item[b)] The vorticity magnitude $|\omega|=\sqrt{\zeta)}$ (black). 
\item[c)] Helicity flux $h_f$ (maroon),
which includes a pressure gradient. 
\item[d)] Enstrophy density dissipation $\epsilon_\zeta$ (red) and 
production $\zeta_p$ (lime). 
\ITN
All four panels appear in figures \ref{fig:T1p2uuoo}, \ref{fig:GuuooT0p4}, \ref{fig:T2p4uuoo} and 
\ref{fig:GuuooT3p6}. For
figures \ref{fig:GuuooT1p2} (Gd05, $t=1.2$), \ref{fig:GuuooT2p4} (Gd05, $t=2.4$) and  
\ref{fig:T3p6uuoo} (r1d015, $t=3.6$, some panels are not shown.
In particular panel b) with $|\omega|$ is not shown because its $s$-profile closely 
follows that for the helicity $h$.

Figures with all, or most, of these six mapped terms are teamed with relevant
three-dimensional helicity-mapped vorticity isosurfaces.  The following markers 
indicate the locations of the primary extrema in three-dimensional space: 
%\ITM\item[{\bf X}] $\omega_m=\|\omega\|_\infty$ $\pmb\bblue{\triangle}$ max($h$)
\ITM\item[$\omega_m$]=$\|\omega\|_\infty$ {\bf X} (black);
max($h$) $\pmb{\triangle}$ (blue);
min($h$) $\pmb{\bigtriangledown}$ (red); max($u$) $\pmb{+}$ (green). 
\ITN
$\bullet$
The additional global extrema from the budget equations are:
% \item  min($\pmb{\textcolor{Dandelion}{\epsilon_h}}$) 
\ITM\item[max($\zeta_p$)] {\LARGE$\pmb{\triangleleft}$} (JungleGreen);% <
~~~~min($\zeta_p$) {\LARGE$\pmb{\triangleright}$} (RedOrange);% >
\item[max($\epsilon_\zeta$)] $\pmb{*}$ (VioletRed);% *
~~~~min($h_f$) $\pmb{\square}$ (Maroon).
\ITN
$\bullet$ To identify relationships between the budget terms on the four panels, sets of
three-fold symmetric dashed vertical lines are added at significant positions to allow
comparisons between panels. The choice of vertical lines changes over time.
\ITM\item At early times when transport along the vortices is most important, the local extrema of
negative helicity transport $\min(h_f)$ positions are the best, and are identified by these marks, 
with vertical lines.:
\item {\bf Maroon} $s_f$ indicate the positions of the local $\min(h_f)$ with
these symbols:
\ITM\item $\star$ star,~%$\pmb{\textcolor{Maroon}{\star}}$, 
\item $\diamond$ diamond,~ %$\pmb{\textcolor{Maroon}{\diamond}}$, 
\item $\circ$ circle. % \\ %$\pmb{\textcolor{Maroon}{\circ}}$. 
\ITN
\item {\bf Yellow} The $s_d$ positions for local $\min(\epsilon_h)$ are important when
reconnection is, or will be, forming.
\item {\bf Turquoise} is used for the $s_o$/$s_o^+$ positions opposing (min or max=+) extrema of 
the helicity flux $h_f$. That is the $s_o$/$s_o^+$ oppose in 3D the $s_f$/$s_f^+$ respectively. 
Each $s_o$/$s_o^+$ is separated from its $s_f$/$s_f^+$-position by 
approximately $\Delta s=L_\omega/2$ along the centerline, where
$L_\omega/2$ is half the length of the centerline trajectory \eqref{eq:vortexlines}.
\ITM\item For algebraic case r1d015: At $t=2.4$ in figures 
\ref{fig:3DT2p4r1d015},\ref{fig:T2p4uuoo} the $s_o(h_f)$ that oppose the $s_f$ are near $s_d$ 
with local $\min(\epsilon_h)$.
\item For $t=3.6$ is transitional with the $s_f$ and $s_d$ in figure \ref{fig:T3p6uuoo}
being equally important as they mark the opposite sides of each developing reconnection site. \ITN
\item {\bf Green} is for the three-fold $s_g$ points opposing the $s_d$ local $\min(\epsilon_h)$ 
points.
\ITM\item For $t=4.8>t_r=4$ the $s_d$ and $s_g$ mark where there is active reconnection.  \ITN
\item For {\bf Lamb-Oseen} $t=1.2$ and 2.4 in figures  \ref{fig:GuuooT1p2}, \ref{fig:GuuooT2p4}
the $s_f$ and $s_o$ mark where reconnection will form.
\item[$\circ$] For Lamb-Oseen $t=3.6$, reconnections are marked by 
pairs of local $s_f^+$ (in cobalt) and $s_o^+$ points in 
figures \ref{fig:GuuooT3p6} and \ref{fig:G3DT3p6}a. 
$\circ$ While the $s_d$ and $s_f$ are co-located and far
from the active reconnection between the $s_f^+$ and $s_o^+$.
%\item[$\circ$] Additional three-fold symmetric positions are added to the figures as needed.
\ITN

These budget maps are used to determine the dynamical interplay between the enstrophy and 
helicity over the period leading to reconnection for the broadest $p_r=1$ algebraic 
case r1d015 and Lamb-Oseen case Gd05. 
For the K-S-R, $p_r=2$ cases in section \ref{sec:KSRdiss}, only the 
essential time evolution and mapped helicity isosurfaces are given.

\vspace{-6mm}
\section{\label{sec:results}Results}
\vspace{-2mm}

The comparisons between helicity-mapped vorticity isosurfaces and the mapped 
centerline budget terms are presented chronologically:
\ITM\item[III-A] Early times for algebraic and Lamb-Oseen ($t=0.4,1.2$). 
\item[III-B] Algebraic mid-reconnection $t$=2.4
and pre-reconnection $t=3.6$, with the first appearance of extended $h\!<\!0$ vortex sheets. 
\item After $t=3.6$, the algebraic and Lamb-Oseen vortical structures and global evolution 
of $Z(t)$ and ${\cal H}(t)$ diverge, as shown by figures \ref{fig:Gd05dm1ZHnus} and \ref{fig:r1ZHnu}. 
\item[III-C] $t\geq3.6$ Lamb-Oseen Gd05. In figure \ref{fig:G3DT4p4} 
reconnection with vorticity bridges, localized sheets, then $t=4.4$ braids. 
\item [III-D] $t\geq4.8$ algebraic reconnection with broad $h<0$ $\omega$-sheets leading to 
wrapping and accelerated enstrophy growth. 
\item[III-E] Finally there is  a short discussion of the K-S-R $p_r=2$ r2d05 case.
\ITN

%\newpage
\vspace{-6mm}
\subsection{Early times ($t=0.4,1.2$) profile dependent evolution and differences. 
\label{sec:early}}
\vspace{-2mm}
To begin, recall that for the $t=1.2$ isosurfaces in figure \ref{fig:3Dr11p2G2pi} 
(cases r1d015, Gd05), the only clear difference between the
frames is the position of the vorticity maximum $\omega_m$.  
Can the centerline budget maps identify any greater differences at early times?
First, the similarities at very early times are given, then the differences.

The centerline maps for the corresponding earliest times in figures 
\ref{fig:T1p2uuoo}, $t=1.2$ algebraic, and  \ref{fig:GuuooT0p4}, $t=0.4$ 
Lamb-Oseen, are similar. While the strongest local $\max(h)$ and local 
$\max(|\omega|)$ are near to one another, other local extrema are associated
with local $\min(h_f$), the vortical helicity flux indicated by dashed maroon 
lines at local $s_f$. Positions of local helicity dissipation minima 
($\min(\epsilon_h)<0$) are near the $s_f$ and the positions of
local compression, $\min(\zeta_p)<0$ are on the $s_f$. Suggesting that
the dominant dynamics at these points is local compression with pinching at
these points on the vortices.

However, starting at $t=1.2$ the centerline dynamics of the two profiles diverge.
\vspace{-1mm}
\ITM\item For {\bf algebraic case r1d015}, the alignments in figure
\ref{fig:T1p2uuoo} persist from $t=0.4$ until the
reconnection time of $t_r\sim4$ is approached. 
\vspace{-1mm}
\item However, for {\bf Lamb-Oseen at $\pmb{t=1.2}$} the corresponding Lamb-Oseen 
budgets in figure \ref{fig:GuuooT1p2} are very different, showing six locations with
roughly equivalent variations of the positive and negative helicity dissipation
$\epsilon_h$ at six significant local $\min(h_f)$ positions, split
into two sets of three, maroon $s_f$ and turquoise $s_o$.
\ITN
\vspace{-1mm}

In figure \ref{fig:GuuooT1p2}a the $s_f$ positions at local $\min(h_f)<0$ (not shown)
are also at the largest dips of $h\!\sim\!0$ and the strongest local $\min(\epsilon_h)$.  
In (b), the $s_f$ are not exactly on local $\min(\zeta_p)$, but on the adjacent 
large positive gradients and local enstrophy dissipation peaks: $\max(\epsilon_\zeta)$.
These $s_f$ can be viewed as one side of the developing reconnection sites.

The turquoise $s_o$ positions that oppose the $s_f$ positions in
figure \ref{fig:3Dr11p2G2pi} are the other side of the developing reconnections. They are also 
secondary local $\min(\epsilon_h)$, secondary local dips in $h$ and near secondary local 
$\min(\zeta_p)$.  Meaning that all six positions (the $s_f$ and $s_o$) are
sitting at or near local compressive $\min(\zeta_p)<0$. 

Having multiple points of local compression at an early time 
has a significant effect upon the the enstrophy growth (or decay). 
At $t=1.2$ and 2.4, the localized pinching enhances the localized dissipation of both 
helicity $\epsilon_h$ and enstrophy $\epsilon_\zeta$, which also suppresses the $\zeta_p$ terms 
needed to enhance enstrophy growth: before that growth has even begun.
A likely source of this localization of the dynamics is the interactions between
the primary vorticity and the oppositely-signed flotsam seen in 
figure \ref{fig:Go2xzT1p2}. That is, the origin of this localized dynamics is the 
amplification of that noise by instability, as previously suggested \cite{Kerr2013a} and 
discussed here in section \ref{sec:Rayleigh}. 

\vspace{-0mm}
The $t=2.4$ Lamb-Oseen centerline budget profiles in figure \ref{fig:GuuooT2p4} show 
some return to normal. They have similarities with the $t=0.4$ Lamb-Oseen profiles in
figure \ref{fig:GuuooT0p4} and the pre-reconnection algebraic profiles for $t\leq3.6$.
While there are only three local $\min(\epsilon_h)$ and $\min(h_f)$, in the right frame there 
still is strong compression with local $\min(\zeta_p)<0$ at all six of the 
former ($t=1.2$) $\min(h_f)$ positions: The three current ($t=2.4$) $s_f$ positions and 
their three $s_o$ opposing positions.  In addition, the magnitudes of 
the enstrophy production $\zeta_p$ and dissipation $\epsilon_\zeta$ terms are 
tempered, being a factor of 5 less than at $t=1.2$.

This localized dynamics is only temporarily stronger than the long-range Biot-Savart interactions: 
Once that dynamics dissipates, the Biot-Savart interactions again control the
large scales and the evolution of the centerline trajectory. However, the dynamics along the
centerlines is permanently affected. When reconnection bridges do form, with some enstrophy 
growth, it is entirely concentrated at the locations in figure \ref{fig:GuuooT1p2}. 
Not over the entire trefoil. With rapid post-reconnection dissipation of the vorticity in the 
bridges, leading to divergent evolution of the enstrophy $Z(t)$ 
and the helicity ${\cal H}(t)$. Explained further in section \ref{sec:Greconnect}.

\begin{figure}[H]
%\graphics[scale=.33,clip=true,trim=60 30 30 0]{figjpgFeb22/XYyl2pi1e3d015nu084T3p6ct8em4azm18el85-headcentr18feb23.jpg} 
\includegraphics[scale=.33,clip=true,trim=60 30 30 0]{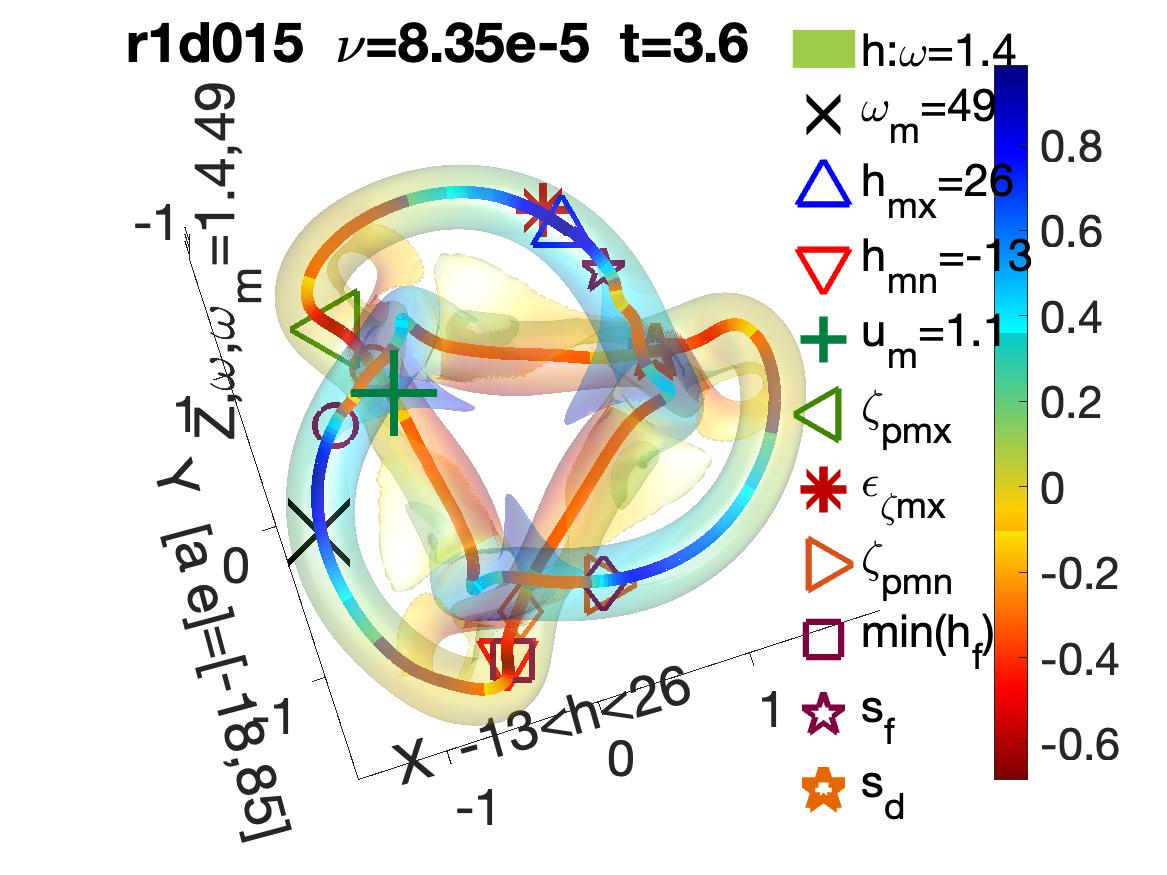} 
\begin{picture}(0,0)\put(-264,240){\large(a)}\end{picture}\\
% png\hspace{-10mm}%\bminic{0.6}
%
\bminic{0.53}\hspace{2mm}%\bminic{0.6}
\includegraphics[scale=.32,clip=true,trim=50 0 390 170]{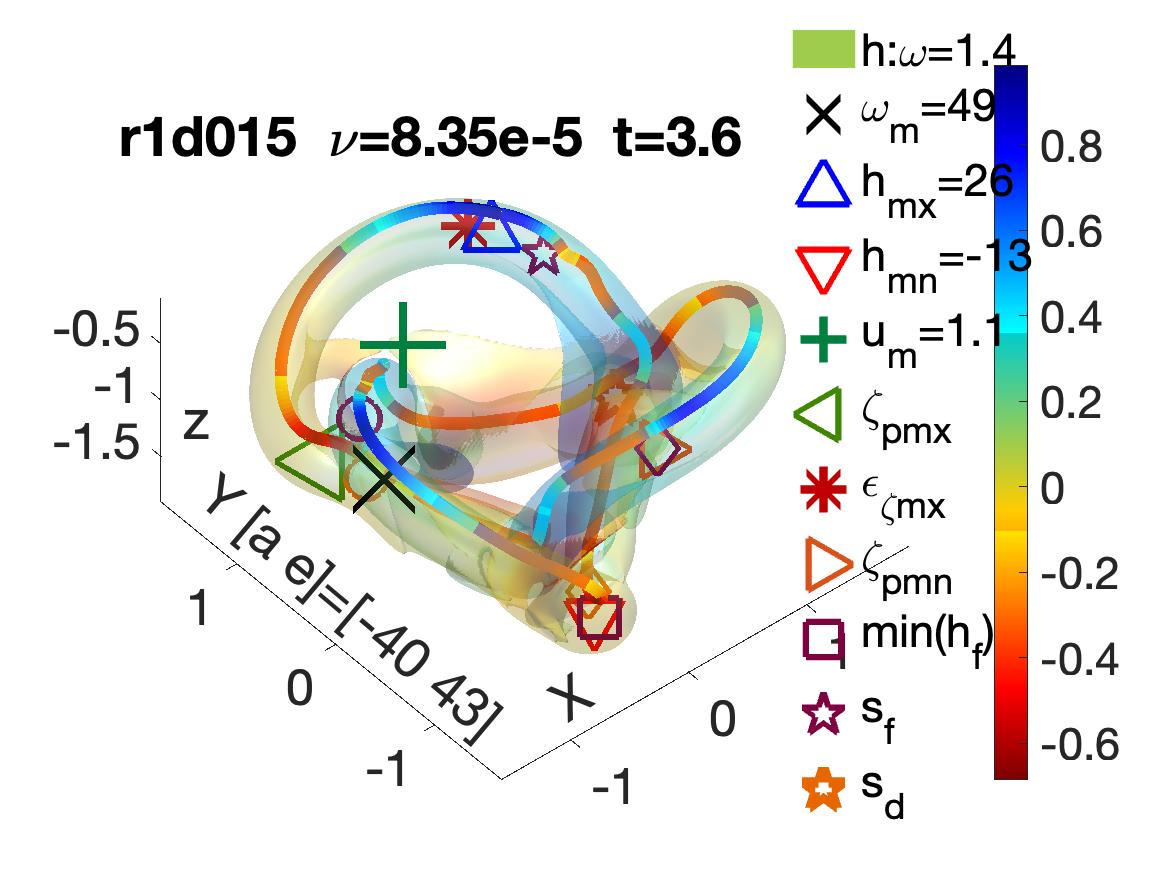}
\begin{picture}(0,0)\put(-189,200){\large(b)}\end{picture}
\emini\bminic{0.36} \vspace{4mm}\hspace{-10mm}
\includegraphics[scale=.27,clip=true,trim=130 0 372 150]{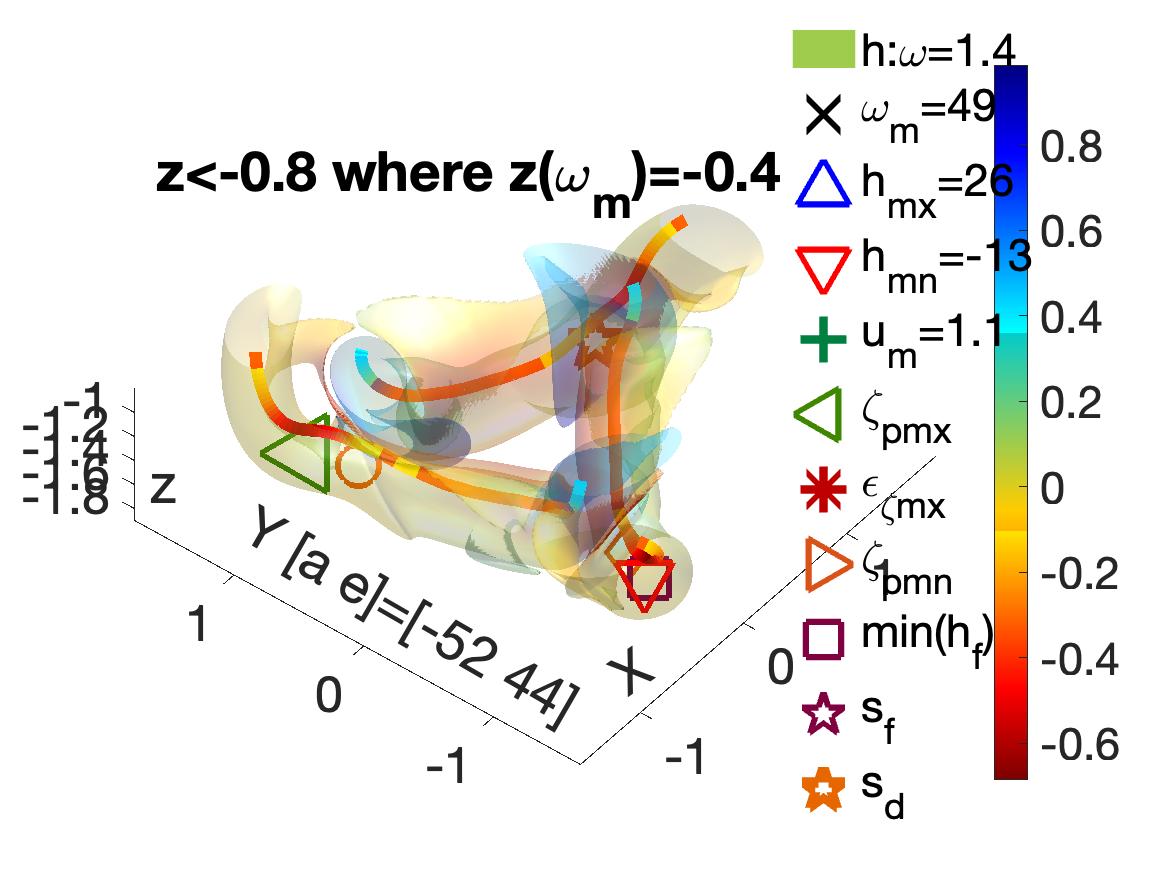}
\begin{picture}(0,0)\put(-30,220){\large(c)}\end{picture} \emini \\
%
% png\includegraphics[scale=0.85,clip=true,trim=0 0 0 0]{T3p6low16mar23.png}
% png\begin{picture}(0,0)\put(-430,190){\large(b)}\end{picture} 
% png\begin{picture}(0,0)\put(-200,175){\large(c)}\end{picture} 
% png\emini\\
\caption{\label{fig:3DT3p6} %14
%For algebraic case r1d015 at $t=3.6$, a color-coded centerline within a mapped helicity, 
%vorticity isosurface from three-perspectives.  
A $t=3.6$ mapped-helicity $\omega$-isosurface for case r1d015 with a color-coded 
centerline from three-perspectives.  
Symbols show the three-dimensional positions of the 
basic $u$, $\omega$ and $h$ extrema as well as extrema from the enstrophy and helicity budget 
equations (\ref{eq:enstrophy},\ref{eq:helicity}). 
Plus the $s_f$ (maroon) positions of local $\min(h_f)$ 
and the $s_d$ (yellow) positions of the local $\min(\epsilon_h)$, which also oppose the $s_f$ 
(the $s_o$ in figure \ref{fig:T3p6uuoo}). 
(a) is a plan view perspective with faint $h\lesssim0$ yellow sheets extending
out from lower reddish ring.  Then two sideviews from the same. (b) shows the entire domain.
(c) shows only $z<-0.8$ with the lower emerging ring, below the {\bf X} position of 
$\omega_m$ at $(x,y,z)$=(-1.37, -0.25, -0.39). The centerline vortex has 
mapped helicity ranging from red ($h=-13$) to blue ($h=26$).
By using a small $\omega\sim1.4\sim0.03\omega_m$ vorticity isosurface, 
a gradation can be seen in the lower $h<0$ zone from a red $h\sim-0.4$ inward facing half to the 
yellow-green $h\lesssim0$ outward half. This is the first step in the formation of the 
yellow negative helicity $h\lesssim0$ vortex sheets at later times. It is rotated to the right
to give some 3D perspective of the yellow lobes on the right and above.}
%This figure is taken from a $1024^3$ calculation.}
\end{figure}
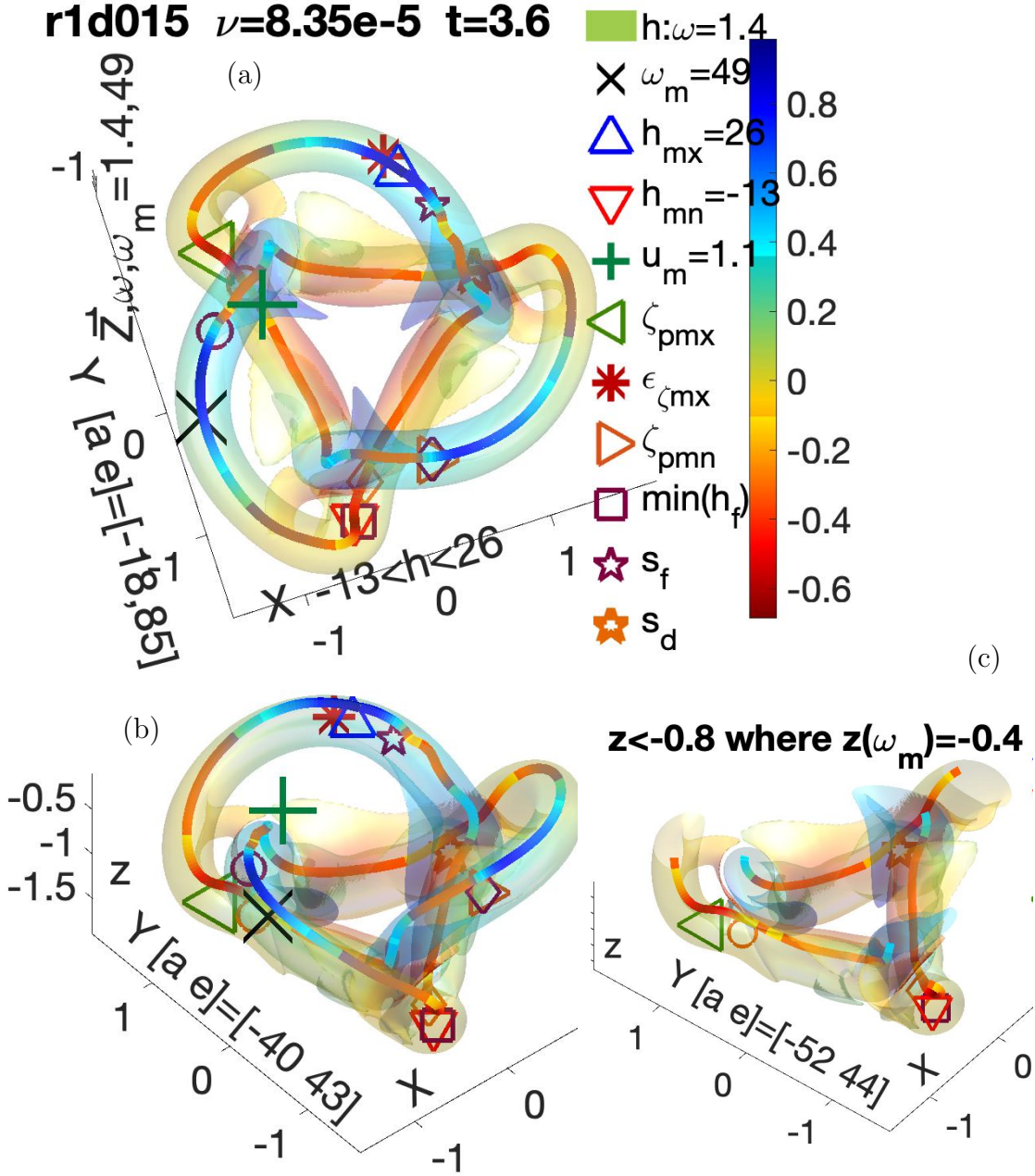

%\newpage
\subsection{Mid-reconnection $\pmb{t}$=2.4, 3.6, with algebraic spawning sheets. \label{sec:mid}}
%Mid-reconnection algebraic as sheets are spawned. 
In the $t\leq3.6$ period before reconnection begins, there are few differences between the inner, 
larger $\omega$ isosurfaces of cases r1d015 and Gd05. However, there are significant differences 
between their pre-reconnection budget profiles. Significant enough that for this mid-reconnection 
phase, the evolution of algebraic case r1d015 and that of Lamb-Oseen case Gd015 are considered 
separately. Algebraic in this section and Lamb-Oseen in section \ref{sec:Greconnect}.

To follow the evolution of  the r1d015 isosurfaces and budgets between $t=1.2$, 2.4 and 3.6, 
three sets of three-fold positions are indicated on each: $s_f$ at local $\min(h_f)$; the $s_d$ 
at local $\min(\epsilon_h)$; and points opposing either the $s_f$ (the $s_o$) or the $s_d$ 
(the $s_g$).  These are in addition to the usual extrema: 
$\max|u|$, $\max|\omega|$, $\max(h)$, $\min(h)$, 
$\min(\epsilon_h)$, $\min(h_f)$, $\max(\epsilon_\zeta)$ and the min and $\max(\zeta_p)$.
Once defined, the $s_f$, $s_d$ and $s_o$/$s_g$ can be used to follow the evolution of the 
isosurfaces and budget profiles of the r1d015 calculation at $t=1.2$, 2.4 and 3.6 as follows:
%figures \ref{fig:3DT2p4r1d015} and \ref{fig:T2p4uuoo}a as intermediate, 
% figures \ref{fig:3Dr11p2G2pi}a, \ref{figT1p2uuoo} and \ref{fig:T3p6uuoo}b,
\ITM\item[$\circ$] At the points of closest approach, the $s_f$  and $s_o$, the isosurfaces are 
drawn together over time. 
\item[$\circ$] At the same time, the $s_d$ and $s_o$ approach one another along 
the centerline until the coincide at $t=3.6$.  
\item[$\circ$] These locations can help identify where there are spans of $\epsilon_h<0$ 
and $h<0$ along the centerline.
So that at $t=2.4$ and 3.6 besides the local $\min(\epsilon_h)<0$ at the $s_d$, 
there are also growing, smaller peaks of $\epsilon_h<0$ next to the $s_f$ and 
%So that at $t=1.2$ and 2.4 besides the local $\min(\epsilon_h)<0$ 
between the $s_f+s_d$ pairs, growing $s$-spans of $\epsilon_h\lesssim0$. On both the 
isosurfaces and the centerlines as in figures \ref{fig:3DT2p4r1d015} and \ref{fig:T2p4uuoo}a
and \ref{fig:T3p6uuoo}a,b.  With some $h\lesssim0$ at the $s_d$. 
\item
At $t=3.6$ the $s_o$ are co-located with the $s_d$. With the $s_g$ nearly co-located with the 
$s_f$, as shown in figure \ref{fig:T3p6uuoo}a. And the spans of $\epsilon_h<0$ and $h\lesssim0$ 
from $t=2.4$ are now concentrated at the $s_d$ points, with $\epsilon_h<0$ and $h<0$ being 
particularly deep at those points.  There is also local $\epsilon_h<0$ at the $s_f$ with 
$\epsilon_h\approx0$ between the $s_f$ the next $s_d$. 
\item[$\circ$] For example $\epsilon_h\approx0$ between $s_f$=6.3 and $s_d$=9.2.
Another $\epsilon_h\approx0$ that started at $t=2.4$ with $20\epsilon_h<-5$ at $s_f$=0.4 to 
$s_d$=2.3 at $t=3.6$ goes to $s_d$=3.3.
\item These small patches of $h<0$ and $\epsilon_h<0$ on spans of the centerlines and inner 
isosurface are not evidence for $h<0$ vortex sheets. 
The patches are even similar to Lamb-Oseen as reconnection at $t=3.6$ in 
section \ref{sec:Greconnect}.  Instead, the patches of $\epsilon_h\lesssim0$ 
%in figure \ref{fig:T3p6uuoo}a between the $s_f$ and $s_d$, 
could be evidence of where  $h<0$ vortex structures are being created.  
\ITN

\bpurp{What can the centerline budget profiles tells us about why at reconnection Lamb-Oseen 
generates localized bridges and braids, while all the algebraic profiles generate broad
vortex sheets?  With interactions between the sheets at the end of section \ref{sec:reconnect} 
showing how they can achieve convergence of the dissipation rate $\epsilon$.}

%To see the r1d015 budget evolution between $t=2.4$ and $3.6$, figures \ref{fig:T2p4uuoo}
%and \ref{fig:T3p6uuoo} include two sets of 
%additional marks: One set is the three-dimensional positions (or nearest centerline points) 
%of these extrema: $\max|u|$, $\max|\omega|$, 
%$\max(h)$, $\min(h)$, $\min(\epsilon_h)$, $\min(h_f)$, $\max(\epsilon_\zeta)$ and the min and 
%max($\zeta_p$). The second, and perhaps better, set for following the evolution of the 
%trefoil between $t=2.4$ and 3.6 on the centerlines are the three-fold positions of: 
%the $s_f$ at local $\min(h_f)$, the $s_d$ at local $\min(\epsilon_h)$ and the $s_o$ at points 
%opposing the $s_f$.

%%\noindent What new insights can the additional two sets of local centerline 
%%positions in figure \ref{fig:T2p4uuoo}, %13
%%the $s_f$ and $s_d$ provide?  Specifically, can their positions on the centerline 
%%budget profiles indicate where and how the vortex sheets with $h<0$ form begin to
%%form?
%%The local $\min(\epsilon_h)\lesssim 0$ positions are the important reference points.
{\bf How the $\pmb{h<0}$ isosurface vorticity forms:}
\ITM
\bpurp{\item Continuing to $t=3.6$ in figure \ref{fig:T3p6uuoo}, the inner,
higher magnitude $\omega=0.2\omega_m=11$ isosurface in the upper-right quadrant 
shows continuity with the centerline Biot-Savart evolution at 
earlier times. This inner structure is not significantly different than the
However, the outer structure is very different.
\vspace{-2mm} }%bpurp
\item {\bf $\pmb{h<0}$ formation.} While at $t=2.4$ there are spans of $h(s)<0$ 
in figure \ref{fig:T2p4uuoo}, this does not translate into signficant
$\pm$ variations of $h$ on the $t=2.4$ isosurface or signs of vortex sheets.
It is not until $t=3.6$ that significant dips of $h< -5$ appear at the $s_d$ 
locations. On both the centerline and the inner (large $\omega$) isosurface in figure
\ref{fig:T3p6uuoo}(a,b). 
\item[$\circ$] What is new in 3D at $t=3.6$ is extensive $h<0$ on parts of the smaller vorticity 
magnitude outer isosurfaces in figure \ref{fig:3DT3p6}. Red for strong $h<0$ along the red-coded 
centerline in the lower ($z<-0.8$) portion of the trefoil. And yellow $h\lesssim0$ helicity 
on the other side of those isosurfaces, with faint signs of shed vorticity.
A trend that continues to later times, as illustrated in figure \ref{fig:3DT4p8} at $t=4.8$.
%\item[$\circ$] A more complete 3D picture of this $t=3.6$ isosurface is given by figure
%\ref{fig:3DT3p6turn} in appendix \ref{sec:3Dr1d015T3p6}
\item {\bf Relation between $\pmb{h<0}$ centerlines and isosurface zones.} The red on the
isosurface is associated with the broader spans of centerline $\epsilon_h(s)\lesssim0$ that 
connect the $s_f$ and $s_d$ local positions. Example: Follow the maroon $s_f$ $\star$ through 
where the loops cross, then down to the yellow $s_d$ $\diamond$. Or from the maroon $s_f$ $\circ$ 
to the yellow $s_d$ $\star$ underneath the maroon $\star$. 
\item[$\circ$] With all corresponding to $\epsilon_h\sim0$ spans between all six 
local $\min(\epsilon_h)$ at the $s_f$ and $s_d$ in figure \ref{fig:T3p6uuoo}a. 
\item[$\circ$] The reddish $h<0$ patches extend over roughly 2/3rds of these spans on the lower 
($z<-0.7$) part of the isosurface.
\item[$\circ$] With the reddish zones smoothly transitioning  into the 
yellowish, more sheet-like outer surfaces. 
\item[$\circ$] This is illustrated further at $t=4.8$ with the red patches in figures 
\ref{fig:3DT4p8} and \ref{fig:T4p8uuoo}. 
\item Further $t=3.6$ figures from different perspectives and different 
cropping levels will appear shortly. Phy. Rev. Fluids (accepted, 2023)
{\it Sensitivity of trefoil vortex knots upon the initial vorticity profile.}
\ITN
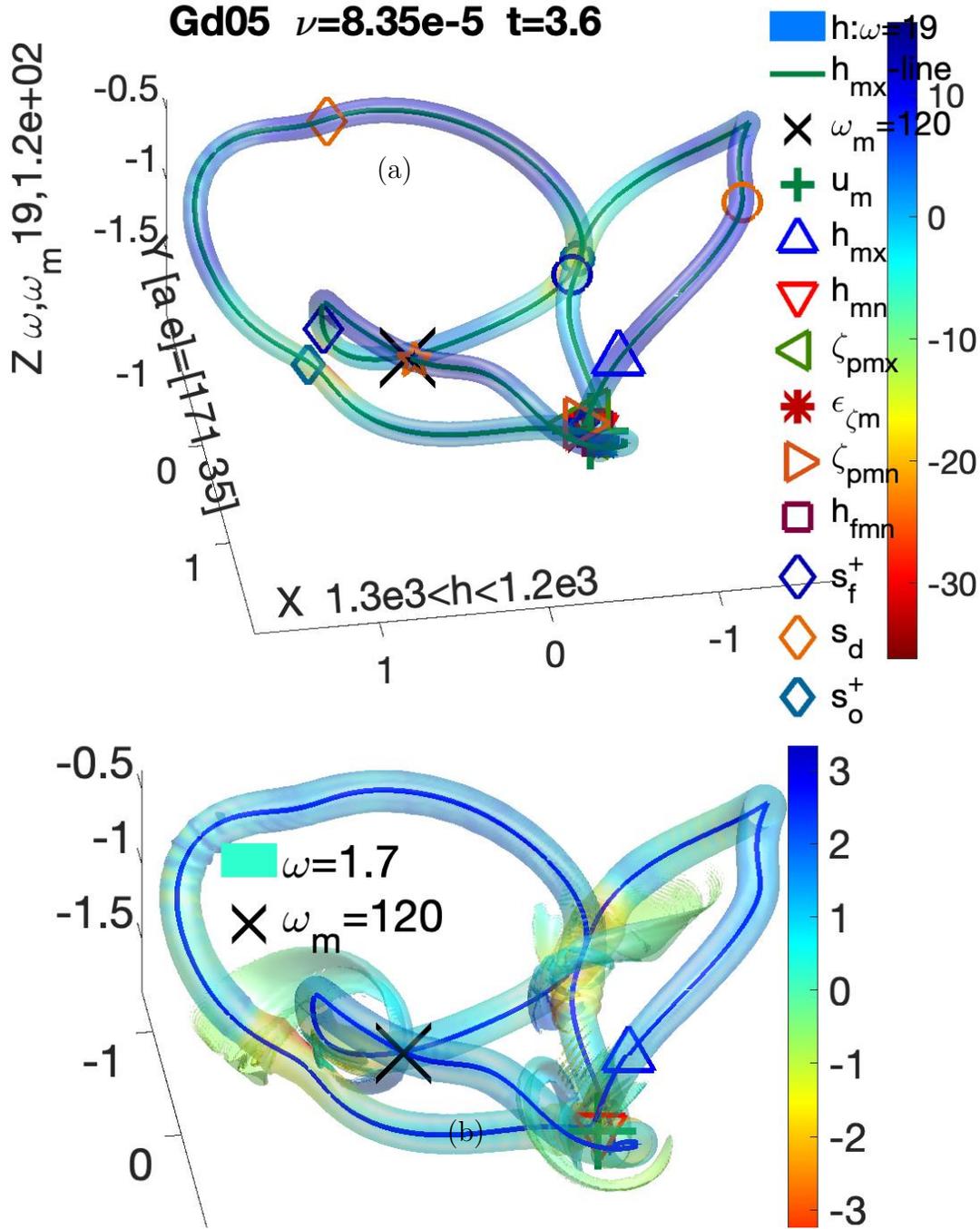
\begin{figure}[H]
%\graphics[scale=.36,clip=true,trim=10 436 308 0]{Gm1e3d5dm1nu0835T3p6uuoosuo20aug22.jpg}
%\emini \bminic{0.5} 
\vspace{-6mm}
\includegraphics[scale=.37,clip=true 10 20 0 0]{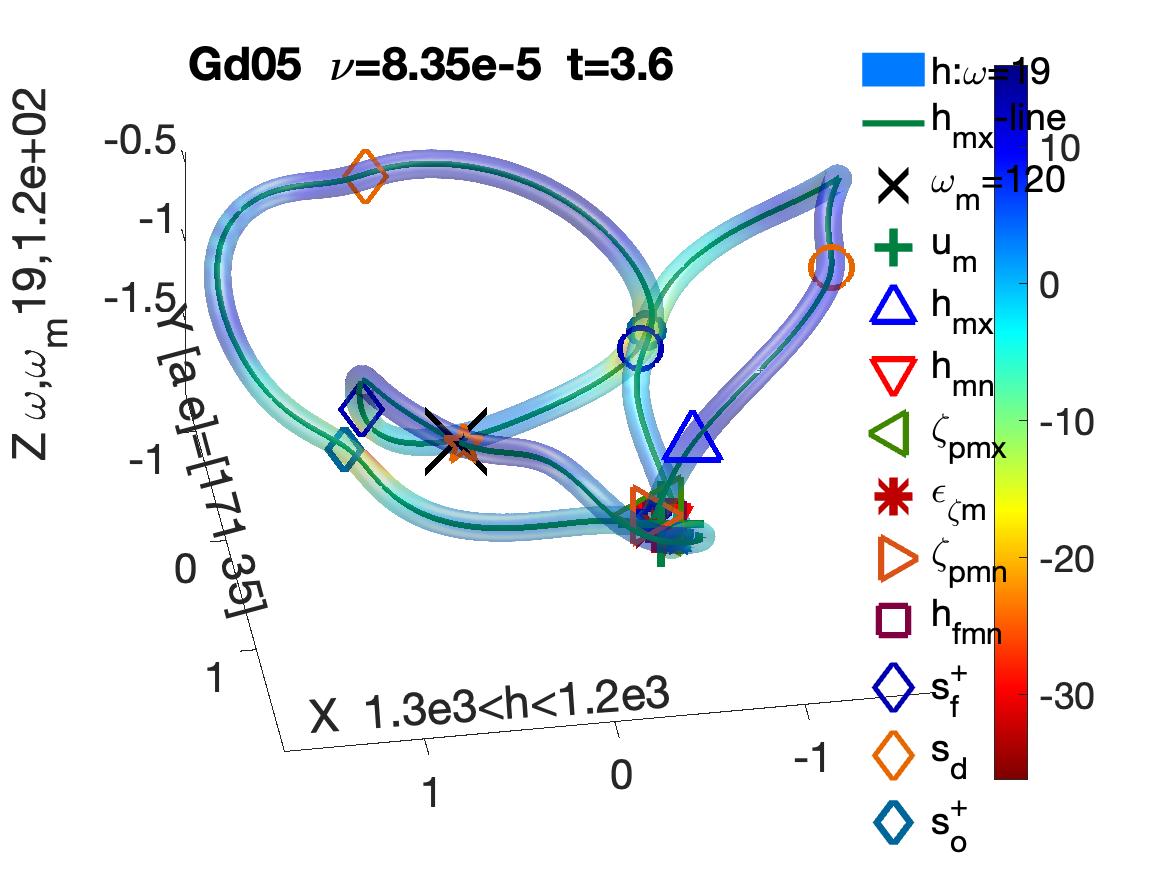} 
\begin{picture}(0,0)\put(-280,236){\large(a)}\end{picture} \\
\includegraphics[scale=.37,clip=true,trim=40 220 150 115]{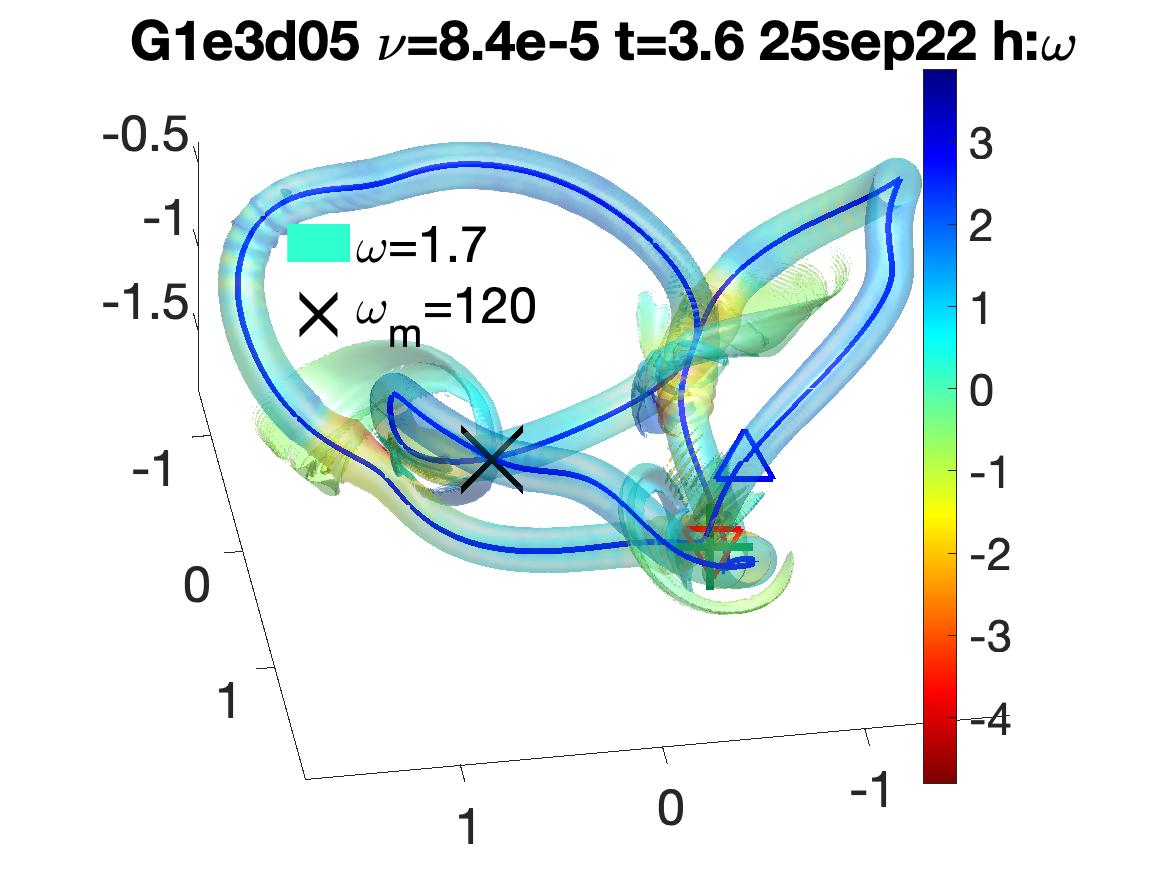}
\begin{picture}(0,0)\put(-180,36){\large(b)}\end{picture} 
\caption{\label{fig:G3DT3p6} %16
Two $t=3.6$ Lamb-Oseen isosurfaces with different vorticity thresholds. 
(a) The primary $\omega=19$ isosurface is similar to the higher-$\omega$ algebraic 
isosurface in figure \ref{fig:T3p6uuoo}. Additional markers indicate the 
three-dimensional locations of the $s_d$ (yellow), local $\min(\epsilon_h)$, $s_f^+$ (cobalt) for 
the local $\max(h_f)$ points and $s_o^+$ (turquoise), points opposing the $s_f^+$ that are also
$\min(h)<0$ and $\min(\zeta_p)$ points.
Reconnection is commencing between the $s_f^+$ and $s_o^+$ points.
The local $s_d$ (yellow), $\min(\epsilon_h)$ sit in strongly positive $h>0$ zones, not
$h<0$ as for the algebraic calculations or Lamb-Oseen for $t\leq2.4$. 
(b) The vorticity of the second isosurface uses very small $\omega=1.7$ to 
show that the outer edges of the isosurface are shedding sheets with slightly negative 
helicity.}
\end{figure}

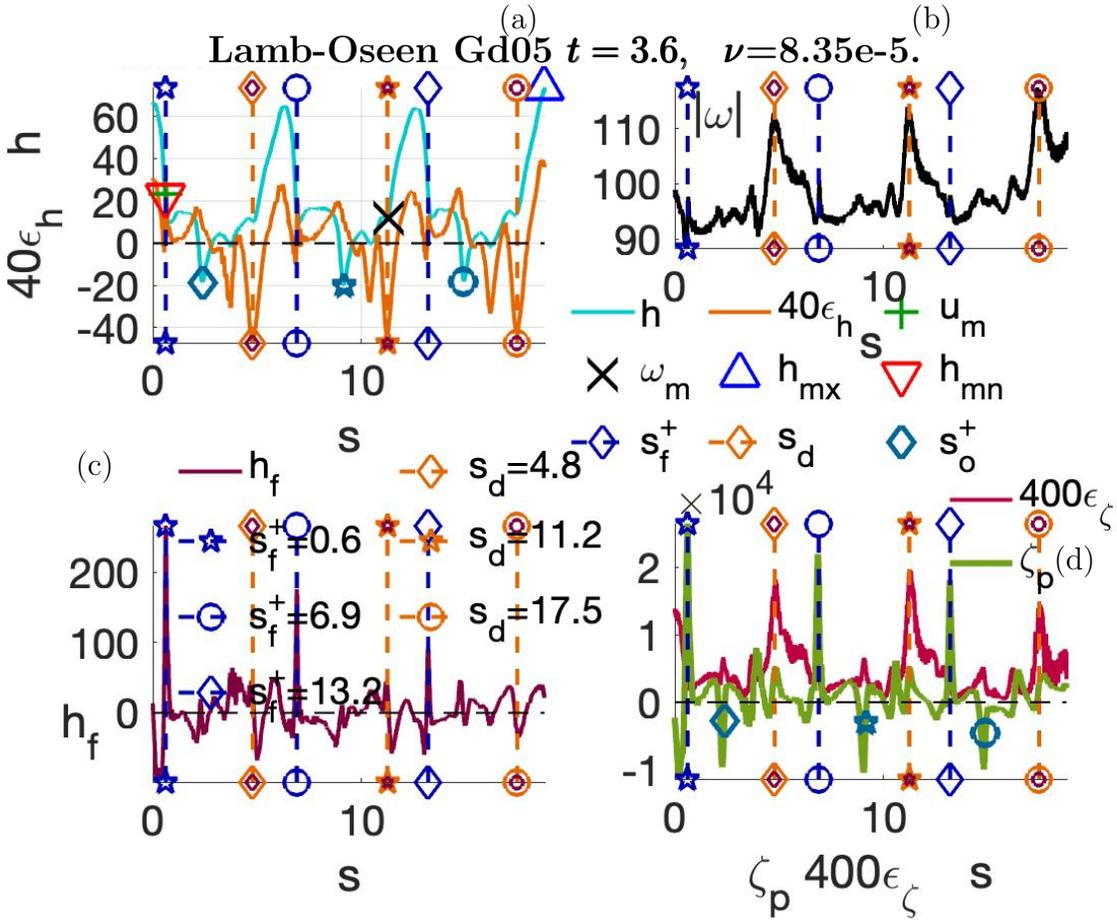
\begin{figure}[H]
\includegraphics[scale=.38,clip=true,trim=12 0 0 67]{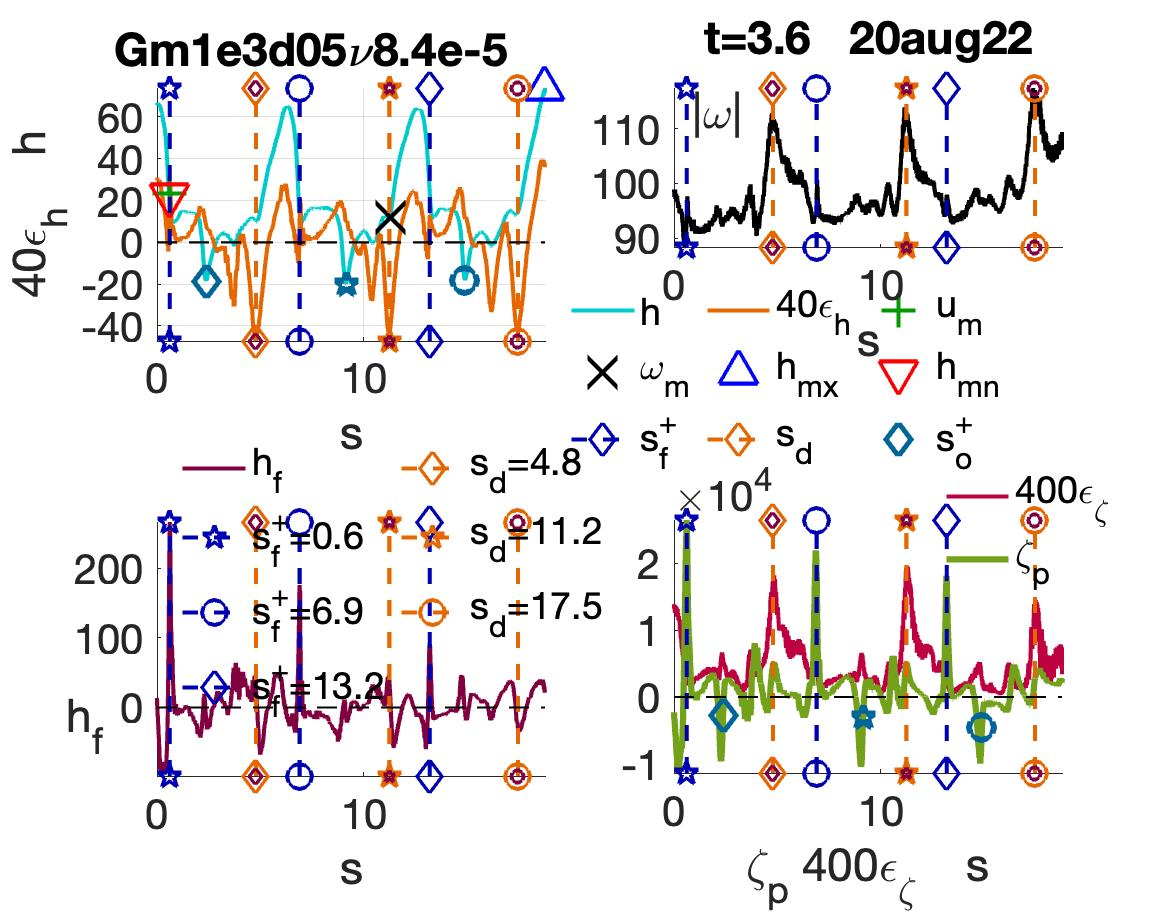} 
\begin{picture}(0,0)\put(-370,326){\Large\bf Lamb-Oseen Gd05 $\pmb{t=3.6,~~\nu}$=8.35e-5.}
\put(-260,339){\large(a)}\put(-104,339){\large(b)}
\put(-420,170){\large(c)}\put(-50,134){\large(d)} \end{picture}
\caption{\label{fig:GuuooT3p6} %17
$t=3.6$ Lamb-Oseen (Gd05) \eqref{eq:Gauss} centerline budget profiles. 
The $s_d$ (yellow/maroon) at local $\min(\epsilon_h)$ and co-located with local 
$\max(\epsilon_\zeta)$ and $\max(|\omega|)$, are in large $h>0$ zones far from 
the reconnections.
The $s_f^+$ (cobalt) are at local $\max(h_f)$ points and co-located with local $\max(\zeta_p)$
and secondary velocity minima.
The $s_o^+$ (turquoise) points oppose the $s_f^+$ and are co-located with  $\min(h)<0$ and 
$\min(\zeta_p)$) points.
Reconnection is commencing between the $s_f^+$ and their opposing $s_o^+$ points.
}\end{figure} 
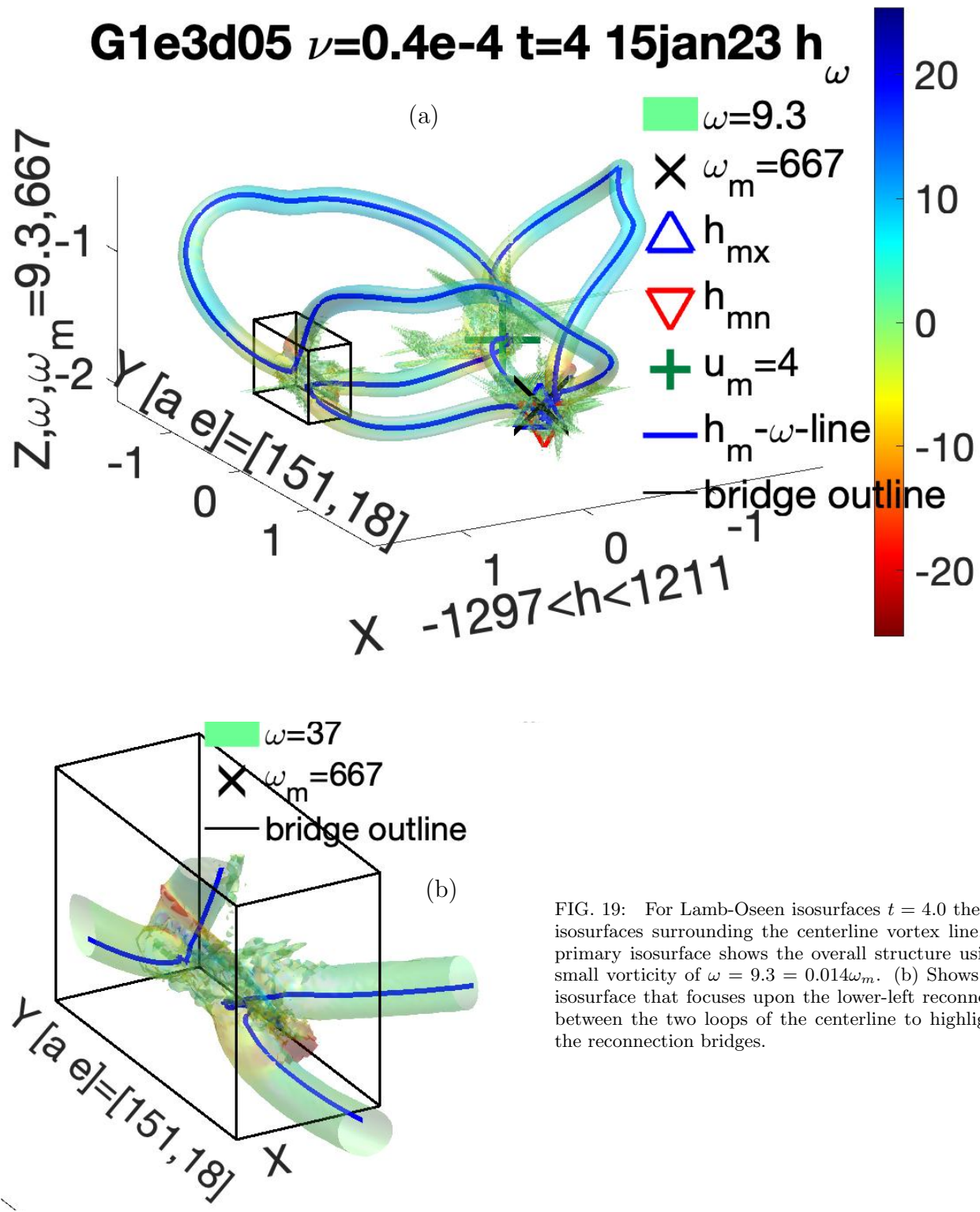
\begin{figure}[H]
\includegraphics[scale=0.42]{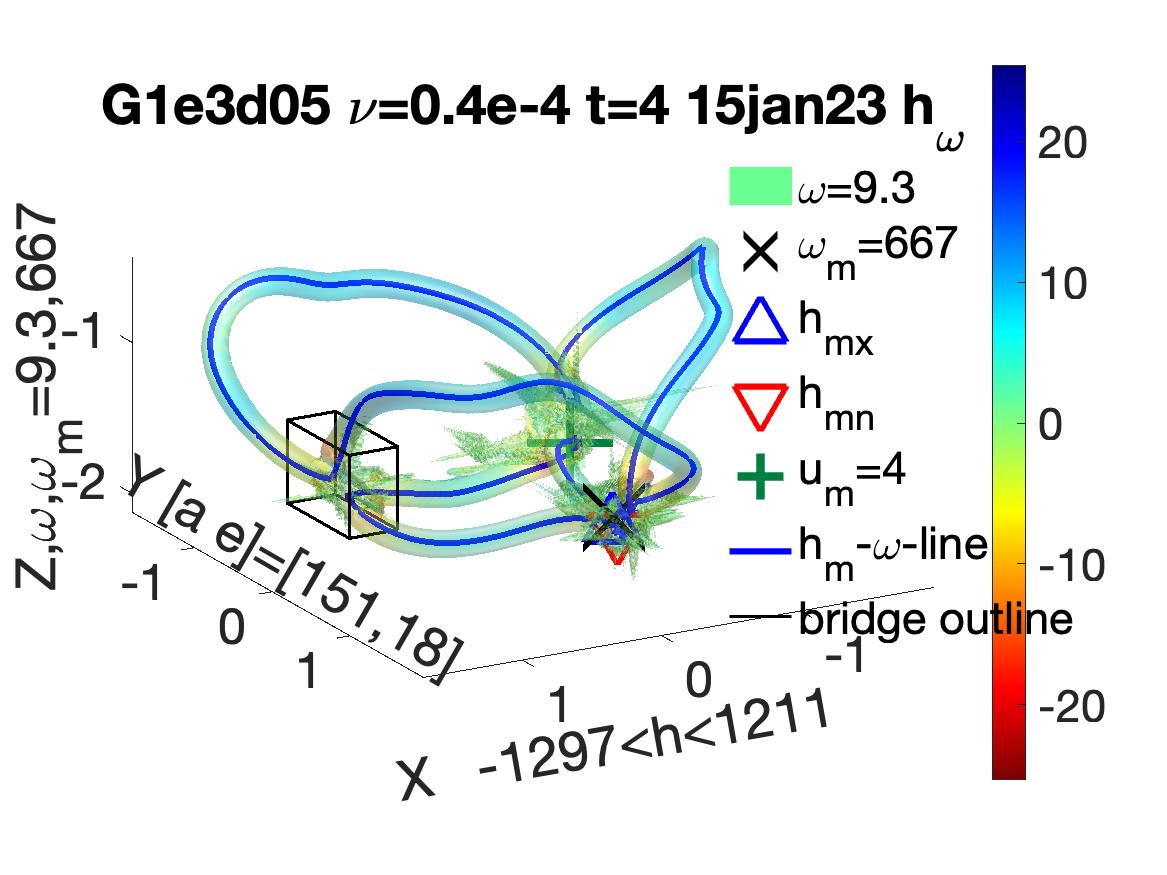}
\begin{picture}(0,0)
%\put(-200,150){\bf\large Gd05  $\pmb{\nu=8.35e-4}$  t=4  h:$\pmb{\omega}$}
\put(-300,286){\large(a)} \end{picture} \\
\bminic{0.52}
\includegraphics[scale=0.42,clip=true,trim=340 210 210 110]{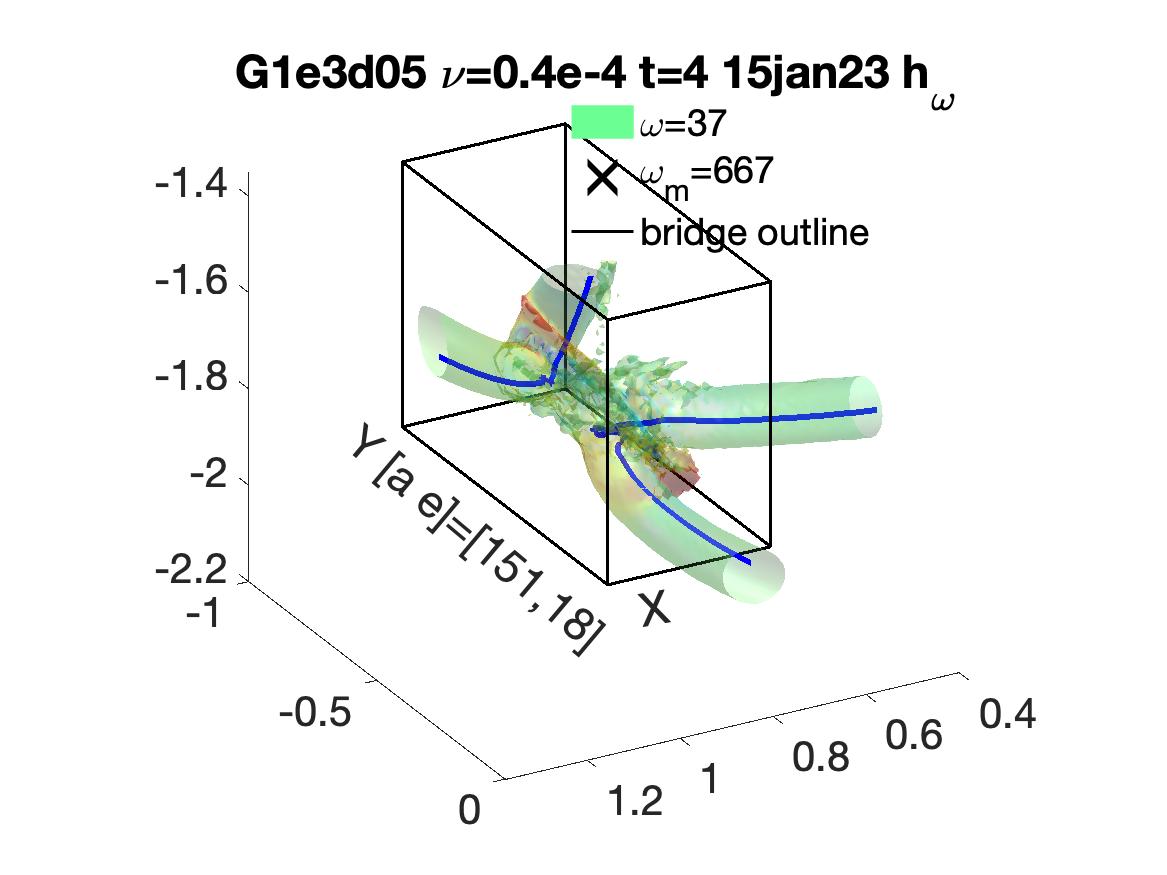}
\begin{picture}(0,0)\put(-60,150){\large(b)} \end{picture} \emini 
\bminic{0.48}
\caption{\label{fig:G3DT4p0} 
For Lamb-Oseen isosurfaces $t=4.0$ there are two isosurfaces surrounding the centerline
vortex line.
(a) The primary isosurface shows the overall structure using a very small vorticity 
of $\omega=9.3=0.014\omega_m$.
(b) Shows a $\omega=37$ isosurface that focuses upon the lower-left reconnection site 
between the two loops of the centerline to highlight one of the reconnection bridges.}\emini 
\end{figure} 
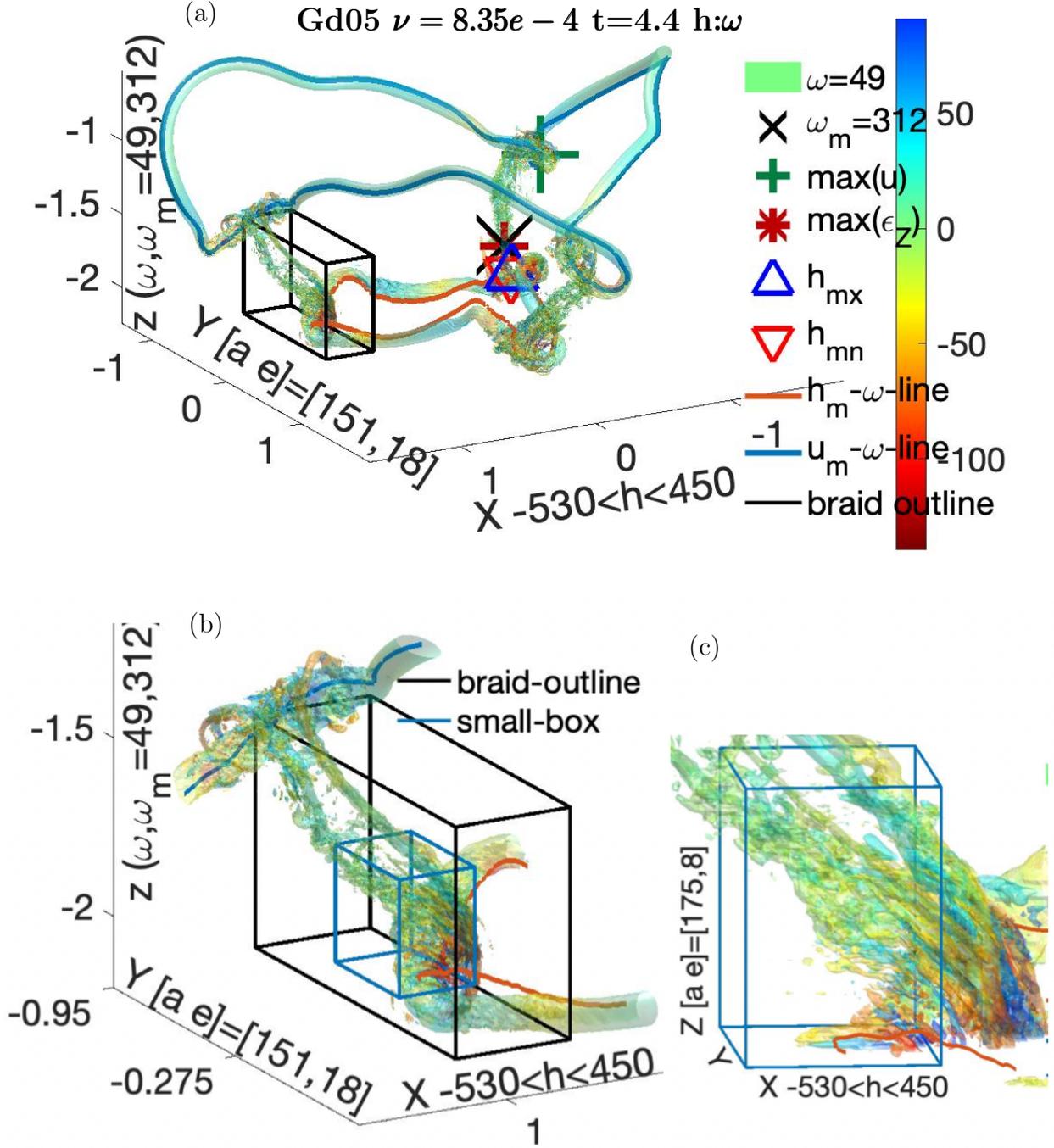
\begin{figure}[H] \vspace{10mm}
\includegraphics[scale=.40,clip=true,trim=0 50 0 190]{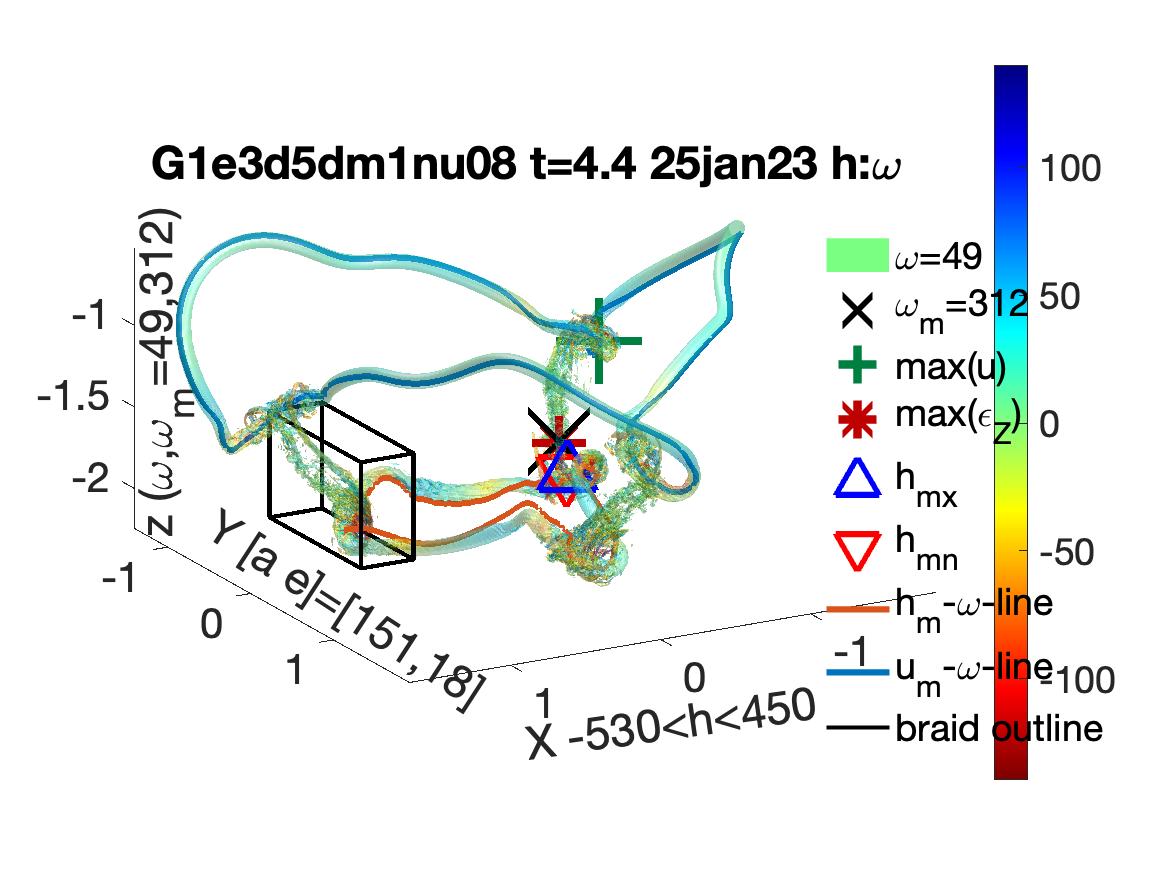} 
\begin{picture}(0,0)\put(-340,250){\bf\Large Gd05  $\pmb{\nu=8.35e-4}$  t=4.4  h:$\pmb{\omega}$}\put(-390,254){\large(a)}\end{picture}\\ 
%\vspace{-52mm} \bminic{1} 
\includegraphics[scale=.90,clip=true,trim=0 0 0 0]{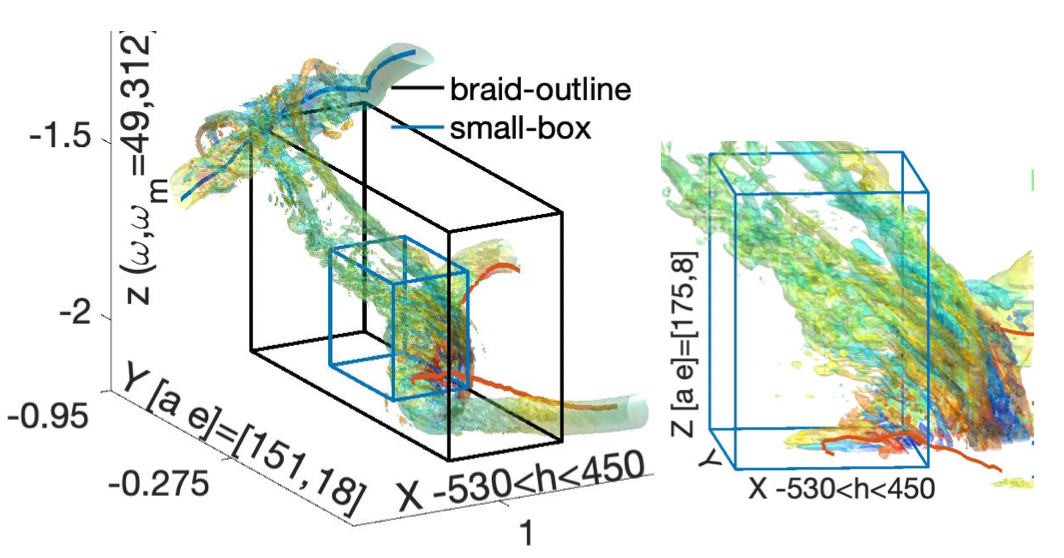}
\begin{picture}(0,0) \put(-400,234){\large(b)}\end{picture}
\begin{picture}(0,0) \put(-180,224){\large(c)}\end{picture}
%\begin{picture}(0,0) \put(-280,224){\large(b)}\end{picture}
%%\bminic{0.55}\graphics[scale=.38,clip=true,trim=80 0 340 180]{Gm1e3d5dm1nu0835-3DhonomT4p4boxleftindat27jan23.jpg} 
%%\includegraphics[scale=.27,clip=true,trim=260 118 310 200]{Gm1e3d5dm1nu0835-3DhonomT4p4boxleftinindat13feb23.jpg}
%\begin{picture}(0,0) \put(-150,222){\large(c)}\end{picture} \emini \\  ~~~~~~~
\caption{\label{fig:G3DT4p4} $t=4.4$ Lamb-Oseen isosurfaces. (a-c) 
Three views the isosurfaces, with the bottom two focusing upon the smallest structures.
(a)  The primary $t=4.4$ isosurface shows the overall structure with 
$\omega=49=0.015\omega_m$(=312) 
to show how braids are forming from bridges, as seen for previous Lamb-Oseen 
calculations. (b) Shows full length of one of the double braids, including 
where it attaches to the new upper and lower vortex rings. Similar to $t=4.29$ of figure 18 
from \cite{YaoYangHussain2021}.
(c) Focuses on one end as that double braid winds around the primary vortex.}  
%Gm1e3d5dm1nu0835-3DhonomT(4p0\&4p4)dat(21\&18)may22.jpfig:G3DT4p4} %18
\end{figure}
\subsection{\label{sec:Greconnect}Gaussian/Lamb-Oseen reconnection: braid formation.}

In section \ref{sec:early}, %III A), 
early divergence  of $t=1.2$ Lamb-Oseen  budget profiles from the algebraic profiles was shown 
respectively in figures \ref{fig:GuuooT1p2} (Gd05) and \ref{fig:T1p2uuoo} (r1d015). %11,12
Section gives the effect of that early divergent dynamics upon Lamb-Oseen as reconnection begins.
Beginning at $t=3.6$ with figures \ref{fig:G3DT3p6} and \ref{fig:GuuooT3p6}. %17,18 
$t=3.6$ is the last time that a single centerline could be identified for case Gd05. 

The Lamb-Oseen analysis ends with the $t=4$ and 4.4 isosurfaces in 
figures \ref{fig:G3DT4p0} and \ref{fig:G3DT4p4}. These show how the trefoil then breaks 
into two vortex rings, connected first by what could be 
described as bridges, then as braids. 

\medskip The two Lamb-Oseen $t=3.6$ isosurfaces in figure \ref{fig:G3DT3p6} 
are:
\ITM\item[(a)] A primary, higher magnitude $\omega=19$ isosurface that 
shows continuity with the earlier Biot-Savart evolution and has minimal 
differences with the $t=3.6$ inner algebraic structure in 
figure \ref{fig:T3p6uuoo}. 
\item[(b)] The lower magnitude $\omega=1.7$ isosurface shows how the Lamb-Oseen profile 
reconnection begins on the outer wings, with sheets shedding with some $h\lesssim0$. 
These sheets with bits of $h\leq 0$ are localized around the reconnection points, 
unlike the broad $h<0$ isosurface zones of the r1d015 algebraic trefoil in 
figure \ref{fig:3DT3p6}. 
\ITN

The $t=3.6$ budget profiles and isosurfaces in figures \ref{fig:G3DT3p6} and 
\ref{fig:GuuooT3p6} have three sets of primary local positional marks. $s_d$, $s_f^+$ and $s_o^+$.
Plus the $s_f$. 
\ITM\item[$\circ$] The $s_d$ in yellow (with embedded maroon $s_f$) are at 
local $\min(\epsilon_h)$+$\min(h_f)$ positions.  The $s_d$ are exactly on local 
$\max(|\omega|)$ and $\max(\epsilon_\zeta)$, the maximum enstrophy dissipation.
\item[$\circ$] The $s_f^+$ in cobalt are at the local $\max(h_f)$ and are coincident 
with local $\max(\zeta_p)$.  Local $\zeta_p>0$ implies stretching, suggesting that 
these positions could be the seeds for the bridges that form during reconnection.
\item[$\circ$] The third set of $s_o^+$ in turquoise are at the points opposing the $s_f^+$.
The $s_o^+$ are also local $\min(h)$ and $\min(\zeta_p)$, local compression, suggesting that 
there is pinching on the trefoil vortex at the other end of the nascent bridges. 
\item All consistent with active reconnection at these positions.
\ITN

$\bullet$ What can the $t=3.6$ markers tell us about the separation of the trefoil into two 
rings? 
\ITM\item[$\circ$] The cobalt $\max(h_f)$ $s_f^+$ points with large $\zeta_p>0$ become one end of
the bridges, with their opposing turquoise $s_o^+$ at the other end. 
\item[$\circ$] The $s_d$ yellow $\min(\epsilon_h)$ points are on what becomes the upper (u)
ring, with magnitudes $h_u>0$ .  
\item[$\circ$] The turquoise $s_o^+$/$\min(h)$ points become the lower ($\ell$) ring, with some 
$h(s_o^+)<0$ appearing on the localized vortex sheets in figure \ref{fig:G3DT3p6}b, such as 
to the left of $\omega_m$ ({\bf X}).
\ITN
$\bullet$ What develops out of this $t=3.6$ state?
\ITM\item At $t=4$ in figure \ref{fig:G3DT4p0}, short, flattened bridges are 
generated as the trefoil is begins to separate into two rings.
\item[$\circ$] The positions of $\omega_m$, $u_m$, $h_{mx}$ and $h_{mn}$ are all on the bridges.
\item[$\circ$] At $t=4.4$, in figure \ref{fig:G3DT4p4}, the new upper (blue) and lower (red)
rings are separating, with each bridge splitting into two braids. 
\item[$\circ$] The positions of $\omega_m$, $h_{mx}$ and $h_{mn}$ are  on the the lower ring
and $u_m$ is on the upper ring.
\item Figures \ref{fig:G3DT4p0} and \ref{fig:G3DT4p4} are roughly equivalent to the $Re=12000$
figures at the same times for a previous trefoil calculations using 
Lamb-Oseen profiles \cite{YaoYangHussain2021}. Including the splitting of each bridge
into two braids.
\item[$\circ$] So providing further Gaussian/Lamb-Oseen graphics and discussion 
in this paper is unnecessary.
\ITN
$\bullet$ Summary of how the Lamb-Oseen budget profiles in figures \ref{fig:GuuooT1p2}, 
\ref{fig:GuuooT2p4} and \ref{fig:GuuooT3p6} can explain the evolution of the global 
enstrophy $Z(t)$ and the helicity ${\cal H}(t)$ in figure \ref{fig:Gd05dm1ZHnus}: %1
\ITM\item[$\circ$] Starting at $t=0$ when $\int{ds}\,\zeta_p\equiv0$, for the spans with local
compression, $\zeta_p<0$, the viscous terms and $\epsilon_\zeta$ are enhanced. Resulting in
$Z(t)$ decreasing for at least short $t\gtrsim0$ times for all cases and viscosities $\nu$. 
\item[$\circ$] Between $t=2.4$ and $3.6$, the global enstrophy production and its
dissipation rate are approximately equal to their centerline integrals: 
$Z_p=\int{dV}\zeta_p\sim \int{ds}\,\Gamma\zeta_p$ and 
$\epsilon_Z=\int{dV}\epsilon_\zeta\sim \int{ds}\,\Gamma\epsilon_\zeta$. 
\item[] With $Z_p$ and $\epsilon_Z$ roughly balancing one another in 
figures \ref{fig:GuuooT2p4},\ref{fig:GuuooT3p6} ($t=2.4$, 3.6), giving 
$dZ/dt=Z_p-\epsilon_Z\approx0$ over the 
temporal span of $2.4\leq t\leq 3.6$. And relatively steady $Z(t)$, enstrophy, over those 
times in figure \ref{fig:Gd05dm1ZHnus}. 
\item[$\circ$] At $t=3.6$ in figure \ref{fig:GuuooT3p6}, at the locations of positive, not 
negative, spikes in $h_f$, there are sharp positive spikes in the enstrophy production$\zeta_p$. 
\item[$\circ$] These spikes of $\zeta_p>0$ continue through $t=4$, generating the
are brief enstrophy spurt in figure \ref{fig:Gd05dm1ZHnus}. This spurt is when the bridges form,
shown in figures \ref{fig:G3DT3p6} and \ref{fig:G3DT4p0}. 

\item[$\circ$] Then as the strong centerline enstrophy dissipation $\epsilon_\zeta$ in 
figure \ref{fig:GuuooT3p6} takes over, the centerline spikes of local $h_f>0$, 
$\zeta_p>0$ and $\omega=\sqrt{\zeta}$ and $\zeta_p$, are dissipated. Along with
the temporally spikes of $Z(t)$ in figure \ref{fig:Gd05dm1ZHnus}.
% that continues as the two new separating rings stretch the new bridges.
\item[$\circ$] For ${\cal H}(t)$, except at $t\sim1.2$ as in figure \ref{fig:GuuooT1p2}, 
its $t\leq3.6$ evolution is dominated by the strongly localized negative helicity dissipation
$\epsilon_h$, which removes $h<0$, thereby leading to increasing ${\cal H}(t)>0$. 
After $t=3.6$, as dissipation removes the small amounts of $h<0$ associated with the bridges,
${\cal H}(t)$ increases further.
\ITN

%\newpage
\subsection{Algebraic reconnection scaling with $h<0$ $\omega$-sheets. \label{sec:reconnect}}

Due to the constraints imposed upon the calculations in this paper, three-fold symmetry and 
a $(2\pi)^3$ domain, it has been a surprise that the algebraic profile cases have generated this: 
Finite-time, finite energy dissipation $\Delta E_\epsilon$ \eqref{eq:dissanom}, as shown in 
figures \ref{fig:r1ZHnu} and \ref{fig:KSRZHr2d1sqnu} by the finite-time convergence of the 
dissipation rates $\epsilon(t)=\nu{Z}$ of the broadest profiles: cases r1d015 and r2d1.
At least for a short range of viscosities.
The evidence for finite $\Delta E_\epsilon$ in the earlier perturbed trefoil calculations 
\cite{KerrJFMR2018} could only be achieved by using very large domains. 

Furthermore, for all of the algebraic profile calculations there are vortex sheets and 
convergent $\sqrt{\nu}Z$, such as in figure \ref{fig:r1sqnuZ} (r1d015) and the examples in 
section \ref{sec:KSRdiss}. %3.5
Although with profile dependent convergent times $t_x>t_r$. 

What are the underlying structures and dynamics that allow the subsequent enstrophy growth to 
accelerate and form finite $\Delta E_\epsilon$ for these cases?  
Figures \ref{fig:3DT3p6} and \ref{fig:T3p6uuoo} %\ref{fig:3DT3p6turn} 
at $t=3.6$ show where, 
and how, the conditions for generating negative helicity vortex sheets originate.
This section extends that analysis to $t=4.8$ to show how the sheets then expand and 
contribute to the enstrophy growth: growth that can lead to finite-time  energy dissipation. 
Skipping the gradual changes at the intermediate times of $t=4$ and $t=4.4$. 
The important differences with the Lamb-Oseen calculation are also highlighted.

\begin{figure}[H]\subfigure{
\includegraphics[scale=.38,clip=true,trim=0 0 0 0]{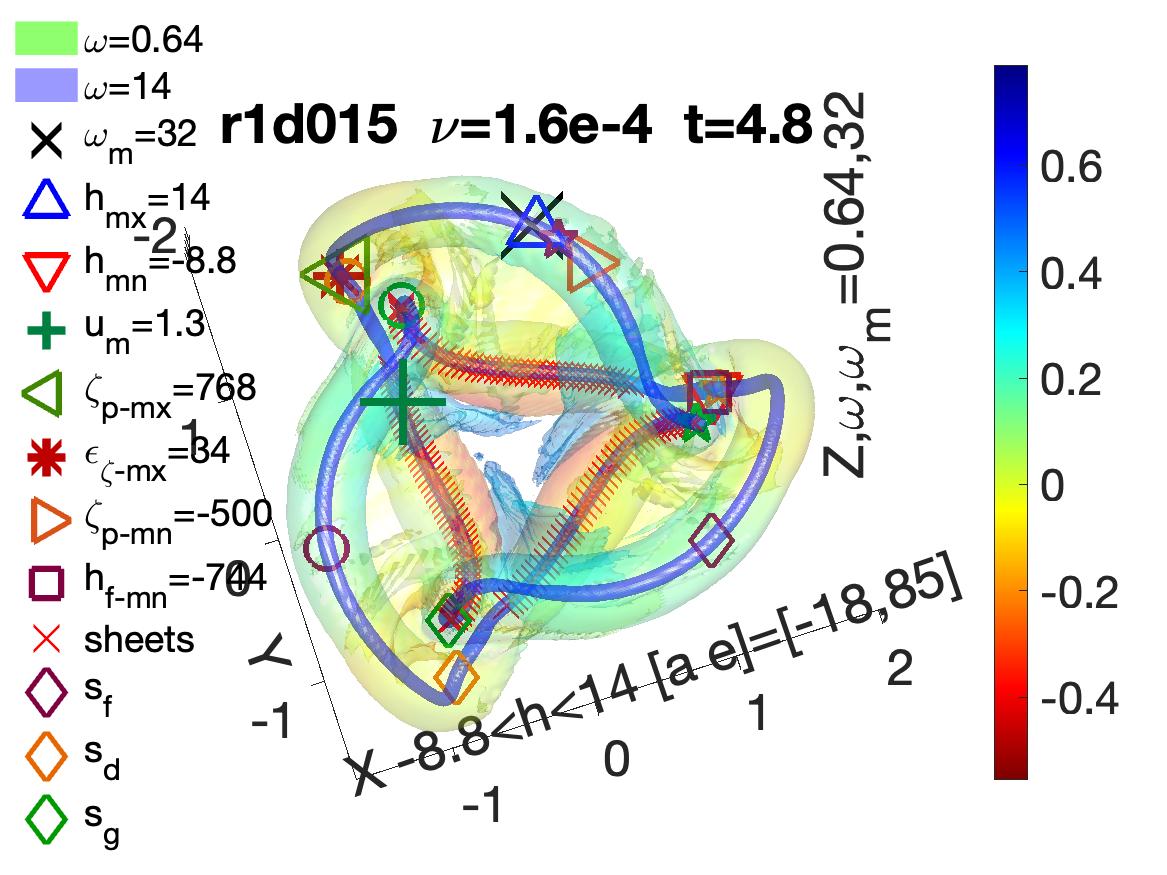}\begin{picture}(0,0)\put(-170,246){\LARGE(a)}\end{picture}} 
\bminic{0.55}\subfigure{
%\vspace{-10mm}\hspace{-8mm}\bminic{0.50}
\includegraphics[scale=.34,clip=true,trim=240 40 0 140]{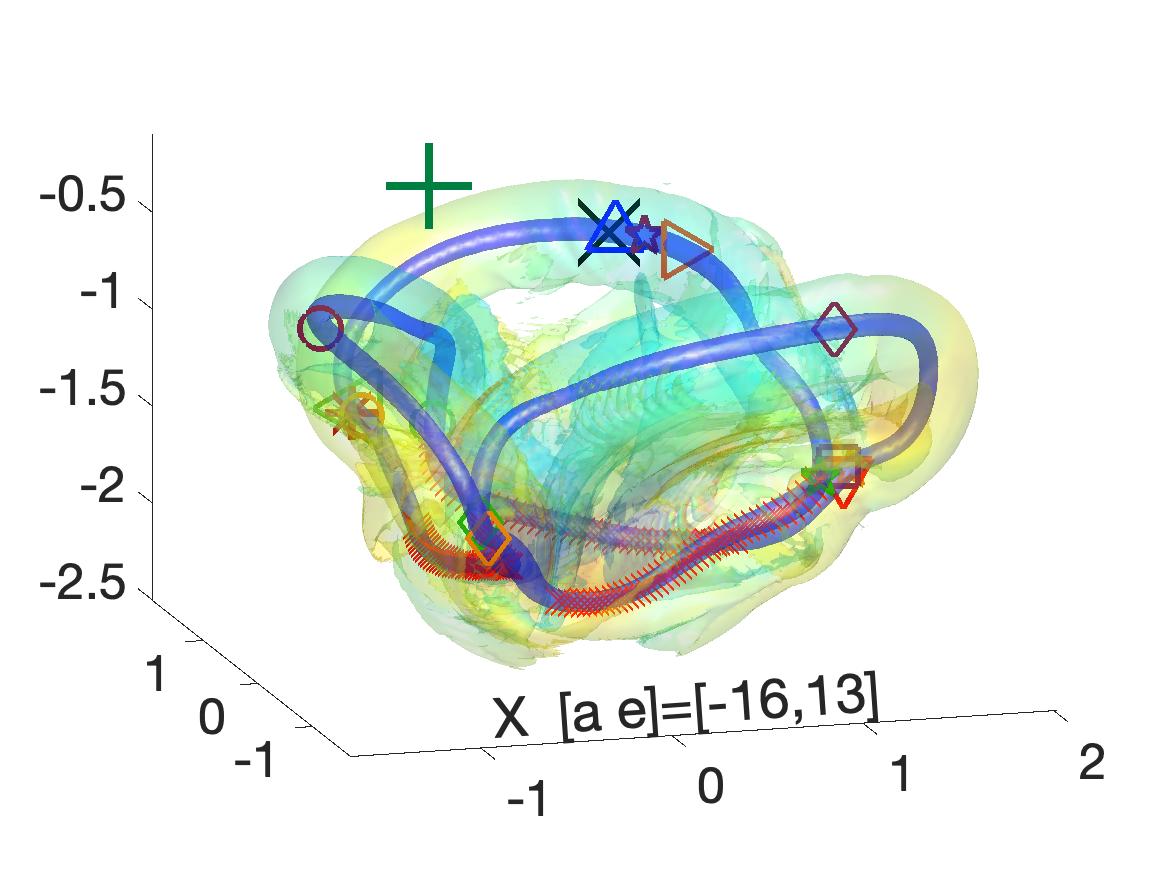}
\begin{picture}(0,0)\put(-94,210){\LARGE(b)} 
\end{picture} } \emini\bminic{0.45}
\vspace{-24mm}
\caption{\label{fig:3DT4p8} Two views of the same $t=4.8$ isosurfaces from the $p_r=1$, 
$r_o=0.015$ (r1d015) calculation from different elevation angles.  (a) (planar view) and 
(b) (side view).
$t=4.8>t_r\sim4$ represents the middle of the initial phase leading that ends with the first 
reconnection at $t_x$=6. $\pmb{\omega}${\bf -isosurfaces}: A blue inner $\omega=14$ surface
and a small $\!\omega\!=\!0.65\!=0.02\omega_m$ isosurface with mapped helicity. 
The positions of $\omega_m$, $\max(h)$, $\min(h)$
and $u_m$ are given along with extrema of terms from the enstrophy and helicity budgets. 
The red hashes indicate where sheets arise from the marked centerline spans of 
$\epsilon_h<0$ in budget figure \ref{fig:T4p8uuoo}a.
Plus three triplets of local positions $s_f$, $s_d$ and $s_g$ at local $\min(h_f)$,
$\min(\epsilon_h)$ and its opposing points. 
The symbols given in the legend are also used in figure \ref{fig:T4p8uuoo}.
%, with (b) a transition observation angle. 
In (a) the overall structure of the lobes is emphasized. %and (c) 
(b) shows that the red hashes are all in the lower portion and represent where
a separate lower vortex ring is forming. The origins and location of the yellow regions are given
in the next figure.  
} \emini 
\end{figure}
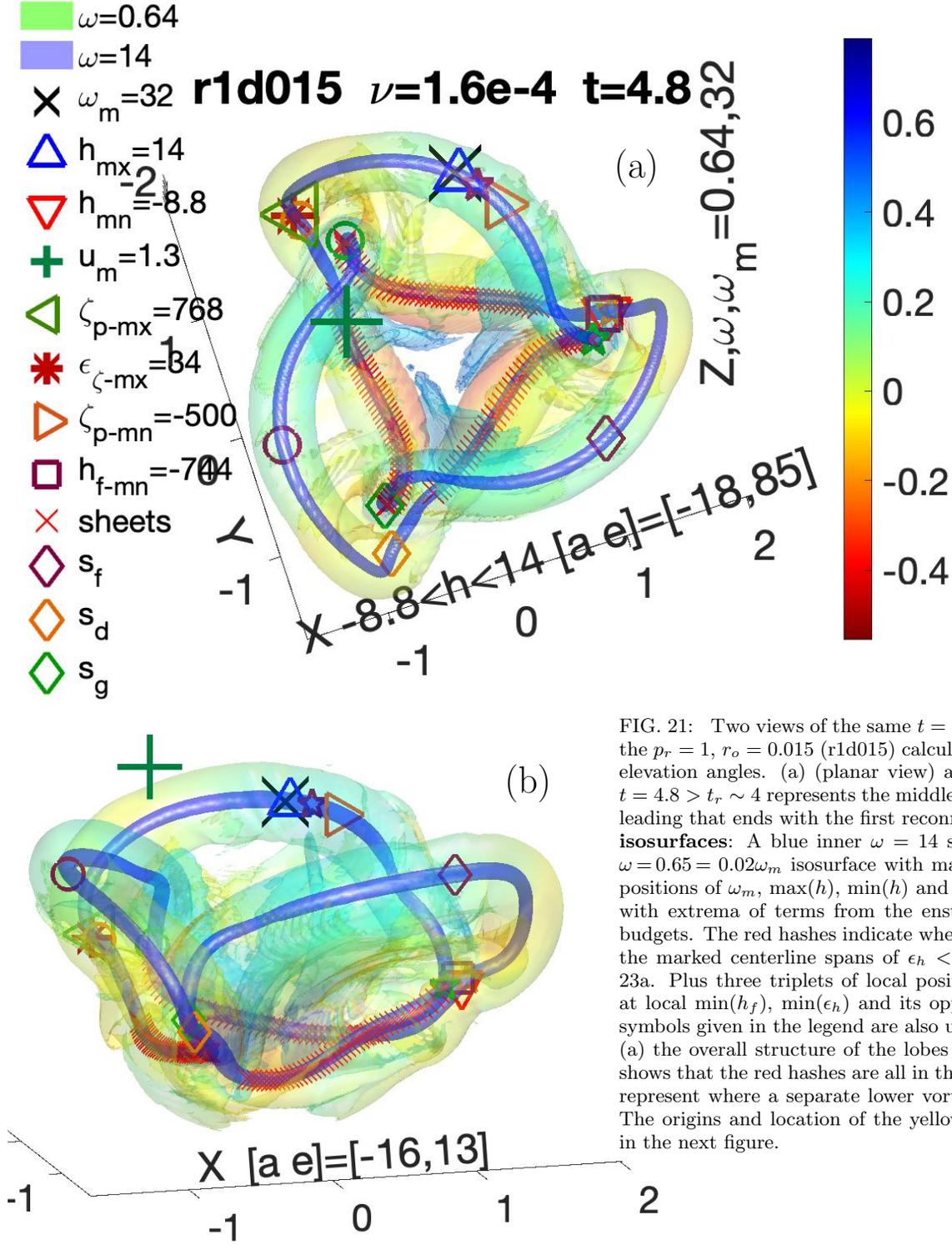

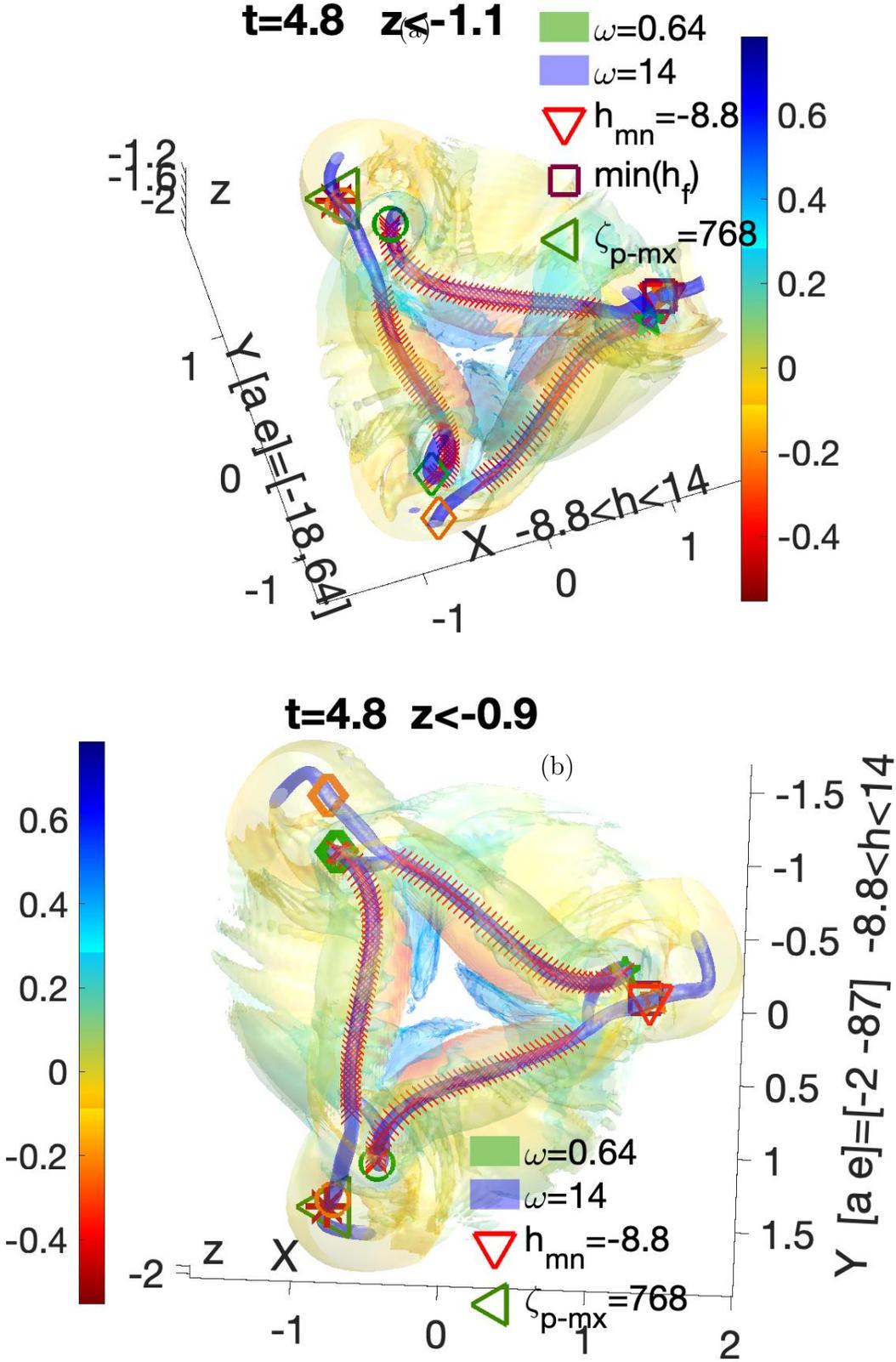
\begin{figure}[H] % scale=.38
\includegraphics[scale=.35,clip=true,trim=0 0 0 0]{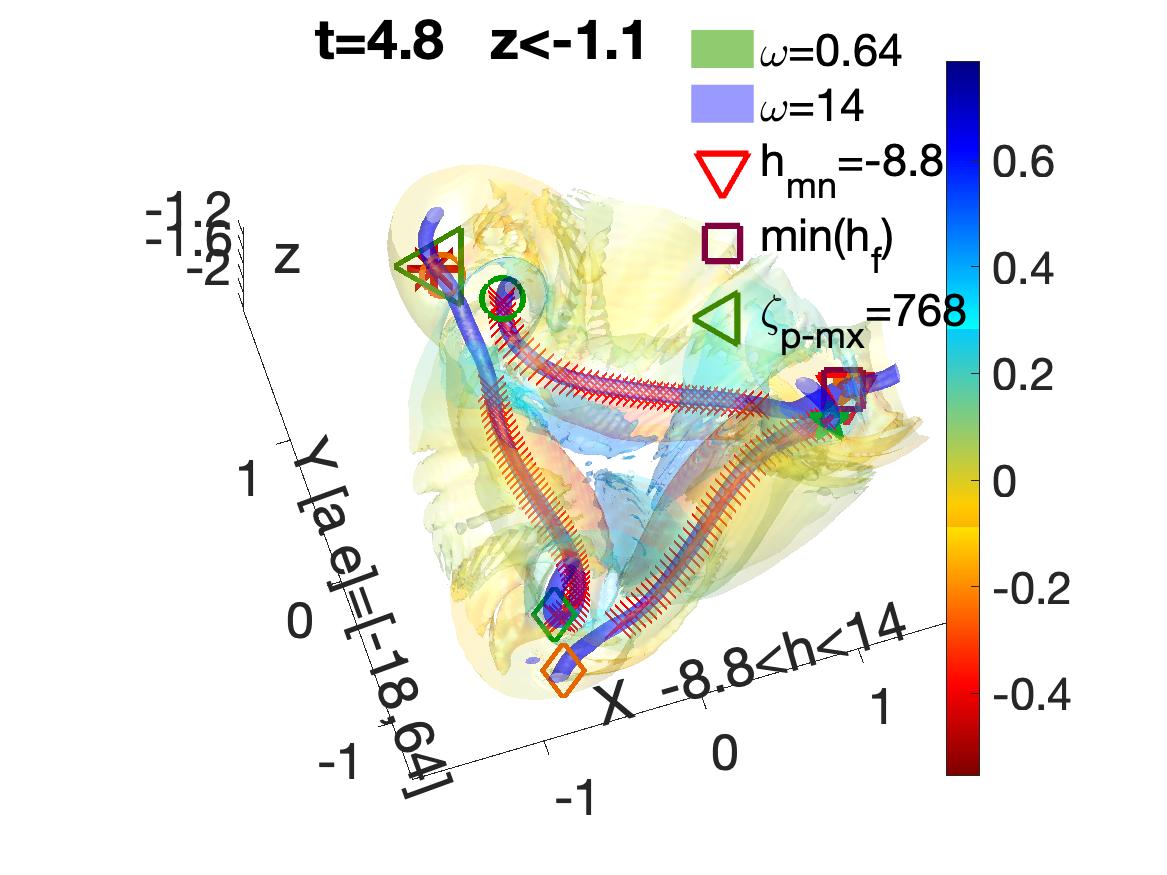}\begin{picture}(0,0)\put(-230,286){\large(a)}\end{picture} 

%\hspace{-30mm}\bminic{0.60}
\includegraphics[scale=.35,clip=true,trim=0 0 0 0]{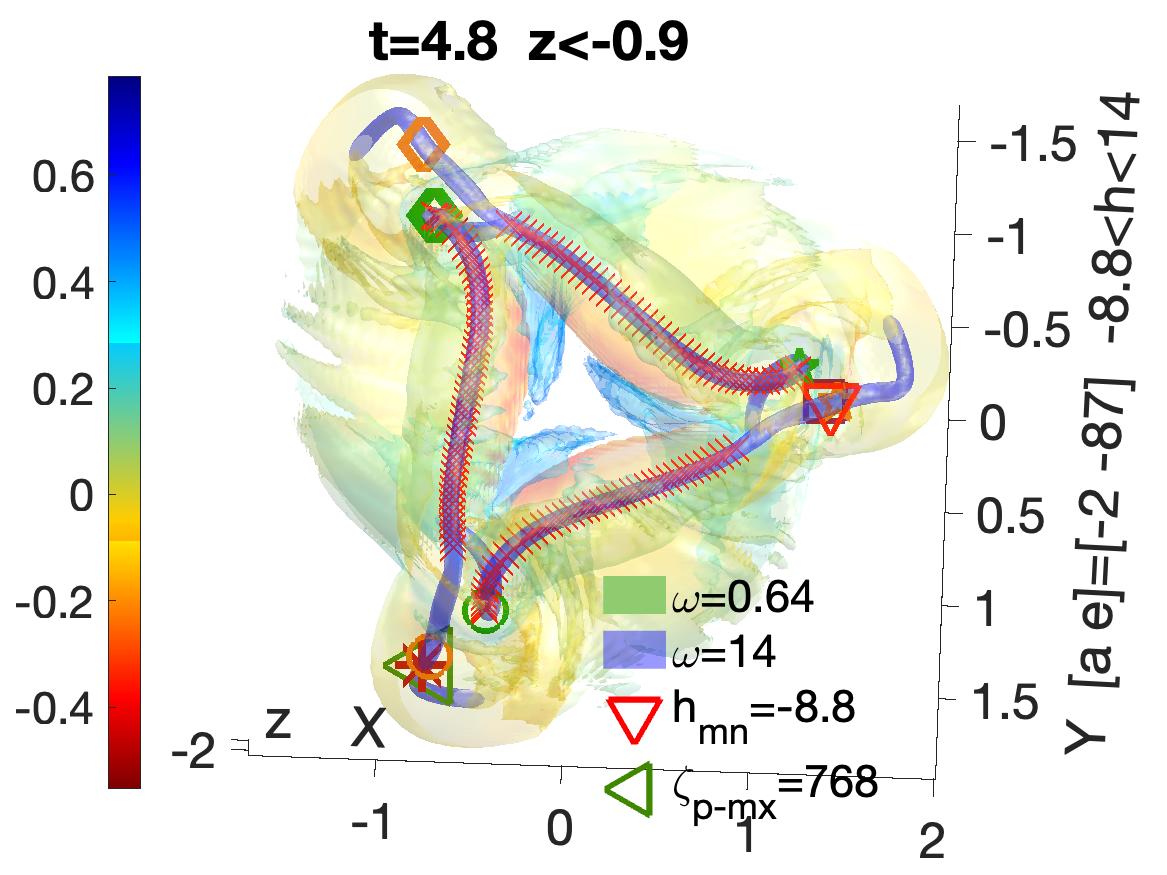} %\emini\bminic{0.5}
\begin{picture}(0,0)\put(-170,266){\large(b)}\end{picture} \\
\caption{\label{fig:3DT4p8low} Two views of the $t=4.8$ lower region for $z<-1.1$ and -0.65
respectively, with each perspective is dominated by yellow $h\lesssim0$: (a) looking down;  
(b) looking up with the domain flipped across a line from
[x y]=[-1 1.5] (green triangle) to the [x y]=[1 -1] corner, with some of the upper $h>0$ zone
included. It is also rotated a bit about the $z$-axis to give a flavor of how the legs of the
lower ring are connecting with the bridge.
Gray is where we are looking through both the lower yellow and upper blue. Some of
the $h>0$ zone is included to show the while the $h\lesssim0$ sheets are being shed from 
the lower $h<0$ centerline, they extend up to the upper $h>0$ blue-marked 
centerline.  The orange $s_d$ and the opposing green $s_g$, both
marked with {\large$\diamond$}'s, are highlighted to show how the legs might be starting to wind 
around each other.  }
\end{figure}

The three-dimensional structure at $t=4.8$ is illustrated in figures \ref{fig:3DT4p8} and 
\ref{fig:3DT4p8low} using several perspectives of two vorticity isosurfaces
and red $h<0$ hash marks. Mapped-$h$ is on the broader isosurface with a lower vorticity: 
$\omega$=$0.64\approx0.02\omega_m$. And a higher vorticity $\omega=14$ monochrome isosurface 
that encases the centerline vortex. With the red hash marks 
indicating the $\epsilon_h\lesssim0$ spans on the centerline from which the sheets are shed. 
Figure \ref{fig:3DT4p8} shows the entire structures from two perspectives. To clearly see
the yellow $h\lesssim0$ sheets, figure \ref{fig:3DT4p8low} lops off upper parts of the trefoil.
\smallskip

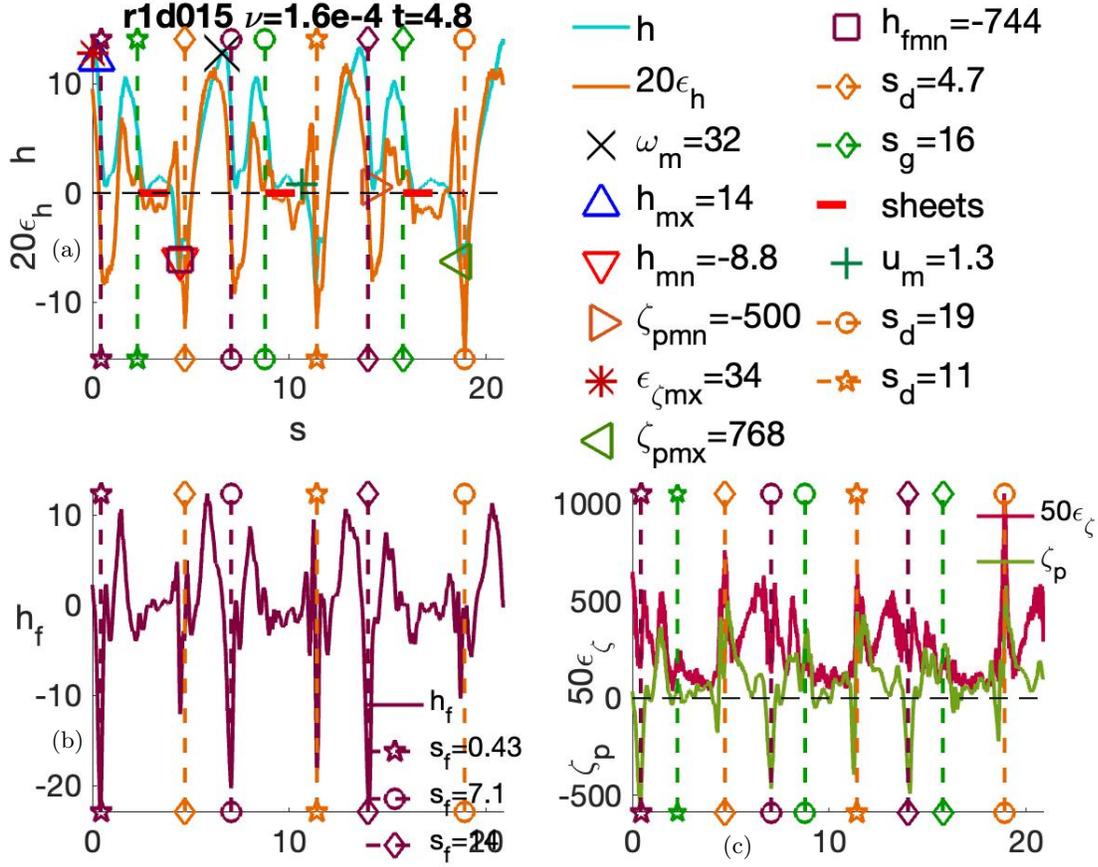
\begin{figure}[H]\vspace{-0mm}
%\bminic{0.76}
\includegraphics[scale=.35,clip=true,trim=20 60 0 0]{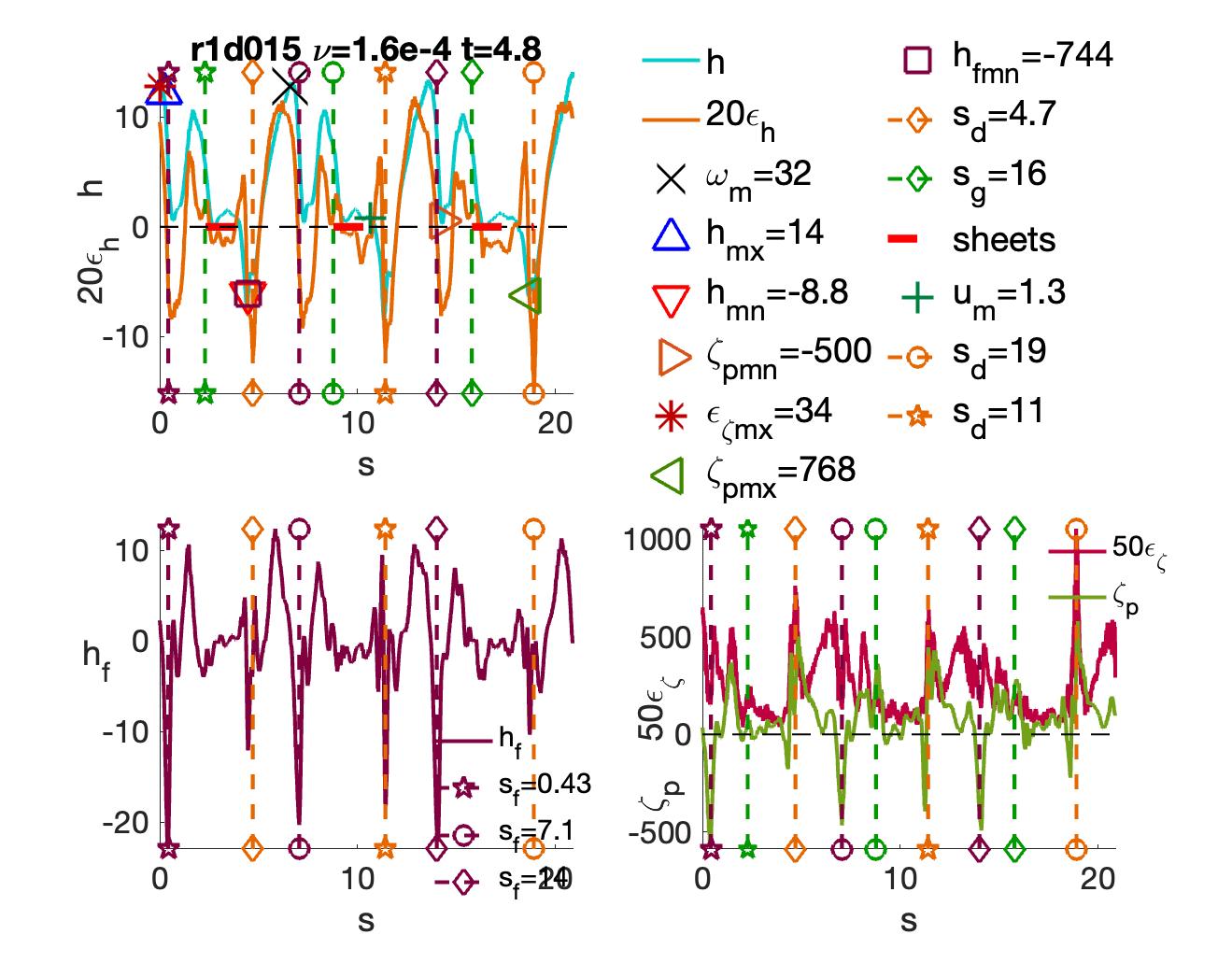} %\emini\bminic{0.25}
\begin{picture}(0,0)\put(-422,232){(a)}
\put(-422,46){(b)}\put(-168,6){(c)} \end{picture}
\vspace{-0mm}\caption{\label{fig:T4p8uuoo} 
Vorticity centerline profiles at t = 4.8 for case r1d015. 
Budget profiles: $h$, $\epsilon_h$, $h_f$, $\epsilon_\zeta$ and $\zeta_p$, with added vertical 
dashed lines for these local positions: $s_f$ (maroon, $\min(h_f)$), 
$s_d$ (yellow, $\min(\epsilon_h)$), and $s_g$ (green) for the $s_d$ opposing points.
The $s_f$ are also at $\min(\zeta_p)$ and at large local enstrophy dissipation 
$\epsilon_\zeta$ positions. The $s_d$ are at secondary $\min(\zeta_p)$ and at 
local $\max(\epsilon_\zeta)$  positions. The $\epsilon_h(s)\lesssim0$ spans over which the 
$h<0$ sheets are being shed are indicated by thick, dashed red lines that are to the right of 
each $s_g$. Reconnection is forming between spans near each $s_d$ and the red hashed patches on 
the opposing loops with green $s_g$ symbols at one end. For example: the yellow diamond at
$s_d=4.7$ and the span next to the green diamond at $s_g=16$}
\end{figure}

{\bf t=4.8 r1d015 centerline budgets} Similar to how figure \ref{fig:3DT3p6} at  $t=3.6$ marks %15
in red the centerline spans with the strongest $\min(\epsilon_h)<0$, for $t=4.8$ in figure
\ref{fig:3DT4p8} marks those spans with with red hashes. Spans whose extent on both the centerline 
in figure \ref{fig:T4p8uuoo} and the isosurfaces is indicated by: one end by the green 
$s_g$, then continues to 2/3rds of the way to a $s_d$ mark from another $s_d-s_g$ pair. 

The maroon $s_f$ positions are no longer part of the ongoing reconnection, but are on a $h>0$ 
zone that is becoming an upper vortex rings. While the red hashes and the $s_d$ and $s_g$ marks 
are becoming part of a lower ring. 

The sideview in figure \ref{fig:3DT4p8}b shows this more clearly.

Further remarks:
\ITM\item[$\circ$] In figure \ref{fig:T4p8uuoo}a the $s_d$ mark the primary $\min(h)<0$ positions 
and in \ref{fig:T4p8uuoo}c the positions of $\max(\epsilon_\zeta)$, enstrophy dissipation.
\item[$\circ$] The $\epsilon_h\lesssim0$ spans with red hashes show that the reconnection between 
the loops is between segments on those loops and is not simply point-to-point as with Lamb-Oseen.
\item[$\circ$] The yellow vortex sheets at $t=4.8$ now encompass almost the entire interior 
within the trefoil. 
\ITN

\ITM\item[{\bf L-O}] Comparing figure \ref{fig:3DT4p8} to Lamb-Oseen in figure 
\ref{fig:GuuooT3p6}, the only similarity is 
that reconnection is forming between a primary marker and its opposing point. However the primary
L-O reconnection markers are not the $s_d$, but the $s_f^+$ at local $\max(h_f)$ points. 
Locations with stretching, $\zeta_p>0$, not compression. Part of the dynamics responsible for why 
the algebraic and Lamb-Oseen reconnection structures are so different.
\item[$\circ$] 
While Lamb-Oseen creates isolated braids that quickly dissipate, and shut down 
enstrophy production, the algebraic profiles shed vortex sheets. 
Sheets whose mutual interactions that can accelerate enstrophy production.
\ITN

In figure \ref{fig:3DT4p8low} the upper, blue $h>0$ zone has been lopped off to reveal the
full extent and nature of the vortex sheets. 

{\bf Centerline budgets and bridge formation.} 
Up through $t=3.6$ the centerline budget profiles have 
largely been used to identify the origins of the divergent evolution between the two types 
of initial vorticity profiles. What can the $t=3.6$
centerline budgets tell us about the dynamics and structures during the next phase?

First question: Why is so little negative helicity ($h<0$) seen on the centerlines?  
Despite the presence of neighboring $h<0$ vortex sheets, 

A likely contributing factor is the spans of strong $\epsilon_h<0$ on the centerlines can
act as sponges that remove centerline $h<0$.
%Explaining why significant $h<0$ on the centerlines is almost never observed. 

Second: What is the local dynamics when the trefoil starts to break into two rings? At $t=3.6$,
the three $s_d$ and the opposing $s_f$-$s_g$ are all locations with local $\min(h_f)$ and 
$\min(\zeta_p)$, indicating local compression and pinching along the vortex lines on both sides of
the developing reconnection bridges. Probably due in part to the interactions between the 
bridges' two ends in three-dimensions.

Third: For how long does this compression/pinch persist? In $t=4.8$ figure \ref{fig:T4p8uuoo}, 
the local $\min(h_f)$ and $\min(\zeta_p)$ diagnostics that foreshadowed reconnection for 
$t\leq3.6$ still have coincident large negative spikes. However these are now located within the 
developing upper ring, far from the three developing reconnections. And unlike at
$t=3.6$, are not adjacent to $s$-spans with significant enstrophy production, 
$\zeta_p>0$. 

Fourth: Even as the compression/pinch dynamics subsides at $t\sim4.8$, the enstrophy continues
to grow. On the centerlines this is because the yellow, local $\min(\epsilon_h)$ $s_d$ points 
still have local enstrophy production maxima, $\max(\zeta_p)>0$. And overall is because for
$t\geq4.8$, most of the enstrophy production is coming from the 
growth of the $h<0$ vortex sheets that that now envelop the lower ring and the bridges that 
connect the upper and lower rings. 

Why is the creation of $h<0$ sheets so important?  Starting with these two reasons. First, 
by creating $h<0$ zones, the vorticity in the $h>0$ zones can grow; this breaks the early, 
pre-viscous, helicity conservation constraint upon vorticity growth. Second, by spreading the 
vorticity into sheets, the enstrophy in figure \ref{fig:r1ZHnu} can continue to grow during the 
first phase of reconnection; unlike the Lamb-Oseen enstrophy in figure \ref{fig:Gd05dm1ZHnus}.
Which sets up the next stage as those sheets begin to interact with one another at $t=6$.

{\bf t=6} The last set of r1d015 isosurfaces are for $t=6$ in figure \ref{fig:3DT6}.
Instead of a finding a centerline vortex, there is a higher vorticity isosurface within the 
low vorticity isosurface. This $t=6$ figure represents when the first phase of reconnection 
ends, defined as the time $t_x$ when the $\sqrt{\nu}Z(t)$ 
converge in figure \ref{fig:r1sqnuZ} and the shedding of $h<0$ sheets has ended.
The views of the isosurfaces at $t=6$ in figure \ref{fig:3DT6} are
similar to those at $t=4.8$ in figures \ref{fig:3DT4p8} and \ref{fig:3DT4p8low}: 
(a) a side view of the entire trefoil; and (b) a plan view of the lower ring,
taken from the subdomain outlined in \ref{fig:3DT6}a. 
With differences.

The side view in figure \ref{fig:3DT6}a shows that the legs of the lower ring have separated 
from the upper ring, with connecting bridges whose inner, large-$\omega$ isosurfaces are
winding around one another. Such as in the upper right, with some wrapping of the helicity-mapped 
isosurface about the core. This has some similarities to how the Lamb-Oseen upper and lower rings 
in figure \ref{fig:G3DT4p0} with connecting bridges at $t=4$. Bridges whose ends then wrap 
about the rings in figure \ref{fig:G3DT4p4}.  Except that for Lamb-Oseen the bridges 
transform into isolated braids in figure \ref{fig:G3DT4p4}. Not broad vortex sheets. 

What the experiments can visualize with Lagrangian markers are only the strongest isolated 
vortices. What those experiments miss are the low vorticity sheets, like those 
at $t\geq4.8$ in figure \ref{fig:3DT4p8low}. 
%eventually leading to accelerated growth of the enstrophy in figures\ref{fig:r1ZHnu,r1sqnuZ}
In this sense, the algebraic large-$\omega$ bridges in figure \ref{fig:3DT6}a, 
are a better representation of recent directly observed experimental 
vortices \cite{KlecknerIrvine2013,ScheeleretalIrvine2014} than Lamb-Oseen bridges, such as in
figures \ref{fig:G3DT4p0} and \ref{fig:G3DT4p4}.

The plan view in figure \ref{fig:3DT6}b shows the beginnings of the next phase, with changes in
the pigmentation on the sheets of the lower ring as they start wrapping around one another. 
The pigmentation changes from the almost all yellow, and some red, at $t=4.8$ in 
figure \ref{fig:3DT4p8low} to pigmentation at $t=6$ in figure \ref{fig:3DT6}b that varies from 
red to yellow to green. Along the leg that runs from lower right to the upper left, there is 
orange ($h<0$) coming out of the bridge in the lower right, yellow ($h\lesssim0$) on the shed 
sheet in the middle, then green ($h\gtrsim0$) on the left that is wrapping around the bridge and 
another sheet.
This variation in color suggests that the sign for the vortical velocity $\bu\cdot\hat{\omega}$ 
is also changing, which implies stretching along the legs. 

Given that these stretched sheets are wrapping around the bridges and their neighboring sheets, 
a configuration has been created with all the elements required to invoke the 
Lundgren model \cite{Lundgren1982} for stretched spiral vortices. This is the only
analytic model that generates the growth of enstrophy required to generate a -5/3 Kolmogorov-like 
spectra. Which also implies the generation of a dissipation anomaly \eqref{eq:dissanom}.  
Work on the details of the responsible inter-sheet dynamics is in progress.

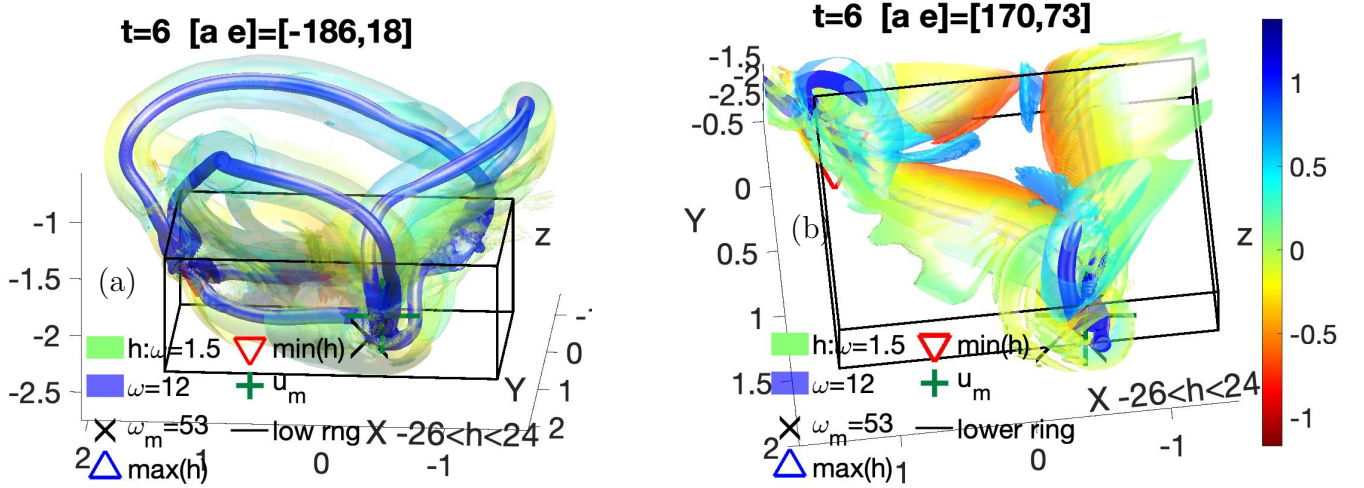
\begin{figure}[H]\bminic{0.5}\subfigure{
\includegraphics[scale=.225,clip=true,trim=0 0 180 0]{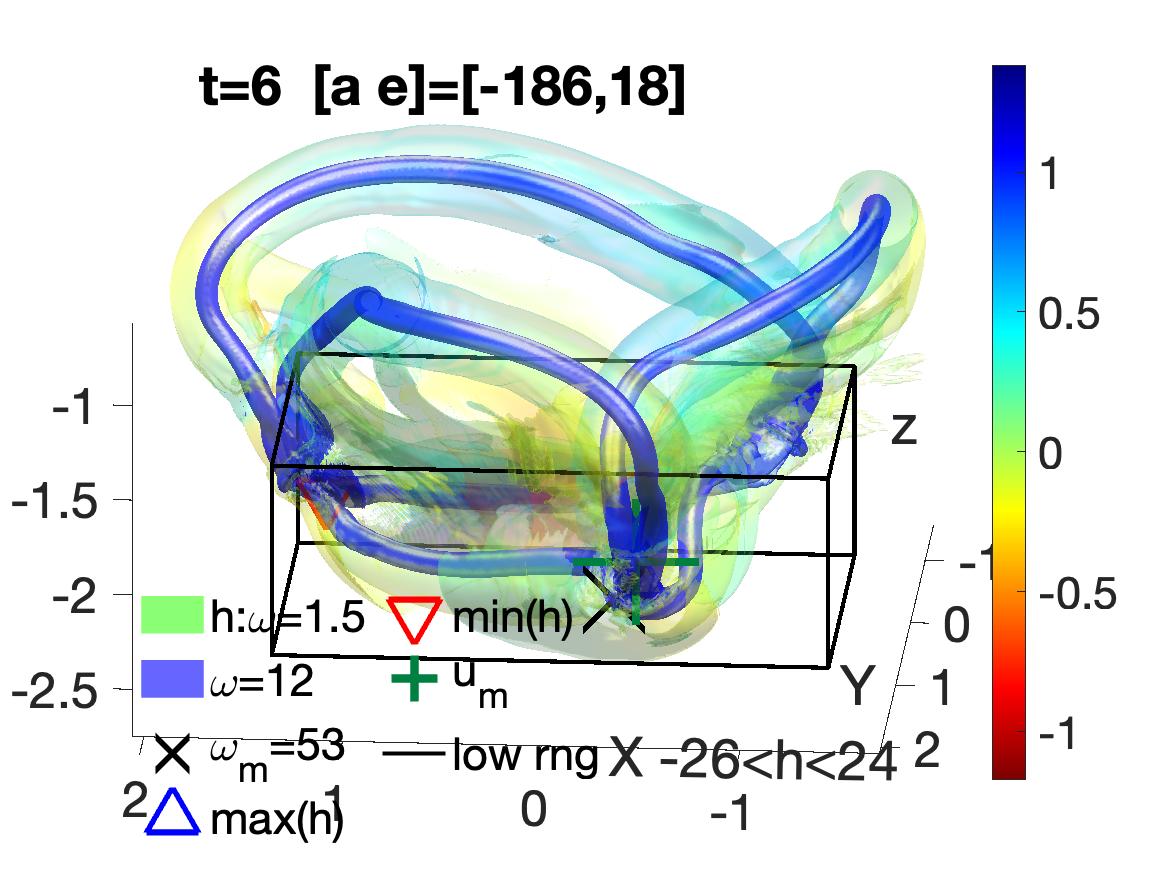}
\begin{picture}(0,0)
\put(-190,80){\large(a)}\end{picture}  }
%\put(-220,80){\large(a)}\end{picture}  }
\emini\bminic{0.5} \subfigure{
%XYyl2pi512d015nu16ks1T6azm39el19-22jul22.jpg
\hspace{-4mm}\includegraphics[scale=.225,clip=true,trim=10 0 0 0]{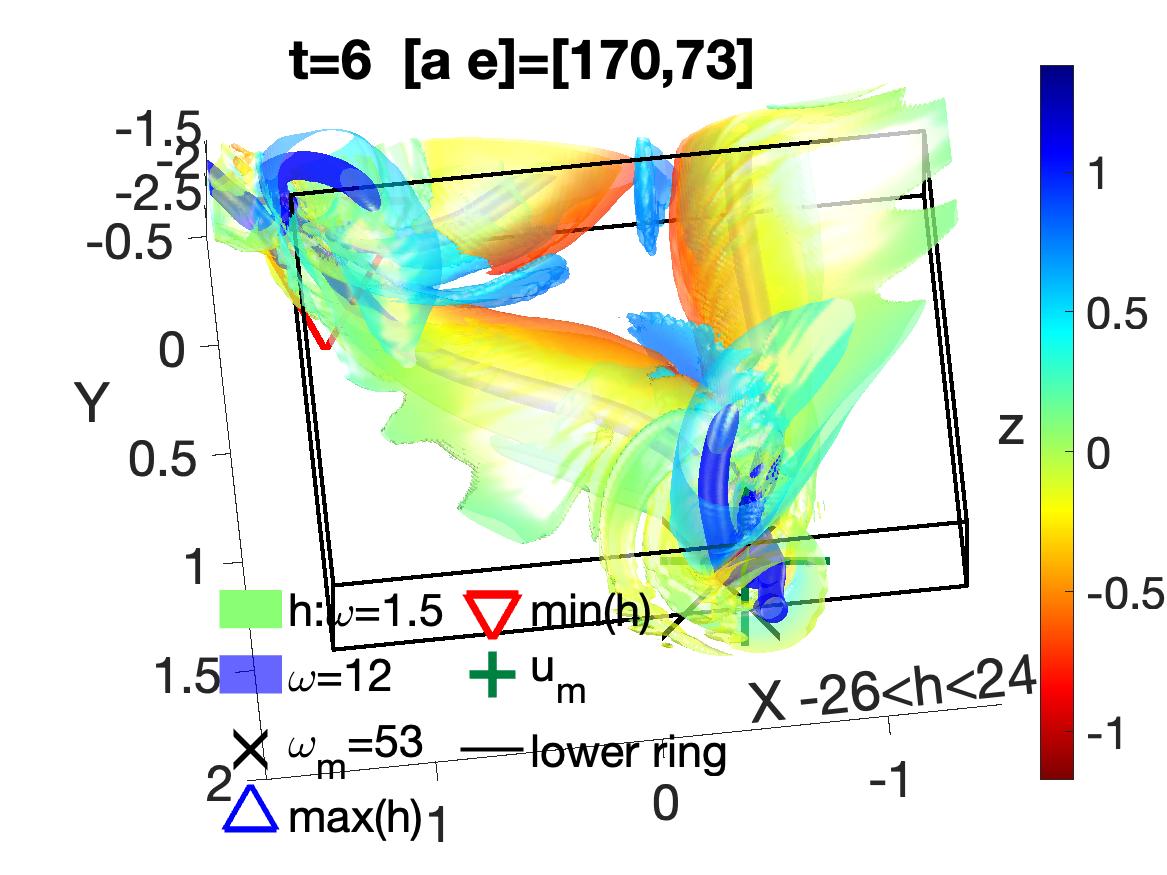}
\begin{picture}(0,0)
\put(-400,204) {\bf\Large Algebraic $\pmb{p_r}$=1 $\pmb{r_o=0.015}$ t=6.} 
\put(-210,100){\large(b)}\end{picture}} \emini
%XYyl2pi512d015nu16ks1T6azm18el85-09aug22.jpg
\vspace{-0mm} \caption{\label{fig:3DT6} Two $t=6$ r1d015 isosurface perspectives at $t_x$, the 
end of the first reconnection, as defined by figure \ref{fig:r1sqnuZ}a. This is when the
dissipation in figure \ref{fig:r1sqnuZ}b begins to accelerate, with
convergence of $\epsilon=\nu Z$ at $t\approx10$. 
There are two isosurfaces: inner $\omega=12$ blue that encases the centerline; 
outer $\omega=15$ with helicity-mapping.
The two perspectives are similar to those at $t=4.8$: (a) is a side view similar to that 
in figure \ref{fig:3DT4p8}; (b) is a cropped plan view, similar to figure \ref{fig:3DT4p8low} 
but with the helicity brightened.  A box is 
drawn on both frames to show where the subdomain in (b) has been taken from the full domain 
in (a). In (a) the dominant structure is the pure blue $\omega=12$ centerline isosurface
with three bridges connecting the separating upper and lower vortex rings. This illustrates what 
direct experimental visualizations of cores are probably observing \cite{KlecknerIrvine2013}.
The plan view shows what those experiments cannot see: lower $\omega$ magnitude $h\lesssim0$ 
vortex sheets.  Two differences with figure \ref{fig:3DT4p8low} are that the sheets shed from the 
legs change pigmentation along their length, and they are wrapping around one another at 
the bridges. The `left' bridge has the $\min(h)$ (red $\pmb\triangledown$) mark. The `right'
bridge has the $\omega_m$ ({\bf X}) and $u_m$ (green {\bf +}) marks.
The color change on the bottom leg is from orange $h<0$ at the {\bf(X,+)} `right' bridge
to green at the `left' bridge. With the `left' green wrapping around the `left' bridge
in the upper left and green from the leg on the right wrapping about `right' bridge and some of 
the y-axis leg.}
%A feature not seen in $t=4.8$ figure \ref{fig:3DT4p8low}, 
%are indications that the sheets are wrapping around each other and the centerline.
\end{figure}
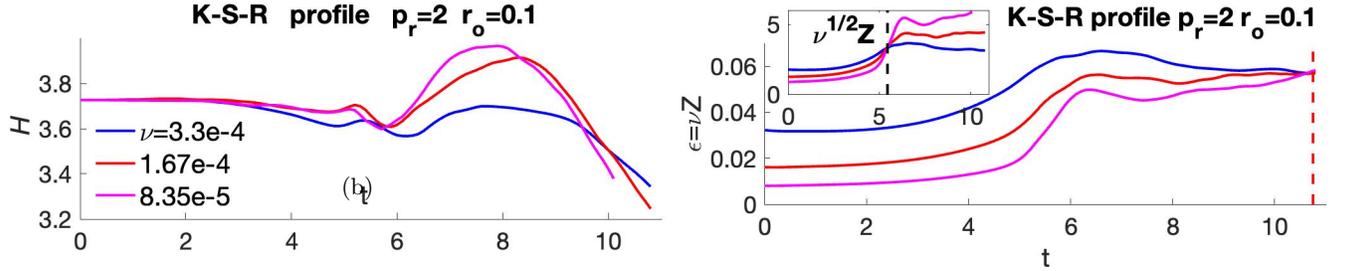
\begin{figure}[H]
\includegraphics[scale=0.24,clip=true,trim=10 0 0 0]{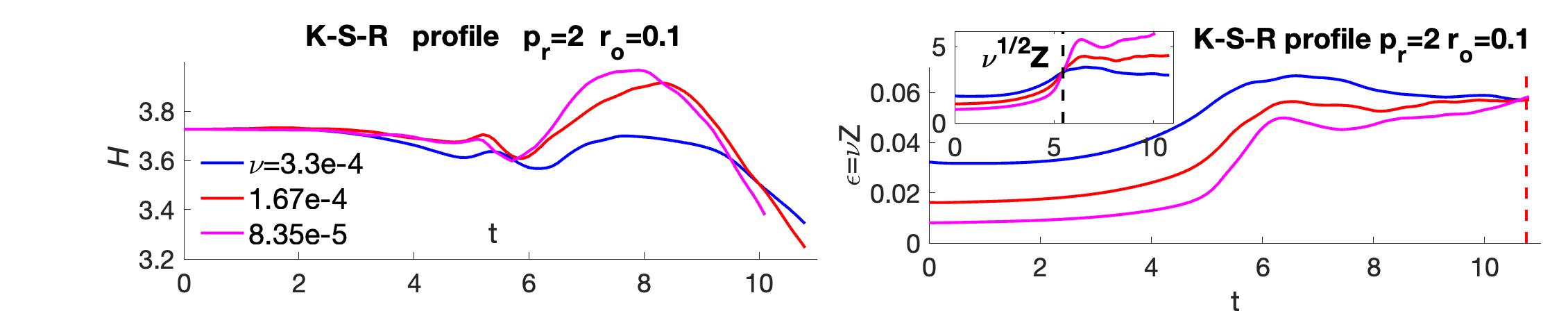}
\begin{picture}(0,0)\put(-80,46){(a)} \put(160,46){(b)}\end{picture} 
\caption{\label{fig:KSRZHr2d1sqnu} For case r2d1, algebraic K-S-R profile 
\eqref{eq:Rosenh} with $p_r=2$ and $r_o=0.1$, evolution of the dissipation rate 
$\epsilon(t)=\nu Z$ (a) with approximate convergence at $t_e=10.75$, convergence of 
the reconnection-enstrophy $\sqrt{\nu}Z(t)$ at $t_x=5.45$ in the inset,
and (b) the helicity ${\cal H}$ for different viscosities. These curves are 
similar to those for
case r1d015 in figures \ref{fig:r1ZHnu} and \ref{fig:r1sqnuZ}.}
\end{figure}
\begin{figure}[H]
\includegraphics[scale=0.24,clip=true,trim=10 180 0 0]{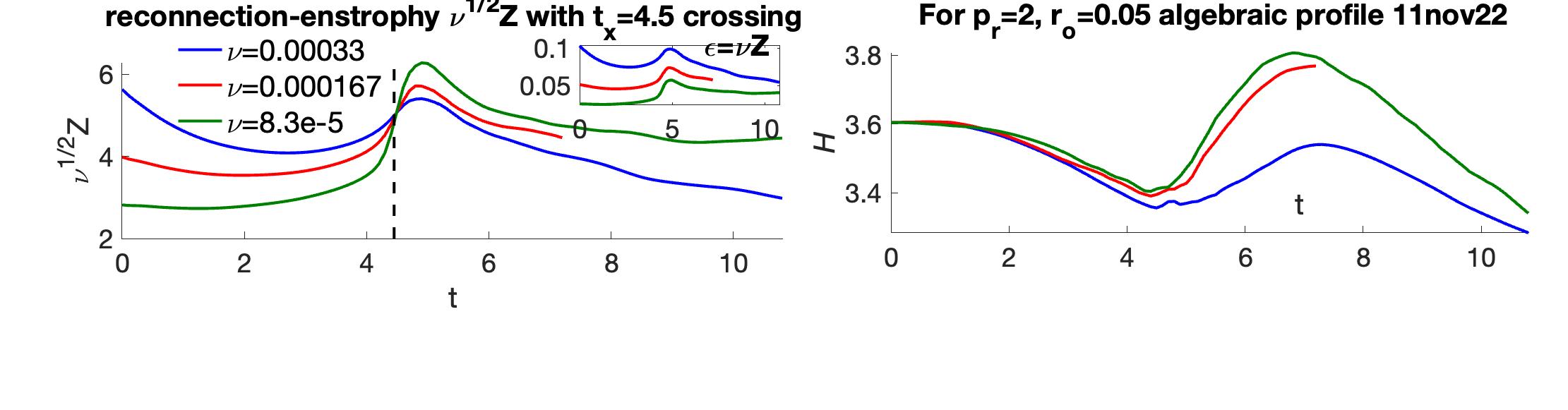}
\begin{picture}(0,0)\put(-80,36){(a)}\put(10,76){(b)}\put(100,36){(c)}
\end{picture} 
\caption{\label{fig:KSRZHr2d05sqnu} For case r2d05, algebraic with $p_r=2$ and 
$r_o=0.05$, for different viscosities: 
(a) Convergence of $\sqrt{\nu}Z(t)$ at $t=4.45$, 
(b) evolution of the dissipation rate $\epsilon(t)=\nu Z$ as an inset, 
and (c) the helicity ${\cal H}$.
Case r1d006 ($p_r=1$, $r_o=0.006$)
has similar $Z(t),~\sqrt{\nu}Z(t)$ and ${\cal H}(t)$ evolution and incipient
vortex sheets because for both, the $(2\pi)^3$ domain is too restrictive when
the core radius is very thin.}
\end{figure}
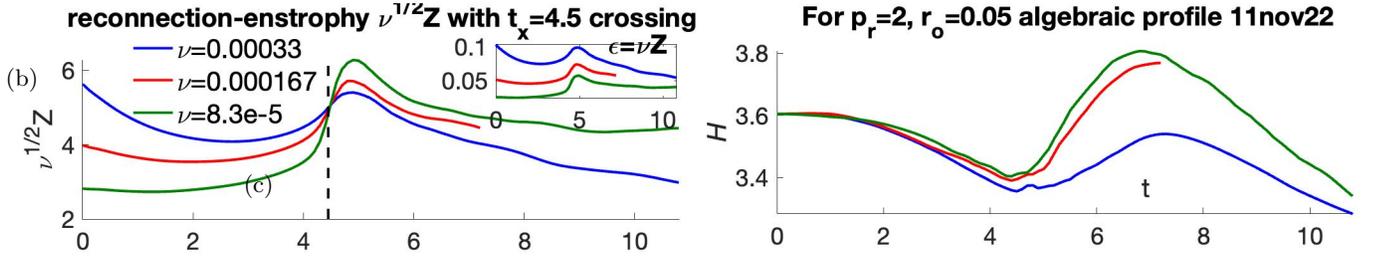

\begin{figure}[H]
\bminic{0.5}\subfigure{
\includegraphics[scale=.22,clip=true,trim=0 0 0 190]{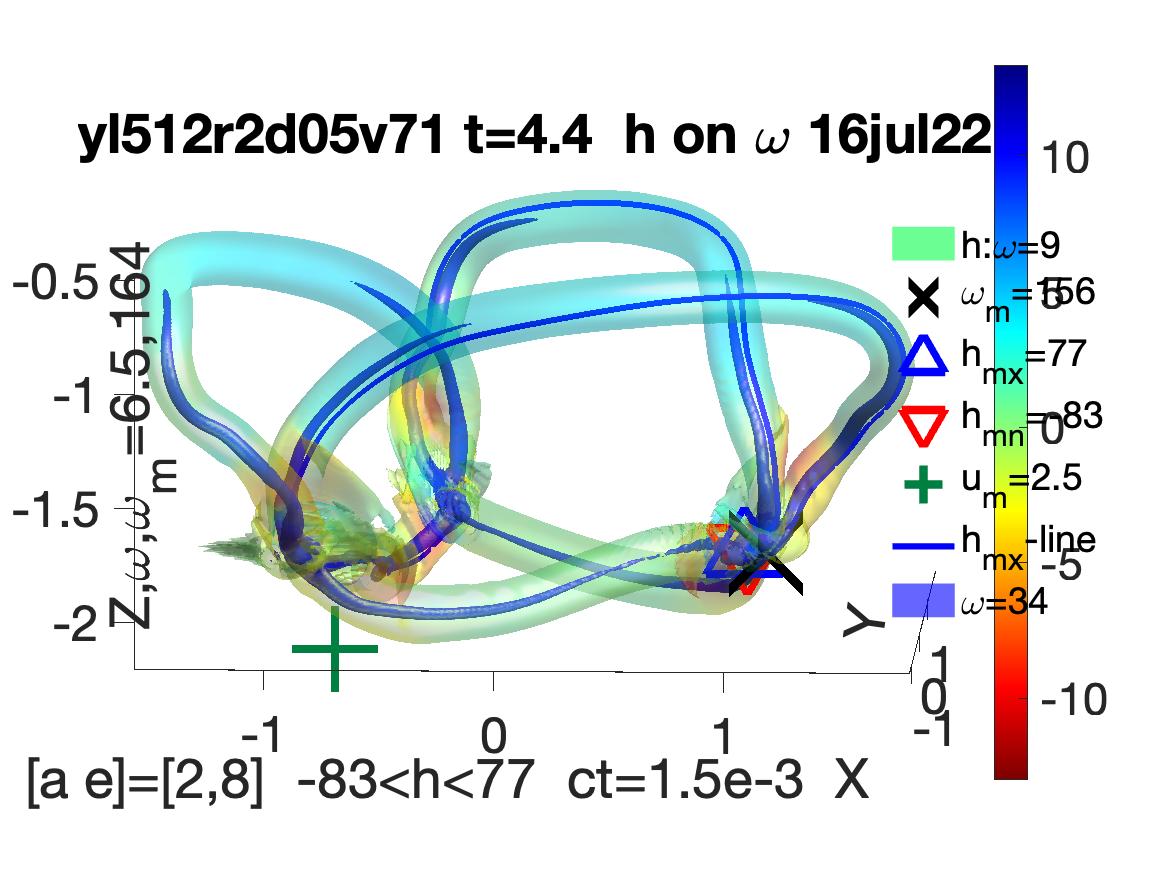}
\begin{picture}(0,0)\put(-250,150){\bf\Large Algebraic $\pmb{p_r}$=2 $\pmb{r_o=0.05}$
 t=4.4.} 
\put(-80,36){(a)}\end{picture} }
\emini\bminic{0.5}\subfigure{
\includegraphics[scale=.22,clip=true,trim=0 0 0 190]{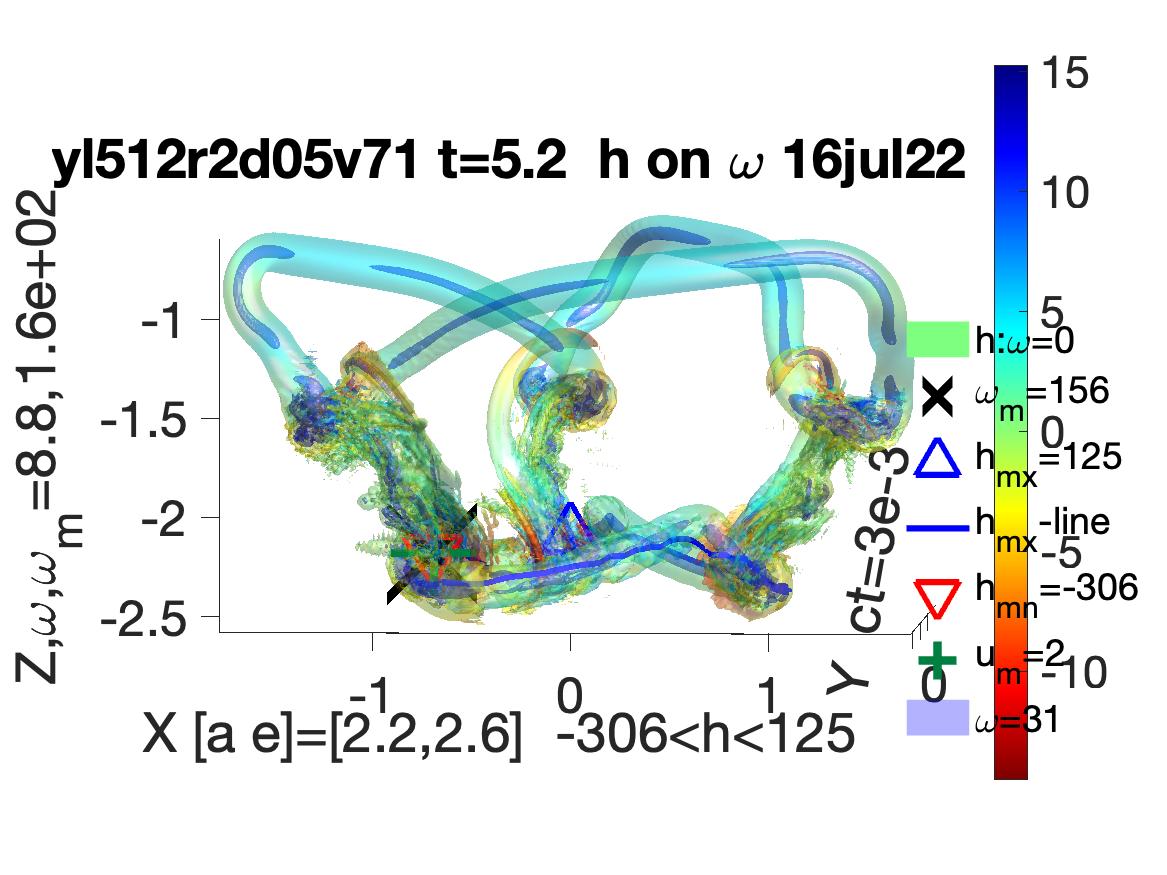}
\begin{picture}(0,0)\put(-180,150){\bf\Large t=5.2}
\put(-80,36){(b)}\end{picture} }
\emini
\caption{\label{fig:r2d05T4p4T5p2} For r2d05 side views at $t=4.4$ and $5.2$. 
(a) At time $t=t_x=4.4$, when the $\sqrt{\nu}Z(t)$ cross, a vortex sheet is being generated. 
(b) Which become connecting bridges at $t=5.2$. High $\omega$ isosurfaces are used
instead of vortex lines to indicate the centerlines.}
\end{figure}
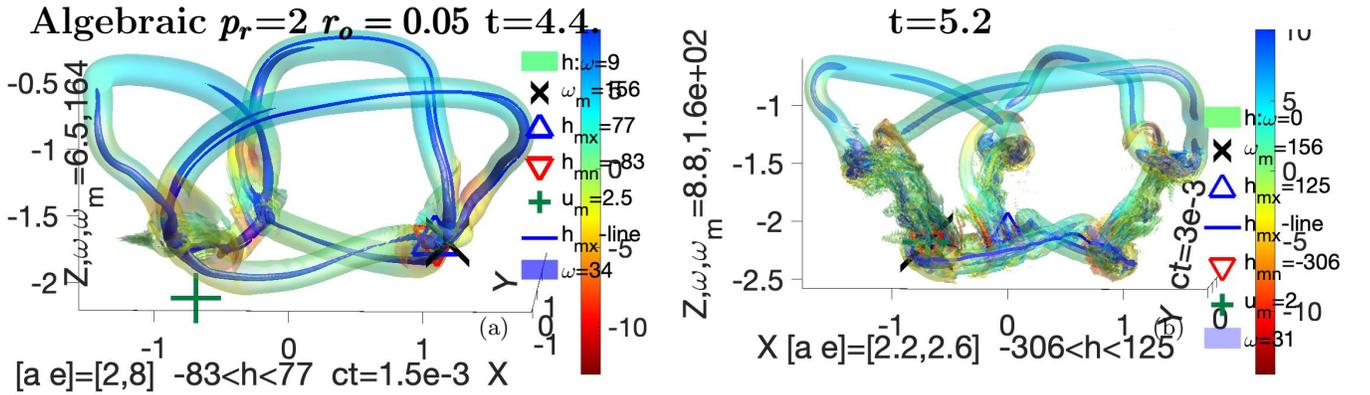

\subsection{Reconnection-dissipation structures for K-S-R $p_r=2$ \label{sec:KSRdiss}}

To finish the cases, a few results from the two K-S-R $p_r=2$ cases r2d1 and r2d05 are
included. Recall that due to stability \eqref{eq:Raystable}, these profiles are stable
unless the azimuthal wavenumber $m$ \eqref{eq:mkmode} is very large. For case r2d1,
the evolution of $Z$, $\sqrt{\nu}Z$ and ${\cal H}$ mirrors that of case r1d015 in 
figure \ref{fig:r1ZHnu}. This includes strong convergence of$\sqrt{\nu}Z$   
at the same time of $t_x\simeq6$, and approximate convergence of the dissipation rate 
$\epsilon=\nu{Z}$ at $t_\epsilon\approx10$, with similar post-reconnection
${\cal H}(t)$ growth, then decay. The evolution of its three-dimensional structures is
also similar.

The calculations with thinner initial algebraic cores (r2d05 and r1d006) behave 
differently. Both generate $\sqrt{\nu}Z$ convergence, but earlier than r2d05 and r1d015,
and both fail to generation dissipation rate $\epsilon$ convergence. And for r2d05, 
the post-reconnection vortex structures in figure \ref{fig:r2d05T4p4T5p2} have similarities with
the Lamb-Oseen braids in figure \ref{fig:G3DT4p4}.

These final results are likely due to the constraints imposed by the three-fold symmetry and 
the confined $(2\pi)^3$ periodic domain. It has previously been shown 
that if the core thickness is thinner \cite{KerrFDR2018} or the Reynolds number is 
higher \cite{KerrJFMR2018}, larger domains are required to get convergence of$\sqrt{\nu}Z$.  
And that by breaking these constraints \cite{KerrJFMR2018}, the calculation can attain the 
accelerated enstrophy growth required for first $\sqrt{\nu}Z(t)$ convergence, then approximate 
convergence of the 
dissipation rates $\epsilon=\nu Z$ by a $\nu$-independent time. Which is not possible for the
final r2d05 and r1d006 calculations due to those constraints.  Full discussion of these questions
using new calculations in larger domains and a wider range of viscosities will be in a paper in 
preparation.

%\newpage
\section{Summary \label{sec:summary}}

\subsection{Concluding remarks. \label{sec:conclude}}

The critical points in this paper are:
\vspace{-0mm}
\ITM\item Demonstrating that the enstrophy and helicity at reconnection depend
upon the initial vorticity profile when vortex knots have the same initial 
trajectory and circulation. 
%\ITM\item The origins of the dependence at reconnection upon the initial vorticity 
%profile or trefoil vortex knots with the same initial trajectory and circulation. 
\vspace{-0mm}
\item Vortex centerline diagnostics that demonstrate how the evolution for different
initial profiles diverges.
\vspace{-0mm}
\item Explaining the structural differences that form during the first reconnection. 
Vortex bridges/braids for the Gaussian/Lamb-Oseen profile and vortex sheets for all 
the algebraic profiles. 
\vspace{-0mm}

\item[$\circ$] Not covered are the interactions between the vortex sheets of the
widest algebraic profiles that lead to $\nu$-independent convergence of $\epsilon$
and finite $\Delta E_\epsilon$ \eqref{eq:dissanom}.\\
That will be the topic of another paper that extends to later times the previous 
calculations of perturbed trefoil knots in domains that grow as the viscosity 
decreases \cite{KerrJFMR2018}.
\ITN

Only the two outlying cases (Gd05 and r1d015) have been discussed in detail. 
For each, these are the critical questions: 
\ITM\item[1)] Is it subject to infinitesimal instabilities? 
\item[2)] How does its $t=0$ stability influence its reconnection-time behavior? 
\item[3)] And does that behaviour allow to finite energy dissipation to form, or not?
\ITN

The answer to 1) comes from recent mathematics \cite{GallaySmets2020} that shows that
initial profiles can be subject to instabilities when the initial state has small, but not 
tiny, perturbations. 
If so, then the mathematics of instabilities upon a columnar vortex \cite{HowardGupta1962},
illustrated in figure \ref{fig:HGJ}, can be used to show that for almost all 
wavenumbers, there is a Richardson number dependent instability \eqref{eq:Jrr}, 
as in figure \ref{fig:Gt0Rht0omyz}. This develops despite the Lamb-Oseen profile being the 
usually successful and favorite choice of the engineering community. 
The resulting instability-induced proliferation of $\omega$=0-contours is illustrated by 
the $t=1.2$ $\omega_y$ 
cross-section in figure \ref{fig:Go2xzT1p2}. A property previously observed for
perturbed anti-parallel vortices \cite{Kerr2013a,BustaKerr2008}. 

In contrast, the regularized $p_r=1$ and $p_r=2$ algebraic profiles \eqref{eq:Rosenh} are almost 
always stable, with a comparison $\omega_y$ cross-section given in figure 
\ref{fig:r1d015T2p4}. 

How can those small $t\gtrsim 0$ differences be the origin of the dramatic post-reconnection 
differences?  New diagnostics are required because with the usual diagnostics of
$Z(t)$ and ${\cal H}(t)$, there are few differences between cases until reconnection truly begins. 

The most that the mapped-helicity isosurfaces can tell us about the dynamics is that around
regions of negative helicity $h<0$, sometimes just spots of yellow or red, viscous 
reconnection develops as the nonlinear timescale of $t_r\sim4$ is approached. What the
isosurfaces cannot explain is why the new structures that are generated are so
different. Bridges and braids for Lamb-Oseen and isosurface sheets for all of the 
algebraic profiles. What is needed is a set of diagnostics that can follow the 
dynamics of the interiors before the enstrophy $Z(t)$ and the helicity ${\cal H}(t)$
diverge after $t\sim t_r$.

2a) The terms from the enstrophy and helicity budget equations 
(\ref{eq:enstrophy},\ref{eq:helicity}) are another set of diagnostics that might provide evidence 
for the early origins for the differences between cases. 
These could be mapped onto isosurfaces, as done for the helicity, or on the centerlines. 
When mapped onto the isosurfaces, their variations are too weak to be useful. 
In contrast, when mapped onto the centerline vortices \eqref{eq:vortexlines}, the variations are 
substantial.

2b) The chosen centerline diagnostics in this paper are $h$, $\epsilon_h$, 
$|\omega|=\sqrt{\zeta}$, $h_f$, $\epsilon_\zeta$ and $\zeta_p$, and are arranged into four 
panels. Plus vertical dashed lines in every panel at positions related to local extrema. 
This includes the positions of local $\min(h_f)$, local $\min(\epsilon_h)$ and 
their nearest positions on the opposite loop of the trefoil.
By following and comparing their extrema between the panels and the isosurfaces, 
a picture of the evolution emerges.

The diagnostics that carry the most information at early times are the centerline
positions  of local $\min(h_f)$, $h$-flux minima \eqref{eq:helicity}. 
At the earliest times shown, $t=1.2$ for r1d015 algebraic profile and $t=0.4$ for 
Lamb-Oseen case Gd05, the local $\min(h_f)$ can be matched with several extrema.
Local minima and maxima of the helicity dissipation $\epsilon_h$ and minima 
of the enstrophy production $\zeta_p$ \eqref{eq:enstrophy}, as given in figures 
\ref{fig:T1p2uuoo} and \ref{fig:GuuooT0p4}. For algebraic case r1d015, from $t=1.2$ to 
when reconnection begins, the relative centerline positions of these extrema are stable. 
Allowing the $h<0$ zones on the new lower ring to gradually shed $h<0$ vortex sheets.

In the period $t=1.2$ to 2.4, the relative positions on the Lamb-Oseen centerline profiles are 
not stable.
Figure \ref{fig:GuuooT1p2} at $t=1.2$ has six roughly equivalent positive and 
negative excursions of $\epsilon_h$ around positions of local compression, local 
$\min(\zeta_p)<0$. 
Likely due to local interactions with the instability-induced, oppositely-signed 
patches shown in figure \ref{fig:Go2xzT1p2}. 
Three are associated with the $s_f$ points. The other three with their $s_o$
opposing points.%10

The Lamb-Oseen $s_f$ points return to something akin to normal for the 
budget curves at $t=2.4$ in figure \ref{fig:GuuooT2p4}. However, the damage has been done 
and when reconnection begins at $t=3.6$, the reconnection structures form only between the
$t=1.2$ extrema points.

3) It is these differences in the respective $t\leq2.4$ budgets that determine whether the 
post-reconnection structures are braids or sheets. And whether finite energy dissipation can form. 
Post-reconnection Lamb-Oseen first generates bridges, as at $t=4$ in figure \ref{fig:G3DT4p0}. 
Then progresses to braids at $t=4.4$ in figure \ref{fig:G3DT4p0}. With only a sort-lived growth 
in the enstrophy $Z(t)$ and energy dissipation $\epsilon(t)$ in figure \ref{fig:Gd05dm1ZHnus} 
before $Z$ and $\epsilon$ decay.

This contrasts with the algebraic profiles that do not have this instability, or any excessive 
local compression. And due to this, the helicity transport $h_f$ is able to 
spread $h<0$ along the centerline. From which $h<0$ vortex sheets can be shed as 
the trefoil self-reconnects, as shown in figure \ref{fig:3DT3p6}a,c % and \ref{fig:3DT3p6turn} 
at $t=3.6$ and figures \ref{fig:3DT4p8} and \ref{fig:3DT4p8low} at $t=4.8$.
Figure \ref{fig:3DT6} at $t=6$ shows how those sheets, when interacting, can allow the enstrophy 
growth to accelerate and convergent energy dissipation rates $\epsilon$ to be achieved.  
Leading to evidence for a dissipation anomaly with finite $\Delta E_\epsilon$ \eqref{eq:dissanom}. 
With the only evidence for bridges or braids from the algebraic calculation
coming from internal higher-$\omega$ isosurfaces, as in
figure \ref{fig:3DT6}.

%\newpage
\subsection{Discussion \label{sec:discuss}}

The centerline budget diagnostics introduced here will next be applied to extensions, 
or variations upon, two existing calculations. First, extensions of the earlier, 
perturbed trefoils in very large domains \cite{KerrJFMR2018} to higher Reynolds numbers and 
later times. Second, versions of recent calculations of interacting orthogonal vortices
\cite{OstilloMetal2021}. For both, approximately convergent $\nu$-independent dissipation 
rates $\epsilon=\nu Z$ develop after the interacting vortices flatten, $\nu$-independent 
convergent $\sqrt{\nu}Z$ is observed at $t_x$  and the sheets wrap around one another. 

On the orthogonal isosurfaces, the mapped helicity 
indicates that within that wrapping, the vortex stretching is vortical. Observations 
that are consistent with  the Lundgren spiral vortex model \cite{Lundgren1982} for 
generating a -5/3 energy spectrum. At the time (circa 1982), a mechanism for creating 
wrapped and stretched vortex sheets within a turbulent flow had not been demonstrated.
Although in retrospect, this is probably what stills \cite{Kerr1985} taken from the 
earliest color, three-dimensional animations of interacting vortices 
are showing.%}bcyan
\bpurp{and need to investigated further. Which, for one of the new perturbed trefoils, is found.}

The recent orthogonal vortices \cite{OstilloMetal2021} were initialized with a 
Lamb-Oseen profile, and did not develop $t\gtrsim0$  negatively-signed ghost vortices. 
Probably because those vortex tubes were not curved, but straight, so were not 
modified by the solenoidal projection as in initialization step 4 in section 
\ref{sec:config}. Meaning, they lacked a perturbation on their outer 
edge similar to that in figure \ref{fig:Gt0Rht0omyz}. %5  
With the only perturbations being inherently numerical and tiny. The additional 
analysis \cite{GallaySmets2020} given after stating the stability function 
$J(\rho)$ \eqref{eq:Jrr} for columnar vortices \cite{HowardGupta1962} says
that tiny perturbations should not generate strong
instabilities. That is, if a Lamb-Oseen profile is applied to straight vortex tubes,
there will not be any instabilities capable of generating negatively-signed ghosts 
like those in figure \ref{fig:Go2xzT1p2} and earlier work \cite{Kerr2013a}.
%analysis that began by finding the stability function $J(\rho)$ \eqref{eq:Jrr} implied by the early analysis 

{\bf Other Lamb-Oseen calculations.} %Twist and coils
In the recent review \cite{YaoHussain2022} of the state of numerical vortex 
reconnection, a reconnection-to-bridges to braids cascade paradigm was presented based 
upon the results from Lamb-Oseen profile calculations, without any examples given of a 
second step in that cascade. 
\bpurp{Neglecting examples with contrived multiple symmetries.}%\bpurp 
Given the contrasting enstrophy evolution of the algebraic 
calculations, how should that paradigm be changed?
\bpurp{What effect would replacing the in the recent review
\cite{YaoHussain2022} with results from calculations using algebraic profiles
have upon reconnection-to-bridges to braids cascade paradigm presented there?}%\bpurp
% The changes are substantial, with the alternative being a two-step process instead of a cascade. First the period until the first reconnection ends at tx. Then the period tx < t≤ tε ≈ 2tx that leads to convergent ε = νZ, with large ε persisting for a finite time, and thus generating finite

The changes are substantial, with the algebraic alternative being a two-step process instead of 
a cascade. First the period that ends at $t_x$ with $\sqrt{\nu}Z(t)$ convergence, generation
of $h<0$ vortex sheets and completion of the first reconnection. Next 
the period $t_x<t\lesssim t_\epsilon\approx  2t_x$ during which the sheets wrap around one
another, leading to convergent $\epsilon=\nu{Z}$. As that large $\epsilon$ persists, 
finite-time, finite $\Delta E_\epsilon$ \eqref{eq:dissanom} forms.

Furthermore, because that review \cite{YaoHussain2022} focuses upon their recent trefoil 
calculation \cite{YaoYangHussain2021} as the latest support for the reconnection-to-braids 
paradigm, it is fair to ask whether the instabilities identified here extend to all the cited 
Gaussian/Lamb-Oseen calculations in that review.

They probably do, going back to the first in 1989 \cite{MelanderHussain89}. The
effects of such instabilities were first
clearly identified for an Euler calculation using an elongated Gaussian 
profile \cite{BustaKerr2008} and were then clarified by 2013 
anti-parallel analysis \cite{Kerr2013a} that shows $t\sim0$ $\omega=0$ contours 
that are  more intense than those in figure \ref{fig:Go2xzT1p2}.
If the authors of that recent review \cite{YaoHussain2022} disagree with the analysis 
behind that conclusion, what would be useful would be a submission to 
Physical Review Fluids that applied the centerline diagnostics introduced here 
to another one of their recent calculations.
\bpurp{It should be noted that that 2021 JFM paper, 
which ignored the 2018 evidence of convergent dissipation rates \cite{KerrJFMR2018}, 
is indirectly commenting upon my published work in JFM \cite{KerrJFMR2018}.}

\bcyan{Some of the consequences of the instability of the Lamb-Oseen profile were
previously noted for perturbed anti-parallel vortices 
simulated with the inviscid Euler equations \cite{BustaKerr2008}, although
not properly described until \cite{Kerr2013a}. That description  emphasized the
importance of the small, oppositely-signed $t=0$ ghost vortices (as in figure
\ref{fig:Go2xzT1p2}) that appear after the raw vorticity field is run through a 
Poisson solver (step 4 in section \ref{sec:config}).  The importance of the ghosts 
is that they redirect the nonlinear growth around the vorticity maximum. 
Because previous attempts to massage the Gaussian profile 
\cite{Kerr1993,BustaKerr2008} failed to remove the ghost vortices, 
in the next anti-parallel Euler paper \cite{Kerr2013c}, the regularized $p_r$=1
algebraic profile (mislabeled Rosenhead) was used.}%\bcyan

\section*{Acknowledgements} I would like to thank the Isaac Newton Institute for 
Mathematical Sciences for support and hospitality during the programme 
{\it Mathematical Aspects of Fluid Turbulence: where do we stand?} in 2022, when work 
on this paper was undertaken and supported by grant number EP/R014604/1. Including
interactions with, among others, A. Leonard and M. Musso. I thank E. Brambley at 
Warwick for clarifying crucial elements during the final preparation. Computing 
resources have been provided by the Scientific Computing Research Technology Platform
at the University of Warwick.

\newpage
\appendix
%\section*{Appendices}
\section{\label{sec:r1d015dm} Results for cut-off Rosenhead profiles.}

In this appendix the evolution of an algebraic profile with a severe cut-off at $\rho_+=0.025$.
The objective is to demonstrate that steep cut-offs can be as much of a problem as the 
chosen profile. The minimum resources were expended and detailed analysis (3D graphics) is
not provided. This case behaves in many ways more like the Lamb-Oseen profile than the other 
algebraic profiles.  There is growth of the enstrophy $Z(t)$ over $t_r=4<t<6$ as in algebraic in 
figure \ref{fig:r1ZHnu}. But growth of $Z(t)$ is then suppressed as ${\cal H}(t)$ grows,
more like Lamb-Oseen in figure \ref{fig:Gd05dm1ZHnus}.

\begin{figure}
\hspace{-10mm}\bminic{0.5}
\includegraphics[scale=0.22,clip=true,trim=20 450 0 0]{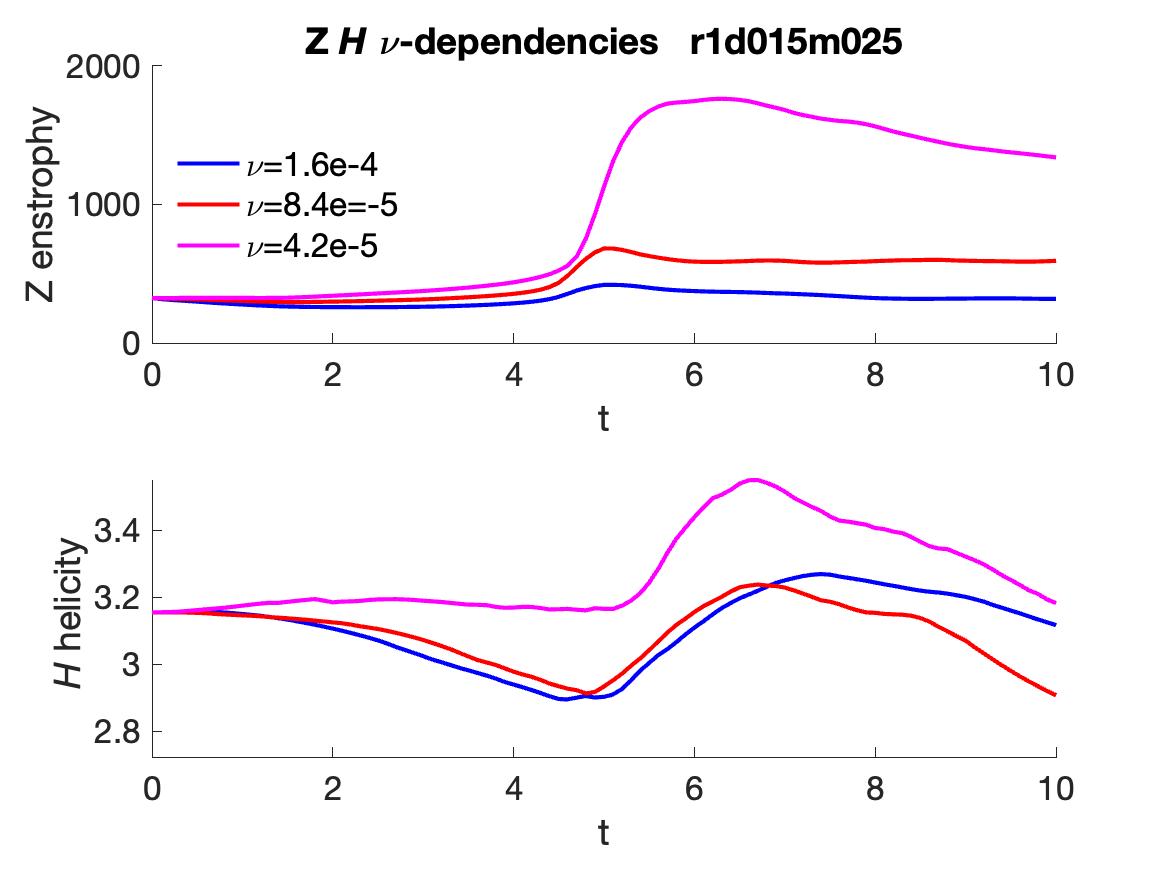}
\emini \bminic{0.45} % Is it t=0 or  t=0.4?
\includegraphics[scale=0.22,clip=true,trim=20 0 0 450]{XYyld015dm025nuZH21mar23.jpg}
\emini
\caption{\label{fig:d015dm025ZHzprof} Cut-off Rosenhead time evolution.
Left: Enstrophy Right: Helicity. Like algebraic in figure \ref{fig:r1ZHnu} the enstrophy
grows as opposing segments are interacting for $t_r=4< t < 6$. However after that, like
Lamb-Oseen in figure \ref{fig:Gd05dm1ZHnus}, $Z(t)$ then decays. And there is more persistent
growth of ${\cal H}(t)$ like Lamb-Oseen.
}
\end{figure}

\begin{figure}
\includegraphics[scale=0.40,clip=true,trim=30 100 0 0]{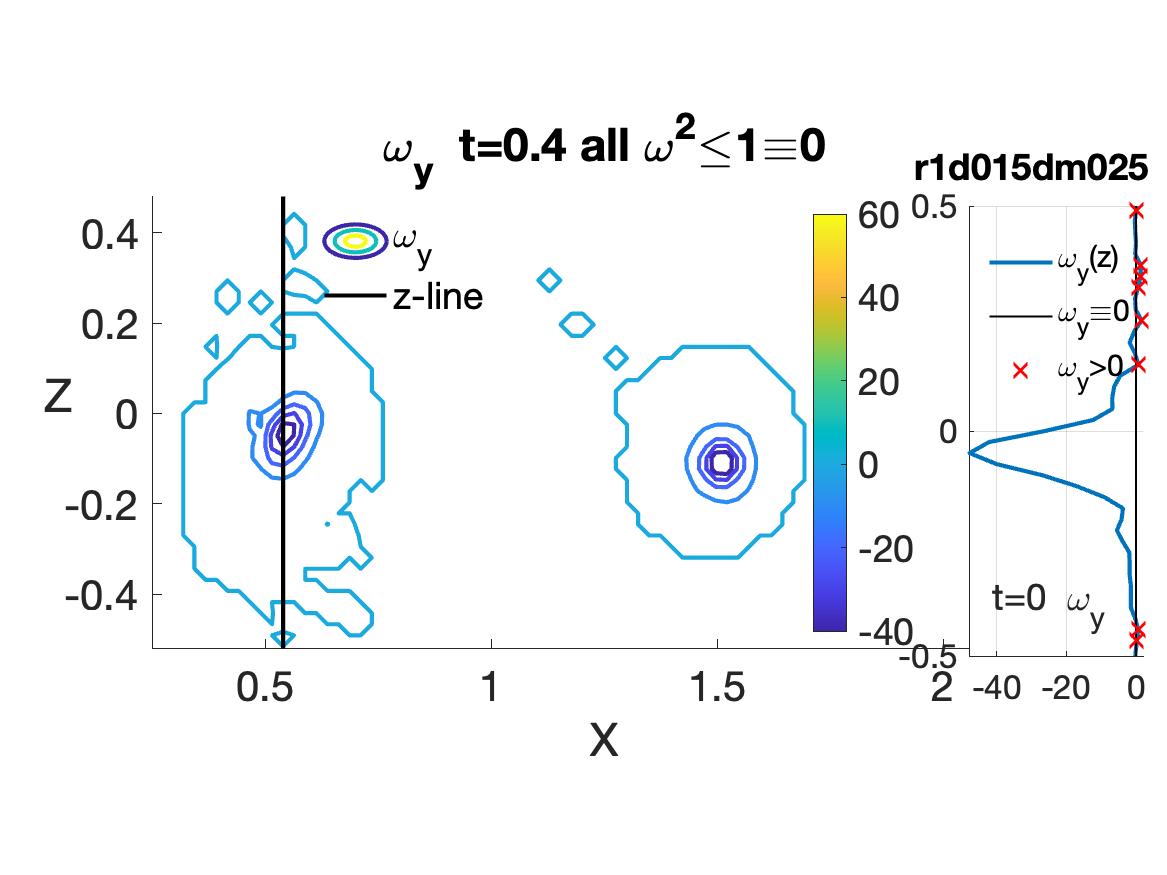}
\caption{\label{fig:d015dm025T0p4} Cut-off $\omega_y$ Rosenhead $x\!-\!z$.
Left: $t=0.4$ cross-section with vertical line through $\min(\omega-y)$. Note several contours
of $\omega_y\equiv0$.
Right: Profile of $\omega_y$ through $t=0$ line through $\min(\omega_y)$. Note
several positions in red where there oppositely signed ($\omega_y>0$) spots. Similiar to
what is found for Lamb-Oseen in figure \ref{fig:3Dr11p2G2pi}.}\end{figure}
%XYyld015dm025nu084m512o2xzT0p4
%\begin{figure}
%\includegraphics[scale=0.40,clip=true,trim=30 100 0 100]{XYyld015dm025nu084m512o2xzT2p8-21mar23.jpg}
%\caption{\label{fig:d015dm025T2p8} Cut-off $\omega_y$ Rosenhead $x\!-\!z$ at $t=2.8$.
%$\omega_y\equiv0$ contours dominate the cross-section.}\end{figure}

\begin{figure}
\includegraphics[scale=0.29,clip=true,trim=30 30 0 30]{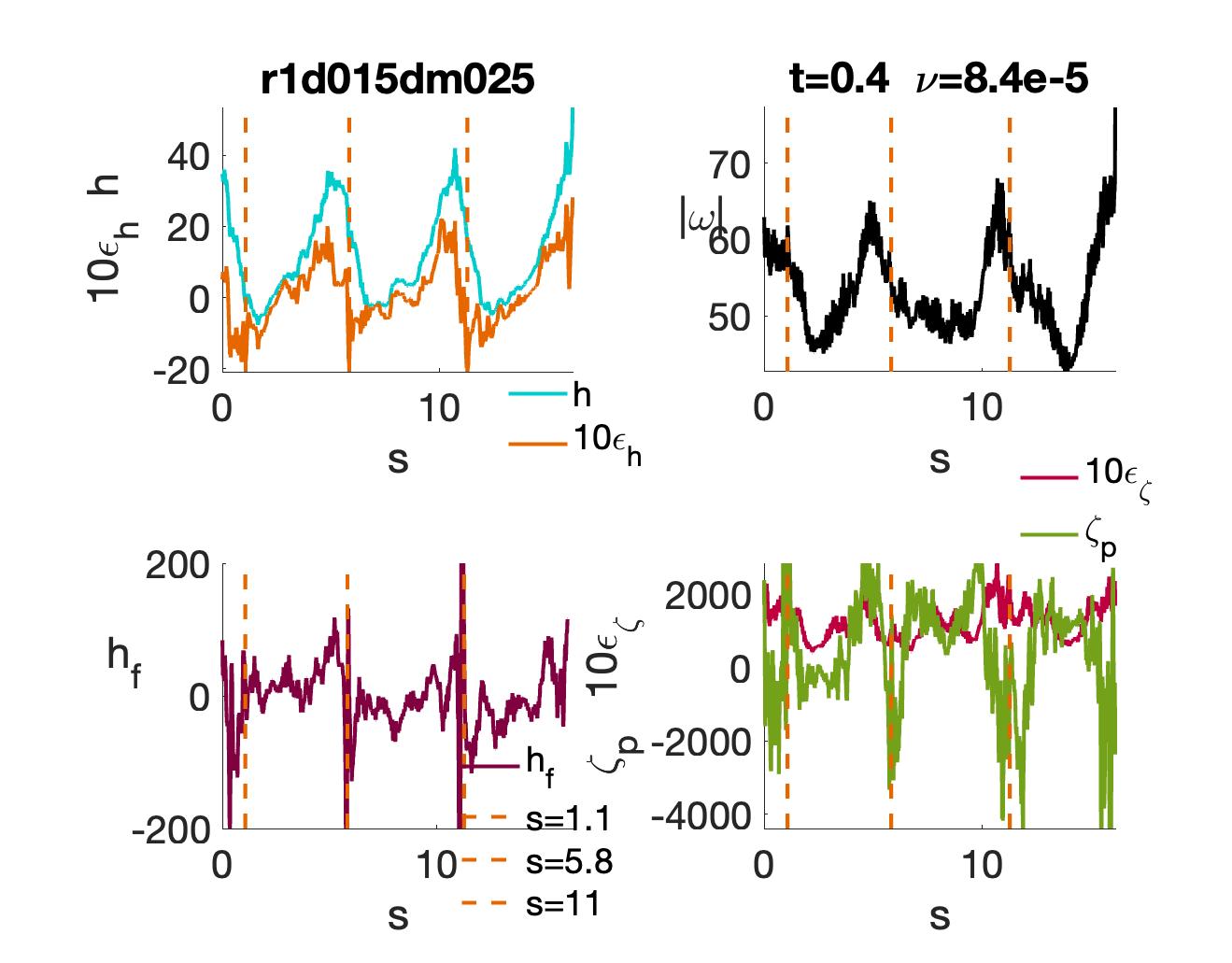} \\
\caption{\label{fig:d015dm025suoT0p4} Cut-off $\omega_y$ Rosenhead $x\!-\!z$ at $t=0.4$
vorticity centerline profiles of $h$, $h_d$, $|\omega|$, $h_f$, $\epsilon_Z$ and $Z_p$.
In most respects this is like early times for both the Lamb-Oseen and
algebraic cases, Lamb-Oseen at $t=0.4$ and untruncated algebraic at $t=1.2$
in figures \ref{fig:GuuooT0p4} and \ref{fig:T1p2uuoo} respectively. 
There are three positions of significance.  Chosen by local negative helicity 
dissipation $\min(\epsilon_h)$ but also at or near local 
$\min(h_f)$ of the helicity flux and local compression $\min(\zeta_p)$.}
\includegraphics[scale=0.29,clip=true,trim=30 30 0 30]{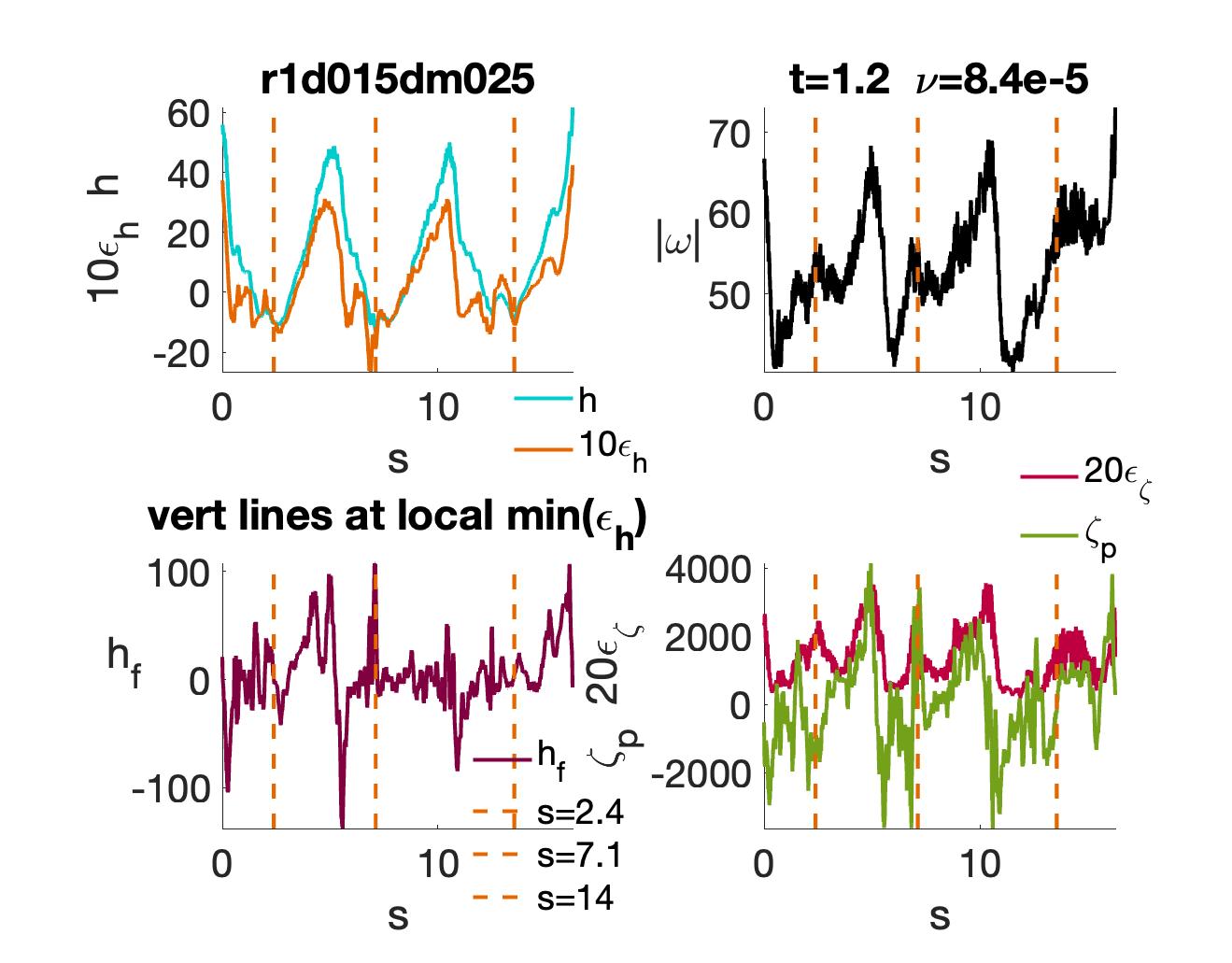}
\caption{\label{fig:d015dm025suoT1p2} Cut-off $\omega_y$ Rosenhead $x\!-\!z$ 
Vorticity centerline profiles at $t=1.2$. This is unlike the regular centerline 
profiles at $t=0.4$ just given and unlike the untruncated algebraic at 
$t=1.2$ in figure \ref{fig:T1p2uuoo}, the relationships between the local
$\min(u_d)$ positions and the other properties are muddled.
However, it is not as extreme as the Lamb-Oseen
six-fold symmetry in figure \ref{fig:GuuooT1p2}.
}
\end{figure}

%\bminic{0.03}\vspace{-90mm}
%\rotatebox{90}{\LARGE\bDF{Role of sub-km turbulence}} \emini 

\end{document}